\documentclass[pdflatex,sn-mathphys-num]{sn-jnl}%

\usepackage{graphicx}%
\usepackage{multirow}%
\usepackage{amsmath,amssymb,amsfonts}%
\usepackage{amsthm}%
\usepackage{mathrsfs}%
\usepackage[title]{appendix}%
\usepackage{xcolor}%
\usepackage{textcomp}%
\usepackage{manyfoot}%
\usepackage{booktabs}%
\usepackage{algorithm}%
\usepackage{algorithmicx}%
\usepackage{algpseudocode}%
\usepackage{listings}%

\usepackage{mathtools}
\usepackage{tikz,enumerate,array}
\usepackage{epsfig}
\usepackage{amsbsy}

\theoremstyle{thmstyleone}%
\newtheorem{theorem}{Theorem}
\newtheorem{proposition}[theorem]{Proposition}% 

\theoremstyle{thmstyletwo}%
\newtheorem{remark}{Remark}%

\theoremstyle{thmstylethree}%
\newtheorem{definition}{Definition}%

	\definecolor{darkerblue}{RGB}{0, 163, 211}
\newcommand{\N}{{\mathbb{N}}}

\newcommand{\R}{{\mathbb{R}}}

\newcommand{\C}{{\mathbb{C}}}

\newcommand{\sigmaess}{{\sigma_{\text{ess}}}}

\newcommand{\ol}{\overline}

\newcommand{\wti}{\widetilde}
\newcommand{\Oh}{O}

\newcommand{\hatt}{\widehat}
\newcommand{\beq}{\begin{equation}}
\newcommand{\eeq}{\end{equation}}
\newcommand{\bdm}{\begin{displaymath}}
\newcommand{\edm}{\end{displaymath}}
\newcommand{\ba}{\begin{align}}
\newcommand{\ea}{\end{align}}
\newcommand{\bpf}{\begin{proof}}
\newcommand{\epf}{\end{proof}}

\newcommand{\la}{\langle}
\newcommand{\ra}{\rangle}

\newcommand{\supp}{\mathrm{supp}\, }               % support
\newcommand{\dist}{\mathrm{dist}}               % distance

\newcommand{\e}{\mathrm{e}}
\renewcommand{\d}{\mathrm{d}}

\newcommand{\veps}{\varepsilon}

\newcommand{\re}{\mathrm{Re}}

\newcommand{\sgn}{{\text{sgn}}}
\newcommand{\id}{\mathbf{1}}                % identity

\newcommand{\ind}{\mathbf{1}}
\allowdisplaybreaks
\newcommand{\calC}{\mathcal{C}}
\newcommand{\calD}{\mathcal{D}}

\newcommand{\locerror}{{L_{\text{err}}}}

\newtheorem{lemma}[theorem]{Lemma}
\newtheorem{corollary}[theorem]{Corollary}

\theoremstyle{definition}

\newcounter{theoremi}[theorem]

\numberwithin{theorem}{section}
\numberwithin{equation}{section}
\numberwithin{figure}{section}

\newcounter{smallenum}
\newenvironment{smallenum}{\begin{list}{{\rm\arabic{smallenum})}}{%
\setlength{\topsep}{0mm}\setlength{\parsep}{0mm}\setlength{\itemsep}{0mm}%
\setlength{\labelwidth}{2em}\setlength{\leftmargin}{2em}\usecounter{smallenum}%
}}{\end{list}}
\begin{document}

\title[Quantum Systems at The Brink]{Quantum Systems at The Brink: Helium--type Systems }

\author*[1,2]{\fnm{Dirk} \sur{Hundertmark}}\email{dirk.hundertmark@kit.edu}
\equalcont{These authors contributed equally to this work.}

\author[3]{\fnm{Michal} \sur{Jex}}\email{michal.jex@fjfi.cvut.cz}
\equalcont{These authors contributed equally to this work.}

\author[4]{\fnm{Markus} \sur{Lange}}\email{markus.lange@dlr.de}
\equalcont{These authors contributed equally to this work.}

\affil*[1]{\orgdiv{Department of Mathematics}, \orgname{Karlsruhe Institute of Technology}, \orgaddress{\street{Englerstraße 2}, \city{Karlsruhe}, \postcode{76131}, \country{Germany}}}

\affil[2]{\orgdiv{Department of Mathematics}, \orgname{University of Illinois at Urbana-Champaign}, \orgaddress{\street{1409 W. Green Street}, \city{Urbana}, \postcode{IL 61801}, \state{Illinois}, \country{USA}}}

\affil[3]{\orgdiv{Faculty of Nuclear Sciences and Physical Engineering}, \orgname{Czech Technical University in Prague}, \orgaddress{\street{Břehová 7}, \city{Praha}, \postcode{11519}, \country{Czechia}}}

\affil[4]{\orgdiv{Institute for AI-Safety and Security}, \orgname{German Aerospace Center (DLR)}, \orgaddress{\street{Wilhelm-Runge-Straße 10}, \city{Ulm}, \postcode{89081}, \country{Germany}}}

\abstract{In the present paper we study two challenging problems for 
helium--type systems. Existence of eigenvalues at thresholds and 
the asymptotic behavior of the corresponding eigenfunctions. 
Since the usual methods for addressing these problems need a 
safety distance to the essential spectrum, they cannot be 
applied in critical cases, when an eigenvalue enters the continuum. 
 
 We develop a method to address both problems and   
 derive \emph{sharp upper and lower bounds} for the 
 asymptotic behavior of the ground state of 
 critical helium--type systems at the threshold of 
 the essential spectrum. 
 This is the first proof of the precise asymptotic behavior of the 
 ground state for this benchmark problem in quantum chemistry. 
 Moreover, our bounds describe precisely how the asymptotic decay of the 
 ground state changes, when the  system becomes critical. }

\keywords{eigenfunctions at the threshold, decay rates of eigenfunctions, weakly bounded states, ground state of atoms}

\maketitle
\section{Introduction}
From the beginning of quantum mechanics, many important questions 
about a quantum system are related to the existence and 
behavior of its bound states. These states are given by the square 
integrable eigenvectors of the operator describing the 
quantum system. In this paper we consider two particle operators
of the form 
\begin{equation}\label{eq:HeliumHamiltonian}
	H_U = P_1^2 + P_2^2 - \frac{1}{|x_1|} - \frac{1}{|x_2|} + \frac{U}{|x_1 - x_2|}\, .
\end{equation}
Here $x_j\in\R^3$ are the positions of the two particles, and $P_j^2$ their kinetic energy, where 
$P_j=-i\nabla_{x_j}$ is the momentum operator of the particle and  $j= 1,2$. The Hamiltonian $H_U$ is 
well-defined and self-adjoint on $\calD(H_U)=H^2(\R^6)$, where $H^2(\R^6)$ is the usual Sobolev space 
of weakly differentiable functions in $L^2(\R^6)$ with square integrable weak derivatives up to second order. 

The operator $H_U$ arises from scaling two--particle helium--type systems described by the operator  
\begin{equation*}
	H= \sum_{j=1}^2\left( \frac{1}{2m} P_j^2- \frac{Ze^2}{|x_j|} \right) + \frac{e^2}{|x_1-x_2|}
\end{equation*}
where  $Z$ is the nuclear charge and $e$ the unit for the 
elementary electric charge. Here we use units, 
where other physical constants such as Planck's constant 
$\hbar$ are set equal to one. 
Using scaling, i.e., a change of length--scale, 
given by $U_s\psi(x)= s^3 \psi(s x)$, $x\in\R^6$, 
which is unitary on $L^2(\R^6)$, one gets with $s=2mZe^2$ and $U=1/Z$
\begin{align*}
  	U_s^* H U_s = 2mZ^2e^4 H_U \, .
\end{align*}
Thus  $H_U$ with $U=1/Z$ is the operator describing helium-type systems such as 
$Li^+$ for $Z=3$, $He$ for $Z=2$, and $H^-$ for $Z=1$.

We denote the ground state energy of $H_U$ by $E_U$. 
More precisely, $E_U=\inf_{\|\psi\|=1}\la \psi, H_U\psi\ra$, 
which is concave and increasing in $U\ge 0$, since 
$H_U$ is linear in $U$ and the two particle Coulomb repulsion is positive. Hence  
the ground state energy $E_U$ is continuous in $U\ge 0$. 
Of course, one should also include the spin of the particles, e.g., electrons, in which case one should consider 
$H_U$ on $L^2((\R^3\times\C^2)^2)$, or the antisymmetric subspace of functions which are antisymmetric 
when permuting the particles. Since the potential does not couple different spins, all bound states of 
$H_U$ on $L^2((\R^3\times\C^2)^2)$ can be classified as two particle bound states with parallel or 
antiparallel spins. In the first case the wave functions is antisymmetric, in the second case it is symmetric 
under permutation of the  particle \emph{positions}. The ground state of $H_U$ on $L^2((\R^3\times\C^2)^2)$ 
lies in the latter, so it is enough to consider $H_U$ on the subspace of $L^2(\R^3\times\R^3)$ which is 
symmetric under permutation of the particle positions, 
see the discussion in \cite[Chapter 4.3]{Thi02}.

The HVZ Theorem \cite{Zhi60}, \cite[Theorem XIII.17]{ReeSim4},  
shows that the essential spectrum 
of $H_U$ is a half--line whose bottom is given by the ground state energy of the system with one 
less electron, i.e., hydrogen whose ground state energy is $-1/4$. 
Thus 
\begin{align*}
	\sigmaess(H_U)=\big[-\frac{1}{4},\infty\big) 
\end{align*}
for any $U\ge 0$. Since usual perturbation theory applies when $E_U<\inf\sigmaess(H_U)=-1/4$, the 
regular first order perturbation theory \cite{ReeSim4}, also called the Feynman-Hellmann formula, gives 
\begin{equation}
	\frac{d}{dU} E_U =\la\psi_U, (\partial_U H_U) \psi_U \ra = \la\psi_U, \frac{1}{|x_1-x_2|}\psi_U \ra >0
\end{equation}
for all $U\ge 0$ with  $E_U<-1/4$. 

Since the work of Stillinger \cite{Sti66}, 
see also \cite{StiSti74},  these helium--type atoms at critical coupling are an 
intensely studied benchmark problem in quantum chemistry. For a review  see, e.g., 
\cite{GraBur15}
and the references therein. In particular, there has been 
interest in the precise value of the critical $U_c$, for which the ground state energy of $H_U$ enters the edge of 
the essential spectrum, that is, $U_c$ is the smallest coupling for which $E_U=-1/4$. Numerically one finds $U_c\simeq 1.1$ \cite{Sti66}. The calculation of 
$U_c$ and the bound state at critical coupling $U=U_c$ continues to serve as an important  
benchmark problem in quantum chemistry.  E.g., going up to order $401$ in perturbation theory, the critical 
value was calculated in the very nice paper  
\cite{BakFreHilMor90} to be approximately $U_c\simeq 1.097 66$, which was further pushed in \cite{Iva95}. 
A variational calculation of $U_c$ was done in \cite{SerKai99}.
Recently  \cite{BusDraEstMoi14}
pushed the calculations to $U_c\simeq 1,097\, 660\, 833\, 738\, 56$. 
Without the Born--Oppenheimer approximation the critical 
coupling, which depends on the nuclear mass, was 
calculated numerically in \cite{King15}.

It is easy to see rigorously that there is a critical 
$1<U_c\le 2$ such that  $E_{U_c}= \inf\sigmaess(H_U)$. 
 As discussed in Appendix \ref{app:basic properties}, $E_U<-1/4$ implies $U<2$ for this two-particle system. So clearly 
 $U_c\le 2$. To see that $U_c>1$ we note that 
the classical result by Bethe \cite{Bet29}, using a test function ansatz due to Hylleras \cite{Hyl29}, 
shows that hydrogen can bind two electrons, i.e., $H_U$ has a ground state with energy $<-1/4$ for $U=1$. 
Hill showed that this is also true without the 
Born--Oppenheimer approximation, as long as the mass of the nucleus is not too small compared to the mass of the lighter 
particles.  So by continuity of $E_U$ in $U$, one knows that 
$E_U$ is below the ionization threshold even for some $U>1$, 
so $U_c>1$. Moreover, the work of Hill \cite{Hil76BO} 
 shows that for $1\le U<U_c$ the operator $H_U$ has 
exactly one bound state with energy below the essential spectrum.  
This also holds for finite nuclear mass \cite{Hil77nonBO}.  

Although the Hamiltonian $H_U$ with $U=U_c$ has been intensely studied with asymptotic and numerical methods,  little is known rigorously.  
The existence of a ground state $\psi_c$ of $H_{U_c}$ at critical coupling 
was proved in \cite{HofOstHofOstSim83} with the help of PDE methods. 
An alternative existence proof was given in \cite{FraLieSei12}. 
General features of the behavior of the ground state energy 
of quantum systems near coupling constant threshold had been discussed, for example, in \cite{KlSi-1, KlSi-2}.  
Unfortunately, still very little is known so far 
concerning \emph{precise quantitative properties} of the ground state $\psi_c$ \emph{at critical coupling}. 

\smallskip

This is the main motivation for our work. We provide sharp upper and lower bounds on the asymptotic behavior of the ground state of helium-type atoms at critical coupling. Moreover, we prove a family of sharp upper and lower bounds which are uniform in the coupling $0\le U\le U_c$. 
Our results provide the first rigorous bounds on the asymptotic behavior of bounds states which correspond to eigenvalues \emph{at the edge of the essential spectrum}. In addition, even in the subcritical case our upper and lower bounds improve on previously known results, since our upper and lower bounds have \emph{the same leading order} and differ only in lower order terms.  
Our main results are

\begin{theorem}[Global upper bound at critical coupling]\label{thm:global pointwise anisotropic upper bound}
	Given parameters $K_1,K_2>0$, $1/6 <\kappa_1<1/2$, and 
	$(3-2\kappa_1)/4<\kappa_2<1$ define 
	\begin{align}\label{eq:F+}
		F_+(r_1,r_2)= 2(U_c-1)^{1/2}r_1^{1/2} - K_1 r_1^{\kappa_1} 
						 	+\frac{1}{2} r_2- K_2 r_2^{\kappa_2}
	\end{align}
	for $r_1,r_2\ge 0$. Then at critical coupling $U=U_c$ the helium atom has a ground state $\psi_c$ with energy 
	$-1/4$ and we have the pointwise upper bound 
	\begin{align}\label{eq:global upper bound pointwise intro}
		\psi_c(x)\le C_u   
					\exp\!\big(\! -F_+(|x|_\infty,|x|_0) \big)
	\end{align}
	for the unique positive ground state where the constant $C_u$ depends only on 
	$\kappa_1$, $\kappa_2$, $K_1$, $K_2$. 
	
	Here $|x|_0=\min(|x_1|,|x_2|)$, respectively $|x|_\infty=\max(|x_1|,|x_2|)$, is the distance of the particle 
	closer to, respectively farther from, the nucleus.
\end{theorem}
The matching \emph{lower bound} is provided by 
\begin{theorem}[Global lower bound at critical coupling]\label{thm:global pointwise anisotropic lower bound}
	Given parameters $1/6<\kappa_1<1/2<\kappa_2<1$, and $K_1,K_2>0$ define 
		\begin{align}
		F_-(r_1,r_2)= 2(U_c-1)^{1/2}r_1^{1/2} + K_1 r_1^{\kappa_1} 
						 	+\frac{1}{2} r_2 + K_2 r_2^{\kappa_2}\, .
	\end{align}
	Then we have the pointwise  lower bound  	
	\begin{align}\label{eq:global lower bound pointwise}
		\psi_c(x)\geq C_l
					\exp\!\big( -F_-(|x|_\infty,|x|_0) \big)
	\end{align} 
	for the unique  positive ground state $\psi_c$ of $H_{U_c}$ at critical coupling where the constant $C_l$ depends only on 
	$\kappa_1$, $\kappa_2$, $K_1$, $K_2$.
	Here again $|x|_0=\min(|x_1|,|x_2|)$ and 
	$|x|_\infty=\max(|x_1|,|x_2|)$.
	\end{theorem}

\begin{remark}
 For $1/6<\kappa_1< 1/2$ we have 
 $1/2< (3-2\kappa_1)/4<2/3$, so the lower bound in Theorem \ref{thm:global pointwise anisotropic lower bound} holds for a slightly larger range of parameters than the upper bound from Theorem \ref{thm:global pointwise anisotropic upper bound}.
 Up to the sign of lower order terms $r_1^{\kappa_1}$ and $r_2^{\kappa_2}$,  our upper and lower bounds from Theorem \ref{thm:global pointwise anisotropic upper bound} 
 and Theorem \ref{thm:global pointwise anisotropic lower bound} perfectly match each other. Thus we get the \emph{precise leading order coefficients} for the asymptotic decay of the ground state for the helium--type systems, unlike previous methods. 
 
 Strictly speaking, the constant $C_u$ in the upper 
 bound \eqref{eq:global upper bound pointwise intro} also depends on $\|\psi_c\|_\infty$. Since the Coulomb 
 potential is in the Kato--class, for a definition see 
 \cite{AizSim82, CycFroKirSim87, Simon-semigroups}, an 
 a--priori bound on 
 $\|\psi_c\|_\infty$ is well--known, see 
 \cite{Simon-semigroups}, as soon as the ground state $\psi_c$ exists. 
 
 In a similar fashion, the constant $C_l$ in the lower 
 bound \eqref{eq:global lower bound pointwise} also 
 depends on local lower bounds for the ground state 
 $\psi_c$. Once the ground state exists it is unique and can be chosen to be strictly 
 positive \cite{Goe77}, up to a global phase. Since the Coulomb potential is in the Kato--class, every eigenfunction has a continuous version \cite{Simon-semigroups}, so by positivity the ground state $\psi_c$ is locally bounded away from zero as soon as it exists.
 
 The existence of a ground state at critical coupling was shown in 
 \cite{HofOstHofOstSim83} by PDE methods. 
 The upper bound from Theorem \ref{thm:global pointwise anisotropic upper bound}, even a much simpler version, 
 see Theorem \ref{thm:isotropic upper bound}, allows for a simple variational proof of \emph{existence} of a 
 ground state of helium--type atoms at critical coupling similar to \cite{FraLieSei12}.
 We can even avoid the Born--Oppenheimer approximation, see Remark \ref{rem:nonBO}. 
\end{remark}

Of course, while being sharp, the above upper and lower bounds hold only at critical coupling. We also have upper and lower bounds which hold uniformly in the  coupling. 
For $\veps$ and $a\ge 0$ we define 
\begin{equation}\label{eq:interpolating F}
    F_{\veps,a}(r) 
     = 
      \Big( \veps +\frac{a^2}{r}\Big)^{1/2}r 
      + \frac{a^2}{\sqrt{\veps}} 
        \big(\ln( \sqrt{\veps r+a^2} +\sqrt{\veps r} ) -\ln a   \big)
\end{equation}
For $0\le U \le U_c$ we define $a_U$ and $\veps_U$ by 
\begin{align}
	a_U\coloneqq (U-1)_+^{1/2}, \quad  \veps_U\coloneqq -\frac{1}{4} - E_U
\end{align}
so $\veps_U$ is the ionization energy, the distance of the energy 
$E_U$ to the edge of the essential spectrum, which is at $-1/4$. 
By the HVZ theorem, this is the minimal energy needed to move 
one particle to infinity. Note that at critical coupling $U=U_c$ the 
ionization energy vanishes.  
Furthermore, we set  
\begin{equation}\label{eq:F_U}
  \begin{split}
	F_U(r)\coloneqq F_{\veps_U,a_U}(r) 
	  = \Big( \veps_U +\frac{a_U^2}{r}\Big)^{1/2}r 
	    + \frac{a_U^2}{\sqrt{\veps_U}} 
        \big(\ln( \sqrt{\veps_U r+a_U^2} +\sqrt{\veps_U r} ) -\ln a_U   \big)
  \end{split}
\end{equation}
and 
\begin{equation}
  \begin{split}
	F^U_\pm(r_1,r_2)\coloneqq &  
		 	F_U(r_1) +\frac{1}{2}r_2 \mp\left( K_1r_1^{\kappa_1} + K_2 r_2^{\kappa_2} \right)\, .
  \end{split}
\end{equation}
Again, we suppress the explicit dependence of $F^U_\pm$ on the parameters 
$\kappa_1,\kappa_2$ and $K_1,K_2$, for notational simplicity. We also set 
$F^U_\pm(x)= F^U_\pm(x_1,x_2)= F^U_\pm(|x|_\infty,|x|_0)$ for $x=(x_1,x_2)\in \R^3\times\R^3$, with a slight 
abuse of notation. 
\begin{remark}\label{rem:interpolating-intro}
  The heuristic reason why $F_{\veps_U,a_U}$ 
  given in \eqref{eq:F_U} captures the asymptotic decay 
  of the ground state $\psi_U$ up to and including 
  criticality will be explained shortly after Theorem 
  \ref{thm:sharp upper and lower bounds all coupling}.
  The precise form of $F_{\veps,a}$ is of no relevance, what is important is that 
  \begin{align}
  	F_{\veps,a}'(r) = \left(\veps_U +\frac{a_U^2}{r} \right)^{1/2}\, .
  \end{align}
One can show that 
\begin{equation}
    \lim_{\veps\to 0} F_{\veps,a}(r) = 2a r^{1/2}\, .  
\end{equation}
In order to get  
the correct value of $F_{\veps,a}$ in the limit $\veps\to 0$,   
one cannot ignore the logarithmic term in \eqref{eq:interpolating F}, 
which for fixed $\veps>0$ is \emph{always of much lower order} 
compared to the first term.   
In the limit of vanishing ionization energy the second term 
gives the \emph{same contribution} as the first term 
and should not be discarded, see the discussion in Remark \ref{rem:interpolating}.  

Thus one recovers $F_\pm$ from $F^U_\pm$ in the limit of  critical coupling, when the ionization energy  vanishes.  Clearly, as long as  the ionization energy  $\veps_U$ 
is  positive, the leading order behavior of $F^U_\pm$ is asymptotic to  $F_U(r_1)\sim(\veps_U+ a_U^2/r_1)^{1/2}r_1\sim \sqrt{\veps_U}r_1$, but it changes 
to  $2 a_U\sqrt{r_1}$ \emph{at critical coupling} $U=U_c$.

\end{remark}
\begin{theorem}[The transition from subcritical to critical: Sharp upper and lower bounds]\label{thm:sharp upper and lower bounds all coupling}
  For any choice of parameters 
  $K_1,K_2>0$, $1/6 <\kappa_1<1/2$, and 
	$(3-2\kappa_1)/4<\kappa_2<1$ 
  there exist positive constants 
  $C_\pm$ depending only on $\kappa_1,\kappa_2, K_1, $ $K_2$, and some local bounds on the ground state $\psi_U$ provided in Proposition \ref{prop:basic properties},   
  such that for the unique positive ground state of the helium-type operator $H_U$ the two--sided pointwise bound 
  \begin{align}\label{eq:sharp upper lower bounds critical range}
    C_-\exp\left( -F^U_-(|x|_\infty,|x|_0) \right)
    	\le \psi_U(x)
    		\le C_+\exp\left( -F^U_+(|x|_\infty,|x|_0) \right)  	
  \end{align}
  holds uniformly in $0\le U\le U_c$. 
  
  If one stays away from the critical parameter, i.e., if for fixed small $\mu>0$ 
  the repulsion parameter $U$ is allowed to vary uniformly 
  in $\mu\le U\le U_c-\mu$, the range of parameters is bigger. 
  More precisely,  assume that $0<\kappa_1<1$ and 
  $1/2< \kappa_2<1$, and $K_1,K_2>0$. Then there 
  exist positive  constants  $\wti{C}_\pm$, depending only on $\kappa_1,\kappa_2, K_1, K_2$, 
  and also $\mu$, such that the two--sided bound  
    \begin{align}\label{eq:one particle sharp upper lower bound subcritical range}
      \wti{C}_-\exp\left( -F^U_-(|x|_\infty,|x|_0) \right)
    	\le \psi_U(x)
    		\le \wti{C}_+\exp\left( -F^U_+(|x|_\infty,|x|_0) \right)  	
    \end{align}
  holds for all  $\mu\le U\le U_c -\mu$. 
\end{theorem}
\begin{remark}
We would like to stress the fact that the constant  
$C_+$ depends only on the parameters 
$\kappa_1,\kappa_2, K_1, K_2$ 
and an a--priori bound on $\|\psi_U\|_\infty$ for 
$0\le U\le U_c$, see part \ref{claim 3} of Proposition \ref{prop:basic properties}. Similarly,  the constant $C_-$ in 
\eqref{eq:sharp upper lower bounds critical range} depends on 
the parameters $\kappa_1,\kappa_2, K_1, K_2$ 
and local lower bounds on the ground state $\psi_U$ 
from part \ref{claim 4} of Proposition \ref{prop:basic properties} which are uniform in $0\le U\le U_c$. 
That is, the constants $C_\pm$ in 
\eqref{eq:sharp upper bound critical range} are indeed 
uniform in $0\le U\le U_c$.  

Similarly as in Theorem \ref{thm:global pointwise anisotropic lower bound}, 
the lower bound in \eqref{eq:sharp upper lower bounds critical range} holds for the slightly larger range of parameters 
$1/6<\kappa_1<1/2<\kappa_2<1$. 
\end{remark}

\smallskip

\noindent 
\textbf{How to guess the function $F_{\veps,a}$:} 
To get an idea what would be the correct ansatz for the decay of the ground state in the regime 
$U\le U_c$, including criticality, one can argue as follows. By the HVZ theorem, the infimum of the essential spectrum is the system with one particle moved to infinity. In our case one is left with one particle, i.e., hydrogen, whose ground state energy is given by $-1/4$ in our units. If $U<U_c$, the ground state energy $E_U$ of the two 
particle system is less than $-1/4$, i.e., when trying to escape to infinity the farthest out particle experiences an energy penalty given by the ionization energy 
\begin{equation}
    \veps_U= -1/4 - E_U>0\, ,
\end{equation}
which is positive.  That is, when trying to escape to infinity, the farthest out particle needs to tunnel into an \emph{energetically forbidden region}. WKB--type arguments then predict that the ground state decays as 
\begin{equation}
    \psi(x)\le C\exp\big(-F(|x|_\infty))
\end{equation}
as long as the inner electron stays close to the nucleus, i.e., $|x|_0$ stays bounded. 
The function $F$ is determined by the energy penalty\footnote{A nice textbook treatment of the WKB method is given in Chapter 9 of \cite{GaPa2}. 
The WKB method also determines a polynomial prefactor in the asymptotic of eigenfunctions, which gives a  logarithmic correction to the exponential weight $F$. The lower order correction terms we use in order to control  errors are much bigger, so we ignore logarithmic corrections for this discussion and concentrate on the leading order term.} it costs to move one particle to  infinity, 
\begin{equation}
    |F'(r)|^2 = \veps_U = -1/4- E_U\, .
\end{equation}
Up to a constant term, $F$ is clearly given by 
$F(r)= \sqrt{\veps_U}r$, so the ground state $\psi$ should decay as 
$\psi(x)\sim \exp(-\sqrt{\veps_U} |x|_\infty)$ when one particle is far away. 

Once one particle is removed from the system, a single hydrogen atom with ground state energy $-\frac{1}{4}$ remains and it is known that the ground state of hydrogen decays asymptotically as $\exp(-\frac{1}{2}|x|)$. 
Thus the decay of the ground state $\psi_U$ of the two particle system, when also the second particle is far away, should be of the form 
\begin{equation}
   \psi_U(x)\sim \exp\big(-\sqrt{\veps_U}|x|_\infty -\frac{1}{2}|x|_0\big) 
\end{equation}
when the system is sub--critical and the ionization energy $\veps_U>0$, 
i.e., $U<U_c$. 
Of course, this bound does not survive the limit $U\nearrow U_c$ 
since $\veps_{U_c}=0$, the ionization energy vanishes at critical coupling! 

However, if $U$ is close to $U_c$ the farthest out particle sees an \emph{additional energy penalty}! 
When $U>1$ and the inner particle is close to the nucleus, the charge of the nucleus, which is one, cannot completely screen the strong repulsion between the electrons. The outer particle, say it is at position $x_1$,  sees the attraction to the nucleus but also repulsion from the second particle. Thus when the second particle at position $x_2$ is close to the nucleus the potential energy of the outer particle is  
\begin{equation}\label{eq:Coulomb boost 1}
    -\frac{1}{|x_1|} + \frac{U}{|x_1-x_2|} \approx \frac{U-1}{|x_1|}
\end{equation}
when $|x_1|$ is large and $|x_2|$ is close to zero. Hence the outer 
particle  gets a \emph{local boost in  energy} of the form 
\begin{equation}
    \veps_U +\frac{U-1}{|x|_\infty}\, .
\end{equation}
That is, tunneling deep into the  
energetically forbidden region becomes more difficult. 
The same heuristic arguments as before then predict that the asymptotic decay of the ground state in the position of the outermost particle should be captured by a function $F$ satisfying 
\begin{equation}
    F'(r)= \sqrt{\text{local energy penalty}} = \sqrt{\veps_U +\frac{U-1}{r}}\, .
\end{equation}
One can integrate this equation by a tedious calculation. A slightly less tedious calculation shows that for $\veps, a>0$ the function 
\begin{equation}
    F_{\veps,a}(r) 
     = 
      \big( \veps+\frac{a^2}{r}\big)^{1/2}r 
      + \frac{a^2}{\sqrt{\veps}} 
        \big(\ln( \sqrt{\veps r+a^2} +\sqrt{\veps r} ) -\ln a   \big)
\end{equation}
satisfies $F'(r) = \sqrt{\veps+a^2/r}$ for all $r>0$. 
Once one particle moved to infinity the remaining system is again a single 
hydrogen atom, so the same heuristic arguments as before predict that the 
full two particle ground state $\psi_U$ should decay, for all $0<U\le U_c$, as 
\begin{equation}\label{eq:asumptotic heuristic}
    \psi_U(x) \sim \exp\big(- F_{\veps_U, a_U}(|x|_\infty) -\frac{1}{2}|x|_0\big)\, 
\end{equation}
with $\veps_U= -1/4 -E_U$ the ionization energy and $a_U=(U-1)_+^{1/2}$.  

In  Remark \ref{rem:interpolating} we will see that, 
\begin{equation}
    \lim_{\veps\to 0+} F_{\veps,a}(r) = 2a\sqrt{r}
\end{equation}
Hence the asymptotic given in \eqref{eq:asumptotic heuristic} has a chance of surviving the limit of the parameter $U$ becoming critical: At critical coupling 
the asymptotic of the ground state $\psi_c=\psi_{U_c}$ should be given by 
\begin{equation}
    \psi_c(x) \sim \exp\big( - 2\sqrt{U_c-1}|x|_\infty^{1/2} -\frac{1}{2}|x|_0\big)\, .
\end{equation}
So at critical coupling, the ground state  of helium type system should decay like a stretched exponential. 
Theorems 
\ref{thm:global pointwise anisotropic upper bound},  
\ref{thm:global pointwise anisotropic lower bound}, and 
 \ref{thm:sharp upper and lower bounds all coupling} show that up to some small lower order correction terms this  
prediction is indeed correct. 
Before we embark on the proof we will illustrate 
in Section \ref{sec:OneParticleThreeDim} the main ideas for making the 
above heuristic rigorous in the one particle case. 

\bigskip
 
To put our results into perspective, let us compare them with previously known results which only addressed the 
subcritical case $U< U_c$.  
The first precise bounds for the 
asymptotic behavior  of eigenfunction of $H_U$ with energy strictly below the essential spectrum 
are due to the groundbreaking works  of  Slaggie and Wichmann for three--body systems \cite{SlaWic62}, 
Ahlrichs for atoms \cite{Ahlrichs}, O'Connor \cite{OCo73}, 
Combes and Thomas \cite{ComTho73},  Deift, Hunziker, Simon,  
and Vock \cite{DeiHunSimVoc78} for multi--particle systems,  culminating 
in the work of Agmon  \cite{Agm82}. The last result provides bounds for asymptotic behavior of bound states 
for general multi-particle systems based on energy methods. 

For two particle systems in the  \emph{subcritical} case, the ionization energy $\veps_U$ is positive and 
Agmon's method yields the upper bound 
\begin{equation}\label{eq:upper bound subcritical}
	|\psi_U(x)| \le C_\delta \exp\left(-c_1|x|_\infty - c_2|x|_0 \right)
\end{equation}
for $c_1= \sqrt{\veps_U}-\delta$ and $c_2=1/2-\delta$ 
for any small $\delta>0$ and some constant $C_\delta$ which diverges for $\delta\to 0$. 
Here $|x|_\infty=\max(|x_1|,|x_2|)$ is the distance to the nucleus of the particle \emph{farther from}, 
and $|x|_0=\min(|x_1|,|x_2|)$ is the distance to the nucleus of the particle \emph{closer to},  
the nucleus. 

Recall that  $\veps_U$ is the ionization energy of the 
two--particle system. After one particle is removed  one is 
left with hydrogen, whose ionization energy is 
$1/4$ in the units we chose. 
So both constants $c_1$ and $c_2$ have a clear physical meaning 
in the upper bound \eqref{eq:upper bound subcritical}. 
Thus, except for reducing the constants by an 
arbitrarily small amount, the upper bound 
\eqref{eq:upper bound subcritical} is exactly what is predicted by WKB--type physical heuristics. 

Using subsolution estimates, it was shown in \cite{HofOstHofOstAhl78} that one can set $\delta=0$ 
at the expense of having polynomial prefactors in the upper bound. 
A matching lower bound for the ground state $\psi$, where 
$U=1/Z=1/2$, of the two particle system describing helium 
has been derived by 
Thomas Hoffmann--Ostenhof in \cite{HofOst80-PhysLett}, 
\begin{equation}\label{eq:lower bound subcritical}
	\psi(x) 
		\ge 
			c_\delta
				\exp\left(
					-(\sqrt{\veps_1}+\delta)|x|_\infty - \frac{1}{2}|x|_0 
				\right) 
\end{equation}
for arbitrary $\delta>0$ and some constant $c_\delta>0$, which goes to zero as $\delta\to 0$. 

Thus, except for decreasing/increasing the 
constants, which are the square roots of the 
successive ionization energies, in the upper/lower bounds by an arbitrary amount, these bounds 
settle the asymptotic behavior of the ground state wave functions 
and they can also be extended to subcritical $U<U_c$ 
where $E_U<-1/4$.  

Unfortunately, these results are \emph{useless at critical coupling}, 
since then first 
ionization energy $\veps_U$ is zero i.e., there is 
\emph{no energy cost} for removing the first particle. 
The only known decay property of the ground state 
$\psi_c=\psi_{U_c}$ of $H_{U_c}$ 
at critical coupling is the result in \cite{HofOstHofOstSim83} 
where they show that a positive solution of the 
Schr\"odinger equation exists and fulfills the upper bound 
\begin{equation}\label{eq:decay H2O2-S}
 |\psi_c(x)|\le C_m (1+|x|_\infty+|x|_0^2)^{-m}	
\end{equation}
for some constants $C_m<\infty$  for any $m>0$ and all 
$x_1,x_2\in\R^3$. Of course, this implies that 
$\psi_c \in L^2(\R^6)$,  so a ground state for this critical 
two-particle system exists. 
Moreover, in the remark in Section 4 of 
\cite{HofOstHofOstSim83}, they note that one can use the 
``Schr\"odinger inequalities" method of 
Ahlrichs and M.\ and T.\ Hoffmann-Ostenhof \cite{HofOstHofOstAhl78} 
to derive sharp upper and lower bounds for the 
\emph{one-particle density} $\rho_c$ of the ground state $\psi_c$, 
\begin{equation}\label{eq:upper and lower bounds on particle density} 
  \begin{split}
  	\sqrt{\rho_c(y)} 
  		&\le 
  			C^+_\delta (1+|y|)^{-3/4+\delta}
  			\exp\left(  
  			  -2(U_c-1)^{1/2} |y|^{1/2}
  			\right) \, ,\\
  	 \sqrt{\rho_c(y)} 
  		&\ge  
  			C^-_\delta (1+|y|)^{-3/4-\delta}
  			\exp\left(  
  			  -2(U_c-1)^{1/2} |y|^{1/2}
  			\right) 
  \end{split}
\end{equation}
for some constants $0<C^\pm_\delta<\infty$,  $\delta>0$ and all $y\in\R^3$. 
Here $\rho_c$ is given by  
\begin{align*}
	\rho_c(y)\coloneqq \int_{\R^3} |\psi_c(y,z)|^2\, dz
\end{align*}
that is, it is the marginal of the ground state probability density 
$|\psi_c|^2$ on $\R^6$, which is symmetric under permutations 
of the particles.

However, all these results says nothing about the asymptotic 
decay of the \emph{full ground state} $\psi_c$. 
We fill this gap by proving sharp anisotropic upper and 
lower bounds on the asymptotic behavior of 
the wave function $\psi_U$ of $H_{U}$ at or below critical 
coupling and also determine rigorously, how the decay 
changes in the whole range $0\le U\le U_c$. 

\begin{remark}\label{rem:nonBO}
We can also treat the case of a finite nuclear mass, i.e., 
avoid the Born-Oppenheimer approximation.  
Consider the following three particle system 
\begin{equation}\label{eq:HeliumHamiltonianRemark}
H_U = \frac{p_1^2}{m} + \frac{p_2^2}{m} +\frac{p_N^2}{M}- \frac{1}{|x_1-x_N|} - \frac{1}{|x_2-x_N|} + \frac{U}{|x_1 - x_2|}
\end{equation}
where $p_N=-i\partial_N$ is the momentum of the nucleus, $M$ is its mass and $x_N$ is its position. The other two particles 
have mass $m$ and their position are given by $x_1$ respectively $x_2$. The domain of the operator \eqref{eq:HeliumHamiltonianRemark} is $\mathcal D(H_{U})=H^2(\R^6)\otimes H^2(\R^3)$. 

For finite nuclear mass, $M<\infty$, and $U_c>1$, the  helium--type three particle operator \ref{eq:HeliumHamiltonianRemark} has a ground state at critical coupling 
in the center of mass frame, see \cite{Gri12}. The energy of this state is  
embedded at the edge of the essential spectrum. As for general Schr\"odinger 
operators the ground state is unique, up to a global phase, and can be chosen to be 
positive. With the help of the methods developed in Section \ref{sec:isotropic} one can derive isotropic upper bounds for the ground state of this system.   
We address the problem of sharp anisotropic upper and lower bounds for this system in a forthcoming work. 
\end{remark}
\textbf{Strategy of the proof:} 
The proof of the upper bound from Theorem 
\ref{thm:global pointwise anisotropic upper bound} is via 
a combination of energy methods, pointwise subsolution type 
bounds, and a subharmonic comparison principle in the spirit of well--known 
comparison principles from elliptic PDEs. 

By now, the derivation of upper bounds on the 
asymptotic decay of eigenfunctions with the help of energy methods 
is standard. However, 
we would like to stress that, following the general approach 
a la Agmon \cite{Agm82}, one looses an epsilon in the 
constants of the leading order terms, which we have to avoid. 
Moreover, the methods of Agmon \cite{Agm82} or  
\cite{DeiHunSimVoc78} and the works before them need a 
\emph{safety distance} of the ground state energy to the 
bottom of the essential spectrum. We do not have such a 
safety distance, since the ground state energy 
at critical coupling is at the edge of the essential spectrum.

Thus we have to overcome several obstacles.  
We need a low regularity version of the subharmonic comparison principle 
which goes back to Agmon \cite{Agm85}. But even having such a low regularity 
version, we cannot apply it directly. The 
comparison principle works well for one or two body problems, but not, 
in general, for multi--particle problems. Even for a restricted 
three body--body problem such as helium--type 
atoms in the infinite nuclear mass approximation, the comparison 
principle is \emph{not directly applicable} because 
in order to control the errors, \emph{both particles} have to be \emph{far 
from the nucleus}. Thus subsolution techniques  \emph{do not work} well in the case when only one particles tries to escape and the other stays close to the nucleus, see 
also Section 4 of \cite{Agm85}, where it was argued that the subharmonic comparison method is not an effective 
tool to derive upper bounds on eigenfunctions of multi--particle Schr\"odinger operators with three or more particles or, as in our case, for a restricted two--body problem with an additional nucleus of infinite mass. 

To overcome this we first prove an anisotropic $L^2$ upper bound 
with the help of energy methods. 
With standard methods, see \cite{AizSim82,Simon-semigroups}, this bound can be easily transformed into a pointwise upper bound. 
Unfortunately, this first upper bound \emph{does not yet yield} the sharp anisotropic upper bound at critical coupling 
since it \emph{does not have the correct asymptotic} in 
a \emph{transition region}, see Remark 
\ref{rem:suboptimal in transition region}.  
Except for this transition region our first anisotropic upper bound has 
\emph{the correct} asymptotic when one particle is far away and the other one close to the nucleus, or both particles are close to each other and far away.  
In a second step, we use this  a--priori information  
as an input for our proof of the sharp global anisotropic upper bound where we bridge the transition region with the help of supersolution bounds.  

Similarly, the proof of the lower bound is done in two steps. First we derive a sharp lower bound on the ground state well within the region where only one particle tries to escape and the other stays relatively close to the nucleus. We call this the  tricky region, the precise definition is given in 
\eqref{eq:regions}. 
In this case the Coulomb repulsion between the particles yields a boost in energy of the form 
\eqref{eq:Coulomb boost 1} but it does not blow up, since the particles cannot get close to each other in the tricky region, which makes working with subsolutions  simpler. We then use this a--priori input to extend the lower bound to the whole region. However, when both particles are not restricted they can come close to each other and  
when the particles are close together their Coulomb repulsion blows up, which provides a serious obstacle for subsolution bounds.

\smallskip

\textbf{Organization of the paper:}
As a warm-up, we explain our main new ideas, which prove why 
a \emph{long-range repulsive} part of the 
potential can stabilize bound states at the ionization threshold  
for a one particle system
in Section \ref{sec:OneParticleThreeDim}. 
We  derive sharp upper and lower bounds on the asymptotic behavior 
of zero energy ground state of such quantum systems.  

The necessary local energy bounds, needed for the proof of 
the  anisotropic $L^2$ upper bound for helium--type systems, are derived 
in Section \ref{sec:local energy bound}. A main new feature here 
is that we do not use 
conical regions to localize the particles, which is the usual 
approach in the study of 
many-particle systems. Instead, to be able to get a sharp anisotropic upper bound, it is important to use paraboloidal regions.

We derive an isotropic upper bound for the ground state in Section 
\ref{sec:isotropic} , which is considerably simpler than the proof 
our sharp anisotropic bounds.  
The proof of the anisotropic upper bound is done in 
Sections \ref{sec:first anisotropic upper bound} and 
\ref{sec:global anisotropic upper bound}, see, in particular, the 
proof of  Theorem \ref{thm:sharp upper bound all coupling most general}, which provides the upper bounds in Theorem \ref{thm:global pointwise anisotropic upper bound} and  \ref{thm:sharp upper and lower bounds all coupling}.

The proof of the lower bounds   
is done with a subharmonic comparison principle. As a first step we derive 
in Section \ref{sec:lower bound tricky region} a lower bound 
in the \emph{tricky region}, where one 
particle tries to escape to infinity and the other one stays 
(essentially) close to the nucleus. 
The global anisotropic lower bound is proven in Section 
\ref{sec:global lower bound}. 
The main difficulty there is to control the singular 
Coulomb repulsion between the two particles 
with terms which are only of lower order. See, in particular, the 
proof of  Theorem \ref{thm:sharp lower bound most general}, which provides the lower bounds of Theorem  
\ref{thm:global pointwise anisotropic lower bound} and \ref{thm:sharp upper and lower bounds all coupling}.  

Certain technical tools are gathered in the appendix. The main 
reason for including them 
is that our exponential weights for the upper bounds and the 
comparison functions used in the proofs of the lower bounds 
lack the usually required  high enough regularity. 
Their derivatives have jumps along codimension one Lipshitz 
surfaces, so standard results cannot, 
or at least not straightforwardly, be applied.

%%%%%%%%%%%%%%%%%%%%%%%%%%%%%%%%%%%%%%%%%%%%%%%%%%%%%%%%%%%%%%%
%%%%%%%%%%%%%%%%%%%%%%%%%%%%%%%%%%%%%%%%%%%%%%%%%%%%%%%%%%%%%%%
\section{One particle case} \label{sec:OneParticleThreeDim}
%%%%%%%%%%%%%%%%%%%%%%%%%%%%%%%%%%%%%%%%%%%%%%%%%%%%%%%%%%%%%%%
%%%%%%%%%%%%%%%%%%%%%%%%%%%%%%%%%%%%%%%%%%%%%%%%%%%%%%%%%%%%%%%
To explain the main ideas of our approach, we consider one particle moving in an external potential. 
This external potential consists of an attractive and a repulsive part with 
Hamiltonian given by 
\begin{equation}\label{eq:OneParticleHamiltonian}
	H = - \Delta - V + W
\end{equation}
where we assume  $V,W \geq 0$ are infinitesimally (form) bounded with 
respect to $-\Delta$. For simplicity, we also assume that $\supp V(x)$, the support of $V$, is compact. 
However, our proof works even for  cases where the support of $V$ is unbounded provided that the repulsion $W$ dominates the attractive part $V$ outside some bounded region. 
We also assume that $W$ goes to zero at infinity, so the essential 
spectrum of the system is given by $\sigma_{\rm ess}(H) = [0,\infty)$, the discrete spectrum of $H$ is below zero. 

In the following we assume that the operator $H=-\Delta -V+W$ is defined 
in the sense of quadratic forms, that is, 
\begin{align*}
	\la \varphi, H\psi \ra 
		= \la\nabla\varphi, \nabla\psi \ra + \la \varphi, (-V+W)\psi\ra 
\end{align*}
for all $\varphi,\psi$ in the standard Sobolev space $H^1(\R^d)$. 
We consider (weak) eigenfunctions $\psi\in H^1(\R^d)$ of $H$, by which we mean that 
\begin{align}\label{eq:weak eigenfunction}
	\la \varphi, H\psi\ra = E\la \varphi,\psi\ra 
\end{align}
for all $\varphi\in H^1(\R^d)$. By density of 
$\calC^\infty_0(\R^d)$ in $H^1(\R^d)$ it is enough to require that \eqref{eq:weak eigenfunction} holds for all $\varphi\in \calC^\infty_0(\R^d)$.  
It is relatively straightforward to see that weak eigenfunctions are in the domain of the operator $H$, see Lemma 2.5 in 
\cite{HunJexLan23-Potential}, but we will not need this.
We are interested in the question whether zero, the edge of the 
essential spectrum, can be an eigenvalue of the system and if so, how does the corresponding eigenfunction decay at infinity?  
A typical example is the situation where a parameter, say the repulsion $W$ or the depth of the well $V$ is tuned in such a way that the ground state eigenvalue hits zero. Does the bound state survive or does it dissipate? 
This is clearly a non-perturbative situation. 

Using previous approaches, such as Agmon's method \cite{Agm82}, one 
can easily show that the eigenvectors corresponding to negative 
eigenvalues decay exponentially. However, these approaches need a 
safety distance to the essential spectrum and yield nothing for 
eigenfunctions at the edge of the essential spectrum! 

We will first discuss how to get upper bounds on eigenfunctions 
without having a safety distance of the corresponding eigenvalue 
to the essential spectrum. Once one has such a type of bound, this 
also implies existence of eigenvalues at the edge of the essential spectrum, 
 Details of this argument are given in Appendix~\ref{app:tight} and \ref{app:basic properties} in a more complicated situation.

%%%%%%%%%%%%%%%%%%%%%%%%%%%%%%%%%%%%%%%%%%%%%%%%%%%%%%%%
\subsection{Taking advantage of long-range repulsion}
%%%%%%%%%%%%%%%%%%%%%%%%%%%%%%%%%%%%%%%%%%%%%%%%%%%%%%%%
In the following we show how to derive upper bounds on the asymptotic 
behavior of zero energy eigenfunctions. 
These upper bounds 
illustrate how the type of asymptotic decay 
of $\psi$ is directly related to the repulsive potential $W$. 

\begin{lemma}\label{lem:one particle upper bound general}
Let $H$ be given as in \eqref{eq:OneParticleHamiltonian} and assume that $V$ has compact support. 
Let  $\psi \in L^2(\R^d)$ be a weak eigenfunction of $H$ 
with energy $E\le 0$. Furthermore, assume that $F\ge 0$ is a 
locally bounded and differentiable function such that 
for some $R>0$  
\begin{equation}\label{eq:one particle bound on F-1}
	|\nabla F(x)|^2 < W(x) \quad \text{for all } |x|\ge R\, .
\end{equation}
Then  
\begin{align}\label{eq:great}
	\int_{|x|\ge R} e^{2F(x)+\ln\big( W(x)-|\nabla F(x)|^2\big)} |\psi(x)|^2\, dx 
		\le C_{F,R} \|\psi\|^2  
\end{align}
with 
\begin{align}\label{eq:constant CFR}
	C_{F,R}=8\sup_{R/2\le |x|\le R} e^{2F(x)}\left(2R^{-2}+ R^{-1}|\nabla F(x)| \right)\, .
\end{align}
\end{lemma}
\begin{remark}
	The point of the bound above is that the right hand side 
	is \emph{uniform} in 
	the eigenvalue $E\le 0$.  It \emph{does not need a safety distance} 
	to the 
	essential spectrum. Moreover, the constant $C_{F,R}$ depends on 
	\emph{local} bounds of $F$, hence it is finite even for 
	unbounded functions $F$. 
	Of course, as $W$ tends to zero at infinity, so does 
	$|\nabla F|^2$. 
	Nevertheless $F$ can still go to infinity under this condition, and 
	the main question is which of the terms in 
	$2F+ \ln\big( W-|\nabla F|^2\big)$  will win this tug--of--war. 
	It is easy to see that the borderline case 
	is a decay of the form $W(x)\sim|x|^{-2}$ at infinity. 
	Any slower decay of $W$ will lead to a growth of $F$ which out--paces 
	the second term $\ln\big( W-|\nabla F|^2\big)$.  
\end{remark}

\begin{proof}[Proof of Lemma \ref{lem:one particle upper bound general}{\rm:}] 
Let $\xi$ be any real-valued bounded and differentiable function 
and use $\varphi=\xi^2\psi$ in the weak form of the 
eigenvalue equation. Then 
$	E\|\xi\psi\|^2 = E\la\xi^2\psi, \psi \ra = \la \xi^2\psi, H\psi \ra$ and since 
$E\|\xi\psi\|^2$ is real we can use the IMS localization formula, see \cite{CycFroKirSim87, Griesemer2004} or Appendix \ref{app:IMS formula}, to get    
\begin{align}
	E\|\xi\psi\|^2 &= \re \la \xi^2\psi, H\psi \ra 
		= \la \nabla(\xi\psi), \nabla(\xi\psi) \ra 
			+ \la \xi\psi, (-V+W)\xi\psi\ra 
			-\la \psi, |\nabla\xi|^2\psi\ra 
\end{align}
Rearranging and dropping the kinetic energy term, which is positive, gives 
\begin{align}\label{eq:trick-1}
	\la \xi\psi, (-V+W-E)\xi\psi\ra  
			\le   \la \psi, |\nabla\xi|^2\psi\ra 
\end{align}
Take $\chi\in \calC^\infty(\R_+)$ with $0\le \chi\le 1$, $\chi(r)=0$ if $r\le 1/2$, $\chi(r)=1$ if $r\ge 1$, and set 
\begin{align}\label{eq:def chi_R}
	\chi_R(x)\coloneqq \chi\left(|x|/R \right) \quad \text{for }x\in\R^d\, .
\end{align}
It is easy to see that a function $\chi$ fulfilling the above 
constraints exists and for which one has  $\|\chi'\|_\infty\le 4$. 
For any such choice we have $\chi_R\in \calC^\infty(\R^d)$ and 
$\|\nabla\chi_R\|_\infty\le 4/R$. 

Now assume that $F$ is bounded and differentiable, for the moment, and use  
$\xi=\xi_R = \chi_R e^{F}$. Using   
\begin{align*}
	\nabla\xi = e^F\nabla\chi_R +e^F\chi_R\nabla F\, ,
\end{align*} 
in \eqref{eq:trick-1} and reshuffling the terms a bit we see 
\begin{align}\label{eq:trick-2}
		\la \xi_R\psi, (-V+W-E-|\nabla F|^2)\xi_R\psi\ra  
			\le   \la \psi, e^{2F}\left( |\nabla\chi_R|^2+ 2\chi_R \nabla\chi_R\nabla F\right)\psi\ra \, .
\end{align} 
Moreover, we can use $0\le \chi_R\le 1$ on the right hand side  
and $E\le 0$ to drop the term $-E$ on the left hand side in \eqref{eq:trick-2} to get 
\begin{align}\label{eq:trick-3}
		\la \xi_R\psi, (W-|\nabla F|^2)\xi_R\psi\ra  
			\le   \la \psi, e^{2F}\left( |\nabla\chi_R|^2+ 2|\nabla\chi_R||\nabla F|\right)\psi\ra \, .
\end{align} 
where we also took $R$ so large that $\chi_R$, hence also $\xi_R$,  
is zero on the support of $V$. 

In the case that $F\ge 0$ is not bounded, we regularize it by considering 
\begin{align}\label{eq:F delta one particle}
	F_\delta = \frac{F}{1+\delta F}\, .
\end{align}
Then $F_\delta \le F$ and 
\begin{align*}
	\nabla F_\delta = \frac{\nabla F}{(1+\delta F)^2}
\end{align*}
so $|\nabla F_\delta|\le |\nabla F|$. Thus \eqref{eq:trick-2} yields 
\begin{align}
	\la \chi_R e^{F_\delta}\psi, (W-|\nabla F|^2)\chi_Re^{F_\delta}\psi\ra  
			\le   \la \psi, e^{2F}\left( |\nabla\chi_R|^2+ 2|\nabla\chi_R||\nabla F|\right)\psi\ra
			\le C_{F,R} \|\psi\|^2
\end{align}
since the support of $\nabla\chi_R$ is contained in the annulus 
$R/2\le |x|\le R$. Since $0\le \chi_R\le 1$ and $\chi_R(x)=1$ for $|x|\ge R$ 
we get 
\begin{align}
	\int_{|x|\ge R} e^{2F_\delta(x)+\ln\big( W(x)-|\nabla F(x)|^2\big)} |\psi(x)|^2\, dx 
		\le C_{F,R} \|\psi\|^2  
\end{align}
and using that $F_\delta$ converges pointwise monotonically to $F$ in the limit $\delta\to 0$ we see that \eqref{eq:great} holds. 
\end{proof}
\begin{remark}\label{rem:subcritical-relaxed}
	Of course, one should not always drop the term $-E$ from the left hand side of \eqref{eq:trick-2}. Keeping it we get the bound 
 \begin{align}\label{eq:great-2}
	\int_{|x|\ge R} e^{2F(x)+\ln\big( W(x)-E-|\nabla F(x)|^2\big)} |\psi(x)|^2\, dx 
		\le C_{F,R} \|\psi\|^2   
\end{align}
 under the condition 
 \begin{equation}\label{eq:one particle bound on F-2}
	|\nabla F(x)|^2 < W(x)-E
\end{equation}
which allows for a larger class of functions $F$, which can 
grow linearly in $|x|$. This leads to exponential 
$L^2$-type upper bounds  when $E<0$, i.e., when one has a 
\emph{safety distance} to the bottom of the essential spectrum. 
How one can easily choose exponential weights $F$ which fulfill 
condition \eqref{eq:one particle bound on F-1}, respectively 
\eqref{eq:one particle bound on F-2}, is shown in Sections \ref{subsec:repulsive-1} and \ref{subsec:repulsive-2}. 
\end{remark}

%%%%%%%%%%%%%%%%%%%%%%%%%%%%%%%%%%%%%%%%%%%%%%%%%%%%%%%%
\subsection{An example with a repulsive Coulomb tail} \label{subsec:repulsive-1}
%%%%%%%%%%%%%%%%%%%%%%%%%%%%%%%%%%%%%%%%%%%%%%%%%%%%%%%%
Assume that $V$ has compact support and $W(x)= a^2/|x|$ with $a>0$ outside some compact set. 
Use 
\begin{align} 
	F(r)= 2a r^{1/2} - Kr^\kappa/2 
\end{align}
for some $0<\kappa<1/2$ and $K>0$ and also set $F(x)=F(|x|)$ by a 
slight abuse of notation. Then $|\nabla F(x)| = |F'(|x|)| $ 
and for any eigenfunction $\psi$ of $H$ with energy $E\le 0$, we have 
\begin{align}
	W(x)- |\nabla F(x)|^2 
		&= 2 a |x|^{\kappa-3/2} -(K\kappa)^2 |x|^{2(\kappa-1)}/4 
			\gtrsim |x|^{\kappa-3/2} \ge |x|^{-3/2}
\end{align}
since $\kappa-3/2>2(\kappa-1)$ iff $\kappa<1/2$. In this case, Lemma \ref{lem:one particle upper bound general} shows that 
\begin{align*}
	x\mapsto e^{2a |x|^{1/2} - K|x|^\kappa/2 - \frac{3}{2}\ln|x|}\psi(x)  
\end{align*}
is in $L^2(\R^d)$ outside of some large enough ball $|x|\ge R$, uniformly 
in the energy $E\le 0$ for normalized eigenfunctions $\psi$. 
Since any fractional power $r^\kappa$ bounds logarithmic terms 
$\ln r$ for large $r$ and $\psi$ is globally $L^2$ we see that 
\begin{align}
	x\mapsto e^{2a |x|^{1/2}-K|x|^\kappa}\psi(x)\in L^2(\R^d)
\end{align}
for any $K>0$ and $0<\kappa<1/2$. 

Using subsolution bounds, or the Harnack inequality for 
ground states, one can get the pointwise upper bound 
\begin{equation}\label{eq:Coulomb tail upper bound pointwise}
 |\psi(x)| \lesssim \exp(-2a |x|^{1/2} +K|x|^\kappa)	
\end{equation}
 from this, see the proof of Corollary 
 \ref{cor:first anisotropic pointwise upper bound}.  

Moreover, Lemma \ref{lem:one particle upper bound general} can also be used to  show 
that if the attractive part $V=V_\lambda$ is tuned in the 
parameter $\lambda$ in such a way 
that $H_\lambda= -\Delta -V_\lambda+W$  converges to some limiting operator $H_{\lambda}$ and $H_\lambda$ 
has a ground state energy $E_\lambda<0$ which converges to 
zero as $\lambda\to \lambda_\text{cr}$, then 
the limiting operator $H_c=H_{\lambda_\text{cr}}$ has a zero 
energy eigenvalue embedded  at the edge of its essential spectrum. 
This follows from tightness and weak convergence arguments 
similarly to the discussion in Appendix \ref{app:tight} and 
\ref{app:basic properties}. 
  
%%%%%%%%%%%%%%%%%%%%%%%%%%%%%%%%%%%%%%%%%%%%%%%%%%%%%%%%%%%%%%
\subsection{Lower bound for zero energy ground states of systems with a repulsive Coulomb tail} \label{subsec:one particle lower bound Coulomb}
%%%%%%%%%%%%%%%%%%%%%%%%%%%%%%%%%%%%%%%%%%%%%%%%%%%%%%%%%%%%%%
Assume that $H=-\Delta -V+ W$ is as above with $W(x)= a^2/|x|$ outside of 
a compact, $V$ has compact support, and $\psi$ is a positive zero 
energy ground state of $H$. Then the upper bound in 
\eqref{eq:Coulomb tail upper bound pointwise} is sharp in the 
sense that we have 
\begin{lemma}\label{lem:lower bound one particle}
  In the above situation, we have the lower bound 
  \begin{align}\label{eq:one particle lower bound} 
  	\psi(x)\gtrsim \exp(-2a |x|^{1/2} - K|x|^\kappa)
  \end{align}
 for all $x\in \R^d$, where $\psi$ is the unique positive zero energy ground state  of $H$. 
\end{lemma}
\begin{proof}
We will use the subharmonic comparison principle. Since $\psi$ is only 
a weak eigenfunction of $H$ in the sense of quadratic forms, we use the 
quadratic form version of the subharmonic comparison 
principle due to Agmon \cite{Agm85}, see also Appendix \ref{app:sub-super-solutions}.   

Define $g= e^{-F}$, which is $\calC^2$ on $|x|> R$ for any $R>0$ and 
assume that $g\in L^2$, or at least on $|x|> R$. Calculating 
\begin{align}
	\nabla g = -g\nabla F\quad \text{and } 
	-\Delta g = g\left( \Delta F -|\nabla F|^2 \right)
\end{align}
we see that if 
\begin{align}\label{eq:one particle basic assumption lower bound}
	W(x)\le |\nabla F(x)|^2 - \Delta F(x) 
\end{align}
 then $Hg\le 0$ on $|x|>R$, or, as quadratic forms
\begin{align}
	\la \varphi, Hg \ra \le 0 
	\quad \text{for all } 0\le \varphi\in \calC^\infty_0(\Omega_R) 
\end{align}
with $\Omega_R= \{|x|>R\}$. Since $\psi>0$ is continuous (here we assume that $V$ and $W$ are Kato--class potentials) it is bounded away from 
zero on $\{|x|\le R\}$, which is compact. 
Since $g$ is bounded, there exist a constant $C>0$ such that 
\begin{align}
  \psi(x)\ge C g(x) \quad \text{for all } |x|\le R+1\, . 
\end{align}
Hence on the boundary layer $R<|x|<R+1$, which is a boundary layer for the boundary $\partial\Omega_R$ in the sense of Definition  \ref{def:boundary layer},  we have  
\begin{align}
  \psi(x)\ge C g(x) \quad \text{for all } R<|x|<R+1  
\end{align}
and then subharmonic comparison principle implies 
\begin{align}
	\psi(x)\ge C g(x) = C\exp(-F(x)) \quad \text{for all } |x|\ge R\, .
\end{align}
By the choice of $C$ this also holds for $|x|\le R$, hence globally. 

We apply the above with the radial choice 
\begin{align}
	F(r) = 2a r^{1/2} +K r^{\kappa},\quad r=|x|
\end{align}
for which one calculates 
$\nabla F(x) = F'(r)\frac{x}{|x|}$ and $\Delta F(x)= F''(r)+ F'(r)\frac{d-1}{r}$, hence 
\begin{equation}\label{eq:subsolution one particle}
  \begin{split}
  	|\nabla &F(x)|^2 - \Delta F(x) -W(x) \\
  	  &= 2a K\kappa r^{\kappa-3/2} 
  	  	    +(K\kappa)^2 r^{2(\kappa-1)} 
  	  	    - a (d-\frac{1}{2}) r^{-3/2} 
  	  	    -K\kappa(\kappa+d-2)r^{\kappa-2}>0
  \end{split}
\end{equation}
since the first term dominates the negative terms for all large enough $r_1$ when $\kappa>0$. 
 This proves \eqref{eq:one particle lower bound}.  
\end{proof}
\begin{remark}
  A similar calculation shows that 
  $ f(x)= \exp(-2a |x|^{1/2}+ K|x|^\kappa)$ is a supersolution at energy zero near infinity. 
  Since eigenfunctions are bounded when the potential is the Kato--class, the subharmonic comparison principle can also be used to see that the 
  matching upper bound 
  \begin{equation}
      \psi(x)\le C \exp(-2a |x|^{1/2}+ K|x|^\kappa)
  \end{equation}
  holds for all $K>0$ and $0<\kappa<1/2$. This is the same upper bound as in \eqref{eq:Coulomb tail upper bound pointwise} and gives an alternative to the derivation of upper bounds with the help of energy 
  methods for one particle systems. For our upper bounds for helium--type systems 
  we eventually need to use a combination of energy methods and the subharmonic comparison principle. 
  
  The lower bound \eqref{lem:lower bound one particle} together with the matching upper bound 
  shows that any zero energy ground state 
  of a one--particle Hamiltonian with a long range Coulomb--type repulsion $W(x)=a^2|x|^{-1}$ obeys for  all $x\in\R^d$ the two sided bound 
  \begin{equation}\label{eq:one-particle critical case sharp upper and lower bound}
  	C_-\exp(-2a |x|^{1/2} - K|x|^\kappa)
  	  \le \psi(x)\le 
  	    C_+\exp(-2a |x|^{1/2} + K|x|^\kappa)
  \end{equation}
 for any $0<\kappa<1/2$ and $K>0$ and some $0<C_-\le C_+<\infty$. 
 In particular, the leading order terms in these bounds are sharp. 
 
 Similar sharp lower and upper bounds can be proven for 
 long range repulsive tails of the form 
 $W(x)= a^2 |x|^{-2\rho}$ for any $0<\rho<1$. In this case $F$ needs to fulfill $F'(r)\sim r^{-\rho}$, hence $F(r)\sim r^{1-\rho}$ and thus any zero energy ground state will have an asymptotic decay of the form
 \begin{align}
 	\psi(x)\sim \exp\left(-\frac{a}{1-\rho} |x|^{1-\rho}\right)\quad \text{for } |x|\gg 1\, .
 \end{align}
 up to lower order correction terms. 
 We leave the details to the interested reader. 
 A detailed study of the one--particle case can be found 
 in \cite{HunJexLan23-Potential}, where for short range 
 repulsive tails a phase transition for the non-existence 
 versus existence of zero energy ground states was proved, 
 depending on the dimension $d\le 3, d=4$, and $d\ge 5$.
\end{remark}

%%%%%%%%%%%%%%%%%%%%%%%%%%%%%%%%%%%%%%%%%%%%%%%%%%%%%%%%
\subsection{Repulsive Coulomb-tail: Interpolating between the  subcritical and critical cases}\label{subsec:repulsive-2}
%%%%%%%%%%%%%%%%%%%%%%%%%%%%%%%%%%%%%%%%%%%%%%%%%%%%%%%%
We can also consider a subcritical situation
\begin{equation}\label{eq:one-particle H-lambda}
H_\lambda = -\Delta - V_\lambda(x) + \frac{a^2}{|x|}
\end{equation}
with ground state eigenvalue $E_\lambda<0$ for 
$0\le \lambda<\lambda_\text{cr}$ and $E_\lambda=0$ for 
$\lambda= \lambda_\text{cr}$. The a--priori bounds above and 
the discussion in Appendix \ref{app:tight} and 
\ref{app:basic properties} show that then 
$H_c=H_{\lambda_\text{cr}}$ has a zero energy bound state. 
Again, let us assume, for simplicity, that $V_\lambda$ has 
a compact support. 
For $\lambda<\lambda_\text{cr}$ the ground states clearly 
have exponential decay and at critical coupling this 
exponential decay switches over to the stretched exponential 
decay with $\lambda=\lambda_\text{cr}$ as discussed  above. 

The question is how the decay rate changes precisely from 
exponential to subexponential as the gap between the essential 
spectrum and the ground state closes for 
$\lambda\nearrow \lambda_\text{cr}$. 
In such a case, one should also take advantage of the fact 
that we can allow 
$|\nabla F|^2< \eta/|x|-E_\lambda$, see Remark 
\ref{rem:subcritical-relaxed}. With $\veps=-E_\lambda$, 
the ionization energy, and choosing $F$ to be radial, one sees 
that one needs to find a function $F=F_{\veps,\eta}$ such that 
\begin{align}
	|F_{\veps,a}'(r)|^2 = \veps +\frac{a^2}{r} 
\end{align}
for large enough  $r$. A slightly tedious calculation shows  that 
\begin{equation}\label{eq:interpolating F 2}
	F_{\veps,a}(r)= \int_0^r\sqrt{\veps+\frac{a^2}{s}}\mathrm d s
	=
	  r \left(\veps+\frac{a^2}{r}\right)^{1/2}
	  +\frac{a^2}{\sqrt{\veps}}
	  \left( \ln\left( 
	          (a^2 +\veps r)^{1/2} +\sqrt{\veps r}\right) 
	     -\ln a
	   \right)\,.
\end{equation} 
which is the reason for the definition \eqref{eq:interpolating F}.  
It is easy to see that the second term in $	F_{\veps,a}(r)$  
is positive and only logarithmically growing in $r$ for fixed $\veps>0$. 
Thus $F_{\veps,a}(r)$ is linearly growing in $r$ and asymptotic 
to $\sqrt{\veps}r$,  which is the decay rate predicted by $WKB$ 
methods when the ionization energy is positive. 

Using Lemma \ref{lem:one particle upper bound general}, we again conclude 
that the ground state corresponding to a negative eigenvalue 
$E<0$ behaves as 
\begin{equation}
\label{eq:limitbehavior upper}
\psi(x)\le C_+ \exp\left(-F_{\veps,a}(|x|)+K|x|^\kappa \right) 
\end{equation}
for some constant $C_+<\infty$ and any $0<\kappa<1/2$, $K>0$. 

One first uses Lemma \ref{lem:one particle upper bound general} to get an $L^2$ upper bound and then transfers this into a pointwise upper 
bound via subsolution bounds of Trudinger, see 
\cite[Theorem C.1.3]{Simon-semigroups}, or the Harnack inequality for the ground state, similar as in the proof of Corollary \ref{cor:first anisotropic pointwise upper bound} below.

To derive a lower bound on the ground state $\psi$, we use the variant 
\begin{align*}
	g=\exp(-F_{\veps,a}(|x|)-K|x|^\kappa )\, 
\end{align*} 
of the function used in Section \ref{subsec:one particle lower bound Coulomb}. 
A straightforward but slightly tedious calculation shows that 
$g$ is also a classical subsolution of 
$H_\lambda= -\Delta-V_\lambda+\eta/|x|$ at energy 
$E_\lambda=-\veps$ outside a large enough ball in $\R^d$,  
i.e., $(H_\lambda-E_\lambda) g(x) \le 0$ for $|x|$ large enough.   
We leave the details of these calculations to the interested reader. 

Exactly as in the proof of Lemma \ref{lem:lower bound one particle} 
one then concludes that the lower bound 
\begin{equation}\label{eq:limitbehavior lower}
	\psi(x)\geq  C_- \exp\left(-F_{\veps,a}(|x|)- K|x|^\kappa \right)
\end{equation}
holds for some constant $C_->0$ and any $0<\kappa<1/2$, $K>0$. 

Collecting the upper and lower bounds one sees that they yield 
sharp upper and lower bounds for one--particle 
quantum systems with a long range Coulomb repulsion. 
\begin{theorem}
  Let $\psi_\lambda$ be a ground state of the one--particle 
  Hamiltonian $H_\lambda $ corresponding to the energy 
  $E_\lambda\le 0$. Then we have the two--sided bound 
  \begin{align}\label{eq:one-particle sharp upper lower bounds uniformly in coupling}
  	C_1\exp\left( F_{|E_\lambda|,\eta}(|x|)-K|x|^\kappa \right)
  	  \le 
  	    \psi_\lambda 
  	   \le 
  	  	C_2\exp\left( F_{|E_\lambda|,\eta}(|x|)+K|x|^\kappa \right)
  \end{align}
  for some constants $0<C_1\le C_2<\infty$ and all $0<\kappa<1/2$, $K>1$. 
\end{theorem}
The constants $C_1,C_2$ may be dependent on the parameter $\lambda$, in general, but will be independent of it for most physically relevant cases, see the end of Remark \ref{rem:interpolating} below.

\begin{remark}\label{rem:interpolating}
  The logarithmic term in \eqref{eq:interpolating F 2} is 
  positive and, for fixed ionization energy 
  $\veps>0$ only logarithmically growing 
  in $r$. In this case, the leading order behavior of $F_{\veps,a}$ is 
  dominated by the first term and is given by 
  \begin{align}
  	  F_{\veps,a}(r)\sim \left(\veps+\frac{a^2}{r}\right)^{1/2}r\sim \veps^{1/2}r
  \end{align}
  for large $r$. 
  
  Clearly, $(\veps+a^2/r)^{1/2}r \to  a r^{1/2}$ in the limit 
  of vanishing ionization energy, $\veps=-E\to 0$, which is exactly 
  half of the leading order term in the estimate 
  \eqref{eq:one-particle critical case sharp upper and lower bound} 
  for the ground state. 
  It is easy to see directly from the definition 
  of $F_{\veps,a}$ as an integral in \eqref{eq:interpolating F 2}, 
  that 
  \begin{align}\label{eq:F limit}
  	F_{0,a}(r) = \lim_{\veps\to 0}F_{\veps,a}(r) = 2a r^{1/2}
  \end{align}
  which is \emph{twice} of the limit of the \emph{leading order term} in 
  the expression of $F_{\veps,\eta}$ on the right hand side 
  of \eqref{eq:interpolating F 2}.  
  Alternatively, we note that the 
  second term in the right--hand--side of \eqref{eq:interpolating F 2} can be written as 
  \begin{align*}
  	\frac{a^2}{\sqrt{\veps}}\ln\left(\left(1+\frac{\veps r}{a^2}\right)^{1/2}+\sqrt{\frac{\veps r}{a^2}}\right)
  	  =  
  	    \frac{a\sqrt{ r}}{t} \ln\left(\left(1+t^2\right)^{1/2}+t\right)
  \end{align*} 
  with $t= \sqrt{\veps r}/a$. Since  
  \begin{align*}
  	\frac{1}{t}\ln\left(\left(1+t^2\right)^{1/2}+t\right)
  	  = 
  	    \ln\left( \Big((1+t+\Oh(t^2)\Big)^{1/t}\right) 
  	  \to \ln e =1
  \end{align*}
  in the limit $t=\sqrt{\veps r}/a \to 0$, we see that
  the logarithmic term in \eqref{eq:interpolating F 2}, 
  which for fixed $\veps>0$ is \emph{always of much lower order} 
  than the first term, gives the \emph{same contribution as the first term} in the limit of vanishing ionization energy 
  $\veps\to 0$ and should not be discarded.

Because of \eqref{eq:F limit} the upper and lower bounds  
 \eqref{eq:limitbehavior upper} and \eqref{eq:limitbehavior lower}
 also hold in the limit of vanishing ionization energy and one 
 recovers our previous subexponential upper and lower 
 bounds for zero energy ground states. 
 Thus the function $F_{\veps,\eta}$ describes precisely the 
 phase transition between exponential and subexponential decay of the 
 ground states of $H_\lambda$ in the critical limit of 
 vanishing ionization energy. 
 
 If the potentials $W$ and 
 $V_\lambda$ are in the so-called Kato--class, 
 for definitions 
 see \cite{AizSim82, CycFroKirSim87, Simon-semigroups}, and 
 $V_\lambda$ is continuous in the Kato--norm, one can show 
 that the constants $C_1, C_2$ in 
 \eqref{eq:one-particle sharp upper lower bounds uniformly in coupling} 
 can be chosen to be \emph{independent of} $\lambda$  
 \emph{up to the critical coupling} by an argument which 
 parallels the one we give for helium--type systems, using  
 a one--particle version of 
 Proposition \ref{prop:basic properties} in the Appendix.  
\end{remark}

%%%%%%%%%%%%%%%%%%%%%%%%%%%%%%%%%%%%%%%%%%%%%%%%%%%%%%%%%%%%%%
\subsection{The fate of zero-energy solutions for short range potentials} \label{subsec:short range potentials}
%%%%%%%%%%%%%%%%%%%%%%%%%%%%%%%%%%%%%%%%%%%%%%%%%%%%%%%%%%%%%%
Without a long--range repulsive tail of the potential  
one can still do a similar analysis as for long--range 
repulsive potentials, using now an additional boost 
coming from the kinetic energy. We say that a potential $V$ 
is \emph{short range} if 
\begin{align}
	|V(x)|\lesssim \frac{1}{|x|^2 \ln^q(|x|)} 
	\quad \text{for } |x|\gtrsim 1 \text{ and some } q>2\, , 
\intertext{or if }
		|V(x)|\le \frac{c}{|x|^2  \ln^2(|x|)} 
		\quad \text{for } |x|\gtrsim 1 \text{ and some } c<\frac{1}{4}
\end{align}
Let $H=-\Delta+V$ be defined via quadratic forms. 
A function $\psi\in H^1_{\text{loc}}(\R^d)$ is a (weak) zero 
energy solution if 
$\la\varphi, H\psi\ra = \la \nabla\varphi,\nabla\psi\ra +\la \varphi, V\psi\ra= 0$ for all 
$\varphi\in\calC^\infty_0(\R^d)$. Clearly, this extends to all 
$\varphi\in H^1(\R^d)$ with compact support. 
\begin{theorem} \label{thm:resonances bound}
  Assume that 
  $\psi\in H^1_\text{loc}(\R^d)$ is a zero energy bound state of 
  a one--particle Schr\"odinger operator $H=-\Delta +V$ with a 
  short--range potential $V$ in dimension $d\ge 3$ and 
  \begin{align}\label{eq:weak decay}
  	\liminf_{L\to\infty} 
  	  \frac{1}{L^2} \int_{L\le |x|\le \alpha L} |\psi(x)|^2\, dx 
  	  =0
  \end{align}
  for some $\alpha>1$. Then 
  \begin{align}\label{eq:resonances bound}
  	x\mapsto  \frac{(1+|x|)^{(d-4)/2}}{\ln(2+|x|)}\psi(x) \in L^2(\R^d)
  \end{align}
\end{theorem}
\begin{remark}
  The condition \eqref{eq:weak decay} poses a weak condition on the decay of 
  $\psi$ at infinity. In particular, $\psi$ does not need to be in $L^2$. 
  The bound \eqref{eq:resonances bound} shows that if $\d\ge 5$, then 
  $\psi\in L^2(\R^d)$, i.e., $\psi$ is a zero energy eigenfunction of $H$. 
  If $d=4$, then $x\mapsto(\ln(2+|x|))^{-1}\psi(x)\in L^2(\R^4)$, i.e., 
  $\psi$ barely misses to be $L^2$ and in three dimensions one sees that 
  $x\mapsto |x|^{-1/2}(\ln(2+|x|))^{-1}\psi(x)\in L^2(\R^3)$. 
  This is consistent with the known behavior of zero-energy resonances of 
  Schr\"odinger operators. A series of sharp criteria  for the 
  existence/absence of zero energy ground states for 
  one--particle Schr\"odinger operators in any dimension 
  can be found in \cite{HunJexLan23-Potential}.  
  
  Theorem \ref{thm:resonances bound} is an extension of results 
  on $L^2$ bounds for zero energy eigenfunctions and resonances of 
  one--particle Schr\"odinger operators in 
  \cite[Theorem 2.1]{BarBitVug19-a}. 
  As already shown in \cite{BarBitVug19-a}, the above bound has 
  the following consequence: Assume that the  potential $V$ is 
  infinitesimally form bounded with respect to $-\Delta$ and that 
  the one--particle Schr\"odinger operators $H=-\Delta+V$, defined 
  via quadratic forms, has a virtual level at zero, that is, 
  $H\ge 0$ and, with $H_\delta= -(1-\delta)+V$, 
  \begin{align*}
  	\inf\sigma(H_\delta)<0
  \end{align*}
  for all $0<\delta<1$. Then in dimensions $d\ge 5$, zero is a 
  simple eigenvalue of $H$. The proof uses 
  Theorem \ref{thm:resonances bound}, or better a slight extension 
  of it to $H_\delta$, 
  and tightness and weak convergence arguments to show that the 
  ground states of $H_\delta$ with $\delta=1/(2n)$, $n\in\N$, yield 
  a sequence which converges strongly to the ground of $H$. Similar 
  arguments are given in Appendix \ref{app:tight} and 
  \ref{app:basic properties}, for the existence of  a ground state for 
  helium--type atoms with a critical repulsion with finite or infinite 
   nuclear mass.
\end{remark}

\begin{proof}[Proof of Theorem \ref{thm:resonances bound}]
  Let $F$ be a function which is bounded and differentiable outside 
  some compact set and whose gradient is bounded by 
  $|\nabla F(x)|\lesssim |x|^{-1}$ for large $|x|$. 
  Let $\wti{\chi}\in \calC^\infty_0(\R^d)$ with $0\le \wti{\chi}\le 1$, 
  $\wti{\chi}(x)=1$ if $|x|\le 1$, and $\wti{\chi}(x)=0$ if $|x|\ge 2$. 
  We use the scaled version $\wti{\chi}_L(x)= \wti{\chi}(x/L)$, which is a smoothed out projection inside a 
  centered ball of radius $\sim L$. 
  
  Use again $\xi=\chi_R e^F$ with $\chi_R$ from 
  \eqref{eq:def chi_R} and now also $\zeta_L=\wti{\chi}_L\xi$. 
  Then $\zeta_L^2\psi$ is in $H^1$ and has 
  compact support. The weak form of the eigenvalue 
  equation and the IMS localization formula again show 
  \begin{align}\label{eq:zero energy resonance-1}
  	0= \la \zeta_L^2\psi, H\psi\ra 
  		= \la \zeta_L\psi, H\zeta_L\psi \ra 
  			-\la \psi, |\nabla\zeta_L|^2\psi \ra \, .
  \end{align}
  One calculates  
  \begin{align*}
  	|\nabla\zeta_L|^2 
  	  &= |\nabla F|^2\wti{\chi}_L^2\chi_R^2e^{2F}  
  	  		+\wti{\chi}_L^2 e^{2F}
  	  			\left( |\nabla\chi_R|^2 +2 \chi_R\nabla\chi_R\nabla F \right) 
  	  		+ \chi_R^2 e^{2F} |\nabla\wti{\chi}_L|^2 \\
  	  &\phantom{=~~}
  	  		+2 \chi_R\wti{\chi}_L \nabla \wti{\chi}_L
  	  			\left(\nabla\chi_R +\chi_R\nabla F \right)e^{2F}\\
  	  &= |\nabla F|^2\zeta_L^2 
  	  		+ e^{2F}
  	  			\left( |\nabla\chi_R|^2 +2 \chi_R\nabla\chi_R\nabla F \right) 
  	  		+  \xi^2 \left(|\nabla\wti{\chi}_L|^2
  	  			+2 \wti{\chi}_L \nabla \wti{\chi}_L
  	  			\nabla F \right)\
  \end{align*}
  where the last equality holds for $L>R$, since then $\nabla\wti{\chi}_L$ and $\nabla\chi_R$ have disjoint supports and $\wti{\chi}_L=1$ on the support of $\nabla\chi_R$. 
  Reshuffling terms in \eqref{eq:zero energy resonance-1} we get
  \begin{equation}\label{eq:zero energy resonance-2}
   \begin{split}	
  	\la \zeta_L\psi, \left(H-|\nabla F|^2\right)\zeta_L\psi \ra 
		&= \la \psi, 
			    e^{2F}
  	  			\left( 
  	  				|\nabla\chi_R|^2 +2 \chi_R\nabla\chi_R\nabla F 
  	  			\right) 
			\psi \ra \\
		&\phantom{=~~} 
		  + \la \psi,   \xi^2 \left(|\nabla\wti{\chi}_L|^2
  	  			+2 \wti{\chi}_L \nabla \wti{\chi}_L
  	  			\nabla F \right)\psi\ra\, .
   \end{split}
  \end{equation} 
  An improvement of Hardy's inequality outside large balls given in 
  \cite{Lun09} shows  
  \begin{align}\label{eq:hardy-improved}
  	\la \nabla\varphi, \nabla\varphi \ra 
  	\ge \la \varphi, 
  			\Big(\big(\frac{d-2}{2}\big)^2 \frac{1}{|x|^2} 
  				+\frac{1}{4|x|^2\ln^2(|x|)}
  			\Big)
  		\varphi \ra 
  \end{align} 
  for any $\varphi\in H^1_0(B_t(0)^c)$ and $t$ large enough. Thus 
  \begin{align}\label{eq:zero energy resonance-3}
  	 \la \zeta_L\psi, \left(H-|\nabla F|^2\right)\zeta_L\psi \ra 
		\ge 
			\la \zeta_L\psi, \frac{c}{|x|^2\ln^2(|x|)} \zeta_L\psi \ra
  \end{align}
  for some $c>0$, if the potential $V$ is short range and if 
  \begin{align}\label{eq:bound on gradient F one particle}
  	|\nabla F(x)| \le \frac{d-2}{2|x|}
  \end{align} 
  for all large $|x|$. 
  Note that if $F$ is bounded then $\xi=\chi_R e^F$ is bounded and 
  if \eqref{eq:bound on gradient F one particle} holds then  
  \begin{align}\label{eq:zero energy resonance-4}
  	 |\la \psi,   \xi^2 \left(|\nabla\wti{\chi}_L|^2
  	  			+2 \wti{\chi}_L \nabla \wti{\chi}_L
  	  			\nabla F \right)\psi\ra| 
  	  \lesssim \|\xi\|_\infty^2 
  	  	\frac{1}{L^2} \int_{L\le |x|\le \alpha L } |\psi(x)|^2\, dx 
  \end{align} 
  since $|\nabla\wti{\chi}_L|^2$ and 
  $|\nabla\wti{\chi}_L\nabla F|\lesssim L^{-2}\id_{\{L\le |x|\le \alpha L\}}$. 
  
  Using \eqref{eq:zero energy resonance-3} and 
  \eqref{eq:zero energy resonance-4} together with Fatou's lemma 
  in \eqref{eq:zero energy resonance-2} one arrives at 
  \begin{equation}
   \begin{split}	
  	\la \xi\psi, \frac{c}{|x|^2\ln^2(|x|)} \xi\psi \ra
		&\le \liminf_{L\to\infty} 
			\la \zeta_L\psi, \frac{c}{|x|^2\ln^2(|x|)} \zeta_L\psi \ra 
			\\
		&\le   \la \psi, 
			    e^{2F}
  	  			\left( 
  	  				|\nabla\chi_R|^2 +2 \chi_R\nabla\chi_R\nabla F 
  	  			\right) 
			\psi \ra \\
		&\le C_{F,R} \int_{R/2\le |x|\le R}|\psi(x)|^2\, dx	
   \end{split}
  \end{equation}
  with the same constant $C_{F,R}$ as in \eqref{eq:constant CFR}, since 
  $|\nabla\chi_R|\le \tfrac{4}{R}\ind_{\{R/2\le |x|\le R\}}$.  
  
  Now let $F(x)= \tfrac{d-2}{2}\ln(|x|)$, so that 
  \eqref{eq:bound on gradient F one particle} holds. 
  In the definition of $\xi$, replace $F$ by $F_\delta$ given in 
  \eqref{eq:F delta one particle} for $\delta>0$, i.e., 
  use $\xi=\chi_Re^{F_\delta}$ in the above argument.  
  Then, since $|\nabla F_\delta|\le |\nabla F|$ we see that 
  \begin{align*}
  	 \la \chi_R e^{F_\delta}\psi, 
  	 	\frac{c}{|x|^2\ln^2(|x|)} 
  	 \chi_R e^{F_\delta}\psi \ra
		\le C_{F,R} \int_{R/2\le |x|\le R}|\psi(x)|^2\, dx	\, .
  \end{align*}
  In addition, $F_\delta$ converges monotonically to $F$ and 
  $e^{F(x)}= |x|^{(d-2)/2}$, so monotone convergence gives 
  \begin{equation} \label{eq:resonances bound 2}
  	\begin{split}
  	 c\int_{|x|\ge R} 
  	 	\frac{|x|^{d-4}}{\ln^2(|x|)}|\psi(x)|^2
  	  \, dx
  	 &=
  	 	\lim_{\delta\to 0}
  	 		\la \chi_R e^{F_\delta}\psi, 
  	 			\frac{c}{|x|^2\ln^2(|x|)} 
  	 		\chi_R e^{F_\delta}\psi \ra \\
	 &\le C_R \int_{R/2\le |x|\le R}|\psi(x)|^2\, dx	\, .  		
  	\end{split}
  \end{equation}
  with constant $C_R$ given by 
  \begin{align*}
	C_R=8\sup_{R/2\le |x|\le R} e^{2F(x)}\left(2R^{-2}+ R^{-1}|\nabla F(x)| \right) 
	\le 8d R^{d-4} 
  \end{align*} 
  Since $\psi$ is locally $L^2$, this proves \eqref{eq:resonances bound}.
\end{proof}
\begin{remark}
  In dimensions $2\le d\le 4$ a repulsive tail of the potential $V$ 
  of the form
  \begin{align}
  	V(x)\ge \frac{\omega}{|x|^2} \quad \text{for } |x|\gg 1 
  \end{align}
 stabilizes zero energy bound states. In this case we can use 
 \eqref{eq:hardy-improved} to see that   
  \begin{align*}
  	\la \nabla\varphi, \nabla\varphi \ra +\la \varphi, V\varphi \ra 
  	\ge \big\la \varphi, 
  			\left(\Big(\Big(\frac{d-2}{2}\Big)^2+\omega\Big) \frac{1}{|x|^2} 
  				+\frac{1}{4|x|^2\ln^2(|x|)}
  			\right)
  		\varphi \big\ra 
  \end{align*}
  for all $\varphi\in H^1$ with support outside a large enough 
  centered ball in $\R^d$. Then the same analysis 
 which yields \eqref{eq:resonances bound 2}, but now with the weight 
 $F(x)= \big(((d-2)/2)^2+\omega\big)^{1/2}\ln |x|$, leads to 
 \begin{align}
     	 c\int_{|x|\ge R} 
  	 	\frac{|x|^{\sigma}}{\ln^2(|x|)}|\psi(x)|^2
  	  \, dx
		\le 
			C  \int_{R/2\le |x|\le R}|\psi(x)|^2\, dx	
 \end{align}
 with $\sigma=  2\big(((d-2)/2)^2+\omega \big)^{1/2}-2$ and some 
 constant $C$ depending on $R$ and $\omega$.  
 Thus $\psi\in L^2(\R^d)$ as soon as $\sigma>0$ and the 
 last condition is equivalent to 
 \begin{align}
 	\omega > \frac{d(4-d)}{4} \, . 
 \end{align}
 Clearly $d(4-d)$ is positive if $d\le 3$, zero if $d=4$, 
 and negative if $d\ge 5$. So a repulsive part stabilizes 
 zero energy bound state in dimensions $d\le 4$ and in dimensions 
 $d\ge 5$ the potential $V$ can be purely attractive and still 
 have zero energy eigenstates. The results of 
 \cite{HunJexLan23-Potential} show that a repulsive part is   
 needed in dimensions $d\le 4$. Dimension $d=4$ is critical in the 
 sense that the repulsive part can be weaker in dimension 
 four than for  $d\le 3$.  
\end{remark}

%%%%%%%%%%%%%%%%%%%%%%%%%%%%%%%%%%%%%%%%%%%%%%%%%%%%%%%%%%%%%%%%%%%
%%%%%%%%%%%%%%%%%%%%%%%%%%%%%%%%%%%%%%%%%%%%%%%%%%%%%%%%%%%%%%%%%%%
\section{Local energy bound for helium--type system in paraboloidal regions}\label{sec:local energy bound}
\begin{figure}
\centering
\begin{tikzpicture}
    \node[anchor=south west,inner sep=0] (image) at (0,0) {\includegraphics[width=0.4\textwidth]{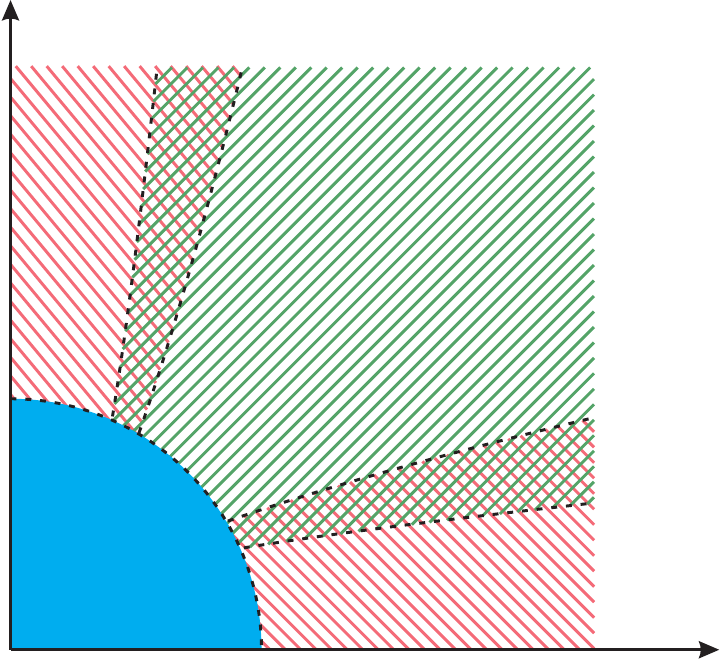}};
    \begin{scope}[x={(image.south east)},y={(image.north west)}]
        \draw (0.97,0) node[below] {$r_1$};
                \draw (-0.04,.33) node[above] {$R_0$};
  \draw (0.38,0.01) node[below] {$R_0$};
                  \draw (-0.04,.91) node[above] {$r_2$};
                          \draw (0.9,0.1) node[above] {\textcolor{red}{$A_1^-$}};
                             \draw (0.9,0.9) node[above] {\textcolor{teal}{$A_2$}};
                             \draw (0.15,0.9) node[above] {\textcolor{red}{$A_1^+$}};
                          \draw (-0.04,0.14) node[above] {\textcolor{darkerblue}{$A_0$}};
    \end{scope}
\end{tikzpicture}
\caption{Sketch of the regions $A_0,A_1,A_2$. 
 In $A_0$ both particles are close to nucleus and in $A_2$ both particles are far away. The paraboloidal region $A_1$, which splits into two disjoint components $A_1^-,A_1^+$, corresponds to the situation where one particle is far away while the other stays relatively close to the nucleus. 
In region $A_1$ we get a local boost in energy, which is important when the system becomes critical. }\label{fig:regions} 
\end{figure}

%%%%%%%%%%%%%%%%%%%%%%%%%%%%%%%%%%%%%%%%%%%%%%%%%%%%%%%%%%%%%%%%%%%
%%%%%%%%%%%%%%%%%%%%%%%%%%%%%%%%%%%%%%%%%%%%%%%%%%%%%%%%%%%%%%%%%%%
In the following sections, we consider a helium--type atom 
consisting of an infinitely heavy nucleus at the origin 
and two indistinguishable electrons. 
In this section, we prove a local energy bound for these systems, 
which is the main tool for the proof of the sharp anisotropic 
upper bounds on the asymptotic behavior of 
the ground state in later sections. 
To obtain suitable energy bounds we need appropriate lower bounds on the potential derived in Lemma \ref{lem: Coulomb lower bound}. The crucial part of the proof is to split the configuration space into several regions sketched in Figure \ref{fig:regions} where the behavior of the system differs. 
The bounds on the potential from Lemma \ref{lem: Coulomb lower bound} are then used to derive local energy bounds for the full quantum system in Proposition \ref{prop:loc energy bound}

Our local energy bound for the helium--type operator $H_U$ on 
$L^2(\R^6)$ is independent of the statistics of the particles 
and one can easily include the spin. The importance of such a local  
energy bound was already emphazised by Agmon in his work on 
non-isotropic upper bounds on the decay of eigenfunctions 
of multiparticle Schr\"odinger operators in  \cite{Agm82}. 

We denote by $x_j$ the position of the $j^{\text{th}}$ particle, 
by $P_j^2$ its kinetic energy, 
where $P_j=-i\nabla_{x_j}$ is its momentum operator, and $j=1,2$.
The Hamiltonian of this system is given by 
\begin{equation}\label{eq:HeliumHamiltonian-recall1}
	H_U = P_1^2 + P_2^2 - \frac{1}{|x_1|} - \frac{1}{|x_2|} + \frac{U}{|x_1 - x_2|}\, .
\end{equation}
It is well-defined and self-adjoint on the Sobolev space 
$\calD(H_U) = H^2(\R^6)\subset L^2(\R^6)$. 
In the following, it will be very convenient to consider 
the quadratic form induced by $H_U$ 
on the Sobolev space $H^1(\R^6)$. With a slight abuse of 
notation, we denote this quadratic form by  
  \begin{align}
	\la\varphi, H_U\varphi\ra 
		\coloneqq 
			\la P_1\varphi, P_1\varphi\ra + \la P_2\varphi, P_2\varphi\ra 
			+ \la \varphi, V_U\varphi\ra 
  \end{align}  
where 
 \begin{align}
 	V_U(x)=V_U(x_1,x_2)= - \frac{1}{|x_1|} -\frac{1}{|x_2|}+\frac{U}{|x_1-x_2|}\, .
 \end{align}
The proofs of our lower bounds in Sections 
\ref{sec:lower bound tricky region} and \ref{sec:global lower bound}  
use the sesquilinear version of this quadratic form. 
Our local energy bound for helium follows from a suitable 
localization in position space and the 
IMS localization formula. 
However, our localization is quite different from the usual 
approach in many-particle physics, where one 
localizes into conical regions, see for example  \cite{FraLieSei12}. 
For our derivation of the sharp upper and lower bounds, 
it turns out to be crucial to use  
paraboloidal regions. 

Recall that for $x=(x_1,x_2)\in \R^3\times\R^3$ we set $|x|=|(x_1,x_2)|= (|x_1|^2+|x_2|^2)^{1/2}$, $|x|_0 = \min(|x_1|,|x_2|)$, and 
$|x|_\infty=\max(|x_1|,|x_2|)$. 
For $R_0 >0$ and $0<\gamma\le 1$ we define the following paraboloidal regions
\begin{equation}\label{eq:regions}
  \begin{split}
    \text{Region 0:~} &  A_0\coloneqq \{ x\in \R^6:\, |x|< 2 R_0 \}    \\
    \text{Region 1:~}  & A_1\coloneqq \{ x\in \R^6:\, |x|> R_0 \text{ and } |x|_0 < 2 |x|_\infty^\gamma  \}. \\
    \text{Region 2:~}  & A_2\coloneqq \{ x\in \R^6:\, |x|>  R_0 \text{ and } |x|_0 >    |x|_\infty^\gamma \}. 
  \end{split}
\end{equation}
We do not write explicitly the dependence of these regions on 
the parameters $R_0$ and $\gamma$.
The inner region $A_0$ cuts out a ball centered around the nucleus. 
It is only necessary in order to 
have a bounded localization  error in the IMS localization formula 
near the nucleus and will be 
irrelevant for our bounds on eigenstates. 
All regions above are invariant under permutation of the particles. 
In order to control 
the localization error, we will eventually choose the parameter $1/2<\gamma<1$.

In the region $A_0$ both particles are close to the nucleus and 
in $A_2$ both particles will be far from the 
nucleus. In these two cases, the local energy bound will be easy. 
The \emph{tricky region} is $A_1$. At critical coupling $U_c$ this 
region is critical in the sense that 
inside this region one particle stays much closer to the nucleus 
compared to the other particle, which can escape to 
infinity with \emph{no apparent energy cost}, 
since the \emph{ionization energy is zero}.  

For large $R_0$ the tricky region $A_1$ is the union of two 
disjoint paraboloid regions
\begin{align*}
  A_{1}^-
  	&= 
  		\{ x\in \R^6:\, |x|>  R_0 \text{ and } |x_2| <   2|x_1|^\gamma \}\, , \\
	A_{1}^+
		&= \{ x\in \R^6:\, |x|>  R_0 \text{ and } |x_1| <   2|x_2|^\gamma \}. 
\end{align*}
which are ``above" and ``below" the diagonal $|x_1|=|x_2|$, respectively.  

The reason why we introduce these paraboloidal regions is that in these regions, especially in the tricky region $A_1$, the full potential has nice lower bounds  
which exploit the interaction term but do not involve it explicitly. 
This is made precise in the following lemma. 
\begin{samepage}
\begin{lemma}\label{lem: Coulomb lower bound}
	Let $V_U$ be the Coulomb potential 
	\begin{align}
		V_U(x)
			= -\frac{1}{|x_1|} - \frac{1}{|x_2|} +\frac{U}{|x_1-x_2|}
	\end{align}
	 on $\R^6$. Then for $0<\gamma<1$ we have 
	 \begin{align}
	 	V_U(x) &\ge -\frac{1}{|x|_0} 
	 				+ \frac{U-1}{|x|_\infty} -\frac{2U }{ |x|_\infty^{2-\gamma}} 
	 				\quad \text{for all } x\in A_1\, ,\label{eq:Coulomb lower bound region 1} \\
	 	V_U(x) &\ge -\frac{1}{|x|_0}-  \frac{1}{|x|_\infty}
					\quad \text{for all } x\in A_2\, \label{eq:Coulomb lower bound region 2}.
	 \end{align} 
\end{lemma}
\end{samepage}
\begin{remark}
  In the \emph{tricky region} $A_1$, where one particle stays close to the 
  nucleus, and the other one can escape, the bound  
  \eqref{eq:Coulomb lower bound region 1} shows that this 
  will have a \emph{local energy cost} of order $(U-1)/|x|_\infty$, 
  which makes the classically forbidden region 
  \emph{stickier}. This is a key input for our a--priori bounds on 
  the decay of ground states at critical coupling.  
  A similar boost was already noted in \cite{FraLieSei12,HofOstHofOstSim83}, however, due to the use of conical regions in \cite{FraLieSei12}, 
  which is quite common in many--body quantum mechanics, 
  their bound for the Coulomb interaction in the tricky region is worse 
  than our bound in Lemma \ref{lem: Coulomb lower bound}: It is 
  \emph{essential to use paraboloidal regions} in order to get the 
  additional positive term  $\frac{U-1}{|x|_\infty}$ with the 
  \emph{sharp constant} $U-1$ in the lower bound for 
  two--particle Coulomb potential in the tricky region. The use of 
  conical regions would not allow us to get sharp upper and lower bound 
  on the  anisotropic decay of the ground state, one always incurs   
  an epsilon loss in the leading order terms when using conical regions.  
\end{remark}

\begin{proof}
  The bound on $A_2=\{ |x| > R_0,\, |x|_0 > |x|_\infty^\gamma\}$ 
  follows by dropping the 
  positive Coulomb repulsion.
  On $A_1$ we have  $|x|_0 < 2|x|_\infty^\gamma$, thus   
  \begin{align*}
  	|x_1-x_2| \le |x_1|+|x_2| \le |x|_\infty +2 |x|_\infty^\gamma  
  \end{align*}
  and 
  \begin{align*}
  	\frac{U}{|x_1-x_2|} \ge  \frac{U}{|x|_\infty+ 2|x|_\infty^\gamma} 
  			= \frac{U}{|x|_\infty} - \frac{2U}{|x|_\infty^{2-\gamma}(1+ 2|x|_\infty^{\gamma-1})} \, .
  \end{align*}
  This implies \eqref{eq:Coulomb lower bound region 1}. 
\end{proof}

In order to use the bounds from Lemma \ref{lem: Coulomb lower bound}, we need to localize the kinetic energy into the different regions $A_0, A_1$, and $A_2$.  
To construct suitable cut--off functions which localize into these  
regions,  take any two functions 
$0\le \tilde{u}, \tilde{v}\in \calC^\infty([0,\infty))$ with 
$\tilde{u}=1$ and $\tilde{v}=0$ on $[0,1]$, $\tilde{u}=0$ and 
$\tilde{v}=1$ on $[2,\infty)$,  $\tilde{u}>0$ on 
$[0,7/4]$, and $\tilde{v}>0$ on $[5/4,\infty)$ and put   
\begin{align}
	u\coloneqq \frac{\tilde{u}}{(\tilde{u}^2 + \tilde{v}^2)^{1/2}}\, ,\, \quad 
	v\coloneqq \frac{\tilde{v}}{(\tilde{u}^2 + \tilde{v}^2)^{1/2}}\, .
\end{align}
Since $\tilde{u}^2 + \tilde{v}^2\ge c$ for some constant $c>0$, 
the functions $u,v$ are infinitely differentiable. 
Moreover, by construction, $0\le u,v\le 1$, $u=1$ and $v=0$ on 
$[0,1]$, $u=0$ and $v=1$ on $[2,\infty)$, and 
\begin{align}
  u^2+v^2 =1 	
\end{align}
Note that one can always find $u,v$ with $ \|u'\|_\infty\le 2$ 
and $\|v'\|_\infty \le 2$.
Given $u$ and $v$ we set 
\begin{equation}\label{eq:localizing functions}
\begin{split}
	\chi_0(x) & \coloneqq u\Big(|x|/R_0 \Big)\, , \\
	\chi_1(x) & \coloneqq v\Big(|x|/R_0 \Big) u\Big(|x|_0/|x|_\infty^\gamma \Big)\, , \\
	\chi_2(x) & \coloneqq v\Big(|x|/R_0 \Big) v\Big(|x|_0/|x|_\infty^\gamma \Big)\, .
\end{split}
\end{equation}
By construction   
\begin{align}\label{eq:quadratic partition of unity}
	\sum_{j=0}^2 \chi_j^2 = 1.
\end{align}
In addition, we clearly have $\chi_0\in\calC^{\infty}_0(\R^6)$. 
This is less obvious for $\chi_1$ and $\chi_2$. 
For large enough $R_0$  the support of $\chi_1$, which is contained 
in the tricky region, separates into two 
disjoint sets (contained in $A_1^\pm$) and on each of these sets 
$\chi_1$ is infinitely differentiable. So we have 
$\chi_1\in\calC^\infty(\R^6)$ for large  $R_0$  and 
a moment of reflection shows that also 
$\chi_2\in\calC^\infty(\R^6)$ for all large enough $R_0$. 
The gradients in $\R^6$ of $\chi_0$ and $\chi_1$ are given by 
\begin{align}
	\nabla &\chi_0(x)
		= R_0^{-1} u'\big(|x|/R_0 \big) \frac{x}{|x|}\, ,\\
	\nabla &\chi_1(x) 
		= R_0^{-1} v'\big(|x|/R_0 \big) u\big(|x|_0/|x|_\infty^\gamma \big) \frac{x}{|x|} \nonumber \\
		&\phantom{=}	+  v\big(|x|/R_0 \big) u'\big(|x|_0/|x|_\infty^\gamma \big) 
				\begin{pmatrix}
					\big(
						\frac{1}{|x_2|^\gamma } \ind_{\{|x_2|>|x_1|\}} 
						-\gamma\frac{|x_2|}{|x_1|^{\gamma+1}}
						 \ind_{\{|x_2|<|x_1|\}}
					\big)\frac{x_1}{|x_1|} \\
					\big(
						\frac{1}{|x_1|^\gamma } \ind_{\{|x_1|>|x_2|\}} 
						-\gamma\frac{|x_1|}{|x_2|^{\gamma+1}}
						 \ind_{\{|x_1|<|x_2|\}}
					\big)\frac{x_2}{|x_2|}
				\end{pmatrix}\, .
\end{align}
Thus, abbreviating $s=|x|/R_0$ and $t=|x|_0/|x|_\infty^\gamma$,  
\begin{align}\label{eq:bound norm gradient chi_0}
	|\nabla\chi_0(x)|^2\le R_0^{-2} \big(  u'(s) \big)^2 
	\le R_0^{-2} \|u'\|_\infty^2 \ind_{\{R_0\le |x|\le 2R_0\}}
\end{align}
and 
\begin{equation}
\begin{split}\label{eq:bound norm gradient chi_1}
	|\nabla\chi_1(x)|^2 
		&=
			R_0^{-2} v'(s)^2 u(t)^2  
			+ 2 R_0^{-1}  v'(s)  v(s) u'(t) u(t) |x|^{-1}
				( t-\gamma t) \\
		&\phantom{=~~}	
			+ v(s)^2 u'(t)^2 |x|_\infty^{-2\gamma} 
				\Big(1  + \gamma^2 t^2\Big) \, .
\end{split}	
\end{equation}
For $\nabla \chi_2$ we note that a similar formula as for 
$\nabla\chi_1$ holds, just with $u$ replaced by $v$. 
Collecting terms we get  
\begin{align*}
	|\nabla&\chi_1(x)|^2 +|\nabla\chi_2(x)|^2 \\
		&= 
			R_0^{-2} v'(s)^2 \big( u(t)^2 +v(t)^2\big)  
			+  2 R_0^{-1}  v'(s)  v(s) \big( u'(t) u(t) + v'(t) v(t)\big) |x|^{-1}(1-\gamma)t \\ 
		&\phantom{=~~}	
			+ v(s)^2 \big( u'(t)^2  + v'(t)^2 \big) 
				|x|_\infty^{-2\gamma} 
				\Big(1  + \gamma^2 t^2\Big)\\
		&= R_0^{-2}  v'(s)^2 
			+  v(s)^2 \Big( u'(t)^2  + v'(t)^2 \Big) 
				|x|_\infty^{-2\gamma} 
				\Big(1  + \gamma^2 t^2\Big)
\end{align*} 
where we also used $u^2+v^2=1$ and $2(u'u + v'v) = (u^2+v^2)'=0$. Hence
\begin{equation}\label{eq:loc error}
 \begin{split}
  \sum_{j=0}^2 |\nabla\chi_j(x)|^2 
  	&= 
  		R_0^{-2} \big( v'(|x|/R_0)^2 +  u'(|x|/R_0)^2 \big) \\
	&\phantom{= ~~}	+  v(|x|/R_0)^2  \big( u'\big(|x|_0/|x|_\infty^\gamma \big)^2  + v'\big(|x|_0/|x|_\infty^\gamma \big)^2 \big) 
				\Big( \frac{1}{|x|_\infty^{2\gamma}}  + \frac{\gamma^2|x|_0^2}{|x|_\infty^{2\gamma+2}}\Big) \, .
 \end{split}
\end{equation}
This yields the following bound on the localization error for localizing into paraboloidal regions. 
\begin{lemma}\label{lem: loc error bound}
	The localization error 
	$ \sum_{j=0}^2|\nabla\chi_j|^2$ for the localizing functions 
	$\chi_j$ constructed above, is bounded from above by 
	\begin{equation}\label{eq: loc error bound}
	  \begin{split}
		\sum_{j=0}^2 |\nabla\chi_j(x)|^2 
			&\le 
				 \|(u')^2+ (v')^2\|_\infty
				\left( 
					R_0^{-2} \ind_{\{R_0\le |x|\le 2R_0\}} 
					+ \frac{1+\gamma^2}{|x|_\infty^{2\gamma}} \ind_{\{|x|\ge R_0\}}  
				\right)
	  \end{split}
	\end{equation}
   for $0<\gamma\le 1$ and  $R_0>0$. 
\end{lemma}
\begin{proof}  
This follows immediately from  \eqref{eq:loc error}  because 
$0\le u,v\le 1$,  $R_0\le |x|\le 2 R_0$ on the support of 
$v'(|x|/R_0)$, $|x|\ge R_0$ on the support of $v(|x|/R_0)$, 
and $ |x|_0\le |x|_\infty$.  
\end{proof}
\begin{remark}  
  In order to make $\|(u')^2+ (v')^2\|_\infty$ small, 
  a convenient choice is to pick  
  \begin{align*}
  	u(s) = 	\left\{ \begin{array}{lr}
  						1, & \text{for } 0\le s\le 1\\
  						\cos(\pi(s-1)/2), &\text{for } 1<s\le 2\\
  						0, &\text{for } s>2
  					\end{array}
 			\right.
  \end{align*}
  and $v=\sqrt{1-u^2}$, in which case 
  $\|(u')^2 + (v')^2\|_\infty = \pi^2/4$. 
  Making this choice for $u$ and $v$ 
  does not yield smooth cut-off function, however.  
  Nevertheless, the set on which $\chi_j$ is not differentiable, 
  is a Lebesgue measure zero Lipshitz hypersurface in $\R^6$, 
  so Lemma \ref{lem:IMS domains ok} shows that  
  $\chi_j\in W^{1,\infty}(\R^6)$  
  and this is all one needs to use a suitable quadratic form version 
  of the  IMS  localization formula, 
  see the discussion in Appendix \ref{app:IMS formula}. 
\end{remark}

Given a wave function $\varphi$ in the quadratic form domain of 
$H_U$, which 
is the Sobolev space $H^1(\R^6)$, we can use 
\eqref{eq:quadratic partition of unity} and the fact that 
$\la\varphi, H_U\varphi \ra$ is real together with the IMS 
localization formula \eqref{eq:IMS2} to get  
\begin{equation}
\begin{split}\label{eq:IMS application 1}
		\la\varphi, H_U\varphi \ra 
		&=  \sum_{j=0}^2 \re \la \chi_j^2\varphi, H_U\varphi\ra 
			= \sum_{j=0}^2 \la \chi_j\varphi, H_U\chi_j\varphi\ra 
				-\sum_{j=0}^2 \la \varphi, |\nabla\chi_j|^2\varphi\ra \, .
\end{split}
\end{equation}
Lemma \ref{lem: loc error bound} allows us to control the 
localization error. To bound the local energies 
$\la \chi_j\varphi, H_U\chi_j\varphi\ra $, where each 
$\chi_j\varphi$ is supported on $A_j$, 
the following bounds will be useful.

In the following, we take any two functions $u$ and $v$ as above. 
Then Lemma \ref{lem: loc error bound} 
shows that there exists a constant $C_l= C_l(u,v)<\infty$ such that for any $0<\gamma\le 1$ the localization error 
is bounded from above by 
  \begin{align}\label{eq:loc error bound applied}
  	\sum_{j=0}^2|\nabla\chi_j(x)|^2 \le \frac{C_l}{\max(|x|_\infty^{2\gamma},R_0^{2\gamma})} 
  	\text{ for all } x\in\R^6\, .
  \end{align}
For this constant $C_l$ we set 
\begin{align}
	W_0(x)&\coloneqq -\frac{1}{2} - \frac{C_l}{R_0^{2\gamma}}\, ,\\
	W_1(x)&\coloneqq -\frac{1}{4} 
	 				+ \frac{U-1}{|x|_\infty} -\frac{2U\, }{|x|_\infty^{2-\gamma}} 
	 				- \frac{C_l}{|x|_\infty^{2\gamma}} \label{eq:W1} \, ,\\
\intertext{and}
	W_2(x)&\coloneqq -\frac{1}{|x|_\infty}-  \frac{1}{|x|_0} 
						- \frac{C_l}{|x|_\infty^{2\gamma}} \, . \label{eq:W2}
\end{align}
Lemma \ref{lem: Coulomb lower bound} and the IMS formula lead to 
\begin{proposition}[Local energy bound] \label{prop:loc energy bound}
	For large enough $R_0$ and all $\varphi$ in the quadratic form 
	domain of $H_U$ 
	\begin{align}
		\la \varphi, H_U\varphi\ra \ge \sum_{j=0}^2 \la\chi_j\varphi, W_j \chi_j\varphi \ra\, .
	\end{align}
\end{proposition}
\begin{remark}
  The localized functions $\chi_j\varphi$, $j=0,1,2$, have the 
  same permutation symmetry as $\varphi$, since the 
  $\chi_j$ are symmetric under permurtation of the particles. 
\end{remark}
\begin{proof}
	Set $\locerror(x)= \frac{C_l}{\max(|x|_\infty^{2\gamma},R_0^{2\gamma})} $. 
	From \eqref{eq:IMS application 1},  
	Lemma \ref{lem: loc error bound}, and \eqref{eq:loc error bound applied} 
	we get 
	\begin{align*}
		\la\varphi, H_U\varphi \ra 
			\ge \sum_{j=0}^2 \la \chi_j\varphi, H_U\chi_j\varphi\ra - \la \varphi, \locerror \varphi\ra 
			= \sum_{j=0}^2 \la \chi_j\varphi, H_U\chi_j\varphi\ra 
				-\sum_{j=0}^2 \la \chi_j\varphi, \locerror \chi_j\varphi\ra 
	\end{align*}
  since $\locerror$ is multiplication by a function and $\sum_{j=0}^2\chi_j^2=1$. 
  So it is enough to show 
  \begin{align}\label{eq: local energy lower bound}
  	\la \chi_j\varphi, H_U\chi_j\varphi\ra
  		\ge \la \chi_j\varphi, \wti{W}_j\chi_j\varphi\ra
  \end{align}
 for each $j=0,1,2$, where $\wti{W}_j= W_j +\locerror$. 
 Using that $\chi_2\varphi$ is supported in the region $A_2$, we 
 can use Lemma \ref{lem: Coulomb lower bound} and immediately 
 get \eqref{eq: local energy lower bound} for $j=2$ by dropping 
 the kinetic energy term 
 $P_1^2+ P_2^2$ in $\la \chi_2\varphi, H_U\chi_2\varphi\ra$. 
 
 For $j=0$ we again drop the Coulomb repulsion term and use  
 \begin{align*}
 	H_U \ge P_1^2-\frac{1}{|x_1|} +P_2^2- \frac{1}{|x_2|}\ge -1/2 
 \end{align*}
 to get \eqref{eq: local energy lower bound}, since the ground 
 state energy of hydrogen is $-1/4$ in the atomic units we use. 

When $j=1$, the particles are localized in the tricky region. 
For large enough $R_0$ the localization  
function $\chi_1$ is a sum $\chi_1=\chi_{1}^-+\chi_{1}^+$, 
where $\chi_{1}^\pm$ are smooth and have 
supports in $A_1^\pm$. In particular, their supports are disjoint 
for all large $R_0$. 
Using Lemma \ref{lem: Coulomb lower bound} and dropping $P_1^2$ we get 
  \begin{align*}
  	\la \chi_1^-\varphi, H_U\chi_1^-\varphi\ra 
  	  &\ge  \la \chi_1^-\varphi, 
  	  		 \left( 
  	  		 	P_2^2-\frac{1}{|x_2| } + \frac{U-1}{|x_1|} -\frac{2U}{|x_1|^{2-\gamma}}
  	  		 \right)\chi_1^-\varphi\ra \\
  	  &\ge \la \chi_1^-\varphi, 
  	  		  \left( 
  	  		 	-\frac{1}{4} + \frac{U-1}{|x_1|} -\frac{2U}{|x_1|^{2-\gamma}}
  	  		 \right)\chi_1^-\varphi\ra \\
  	  &= \la \chi_1^-\varphi, \wti{W}_1\chi_1^-\varphi\ra 
  \end{align*}
  since $P_2^2- \frac{1}{|x_2|}\ge -1/4$. Similarly one sees 
   \begin{align*}
  	\la \chi_1^+\varphi, H_U\chi_1^+\varphi\ra 
  	  &\ge  \la \chi_1^+\varphi, \wti{W}_1\chi_1^+\varphi\ra \, .
  \end{align*}
  Since the supports of $\chi_1^-$ and $\chi_1^+$ do not overlap 
  when $R_0$ is  large enough, we get  
  \begin{align*}
  	\la \chi_1\varphi, H_U\chi_1\varphi\ra 
  	  &= \la \chi_1^-\varphi, H_U\chi_1^-\varphi\ra + \la \chi_1^+\varphi, H_U\chi_1^+\varphi\ra \\
  	  &\ge  \la \chi_1^-\varphi, \wti{W}_1\chi_1^-\varphi\ra + \la \chi_1^+\varphi,  \wti{W}_1\chi_1^+\varphi\ra 
  	  	=   \la \chi_1\varphi, \wti{W}_1\chi_1\varphi\ra 
  \end{align*}
  which is \eqref{eq: local energy lower bound} for $j=1$. 
\end{proof}

%%%%%%%%%%%%%%%%%%%%%%%%%%%%%%%%%%%%%%%%%%%%%%%%%%%%%%%%%%%%%%%%%%%%%
\section{Isotropic upper bounds on the asymptotic decay of bound states}\label{sec:isotropic}
%%%%%%%%%%%%%%%%%%%%%%%%%%%%%%%%%%%%%%%%%%%%%%%%%%%%%%%%%%%%%%%%%%%%%
Before we start the proof of the sharp anisotropic upper bound in the next section, we give a proof of 
a  simple isotropic bound. Such an upper bound is already quite useful, since it is \emph{uniform} in the 
energy $E\le -1/4$, i.e. up to the bottom of the essential spectrum. 
It allows for an easy proof of the \emph{existence of a bound state} at critical coupling: The infimum of the 
spectrum of $H_{U_c}$ is a simple eigenvalue, see Appendix \ref{app:basic properties}, even 
though it is embedded at the edge of the essential spectrum of $H_{U_c}$.  

Our main tools are the local energy bounds from  Section \ref{sec:local energy bound} and the 
quadratic form version of the IMS localization formula from Appendix  \ref{app:IMS formula}.
For the simple isotropic upper bound on the asymptotic decay of eigenfunctions of $H_U$ we use 
the weight function 
\begin{align}\label{eq:F isotropic}
  F_1(r)= 2(U-1)^{1/2}r^{1/2} -K r^\kappa
\end{align}
    for $1<U\le U_c$, $K>0$, and $1/6<\kappa<1/2$. The reason why 
    we have to take $\kappa>1/6$ in the isotropic upper bound will follow from the proof, in particular,  \eqref{eq:why kappa > 1/6}.  
    Note that the leading order term in 
  $F_1$ is given by $2(U-1)^{1/2}r^{1/2}$, the other term is a \emph{lower order correction} for any 
  $K>0$ and $1/6<\kappa<1/2$. 
    
    When $x=(x_1,x_2)\in\R^3\times\R^3$, $|x|_\infty=\max(|x_1|,|x_2|)$, we will identify  
  $F_1(x)= F_1(|x|_\infty)$, by a slight abuse of notation.  
For simplicity of notation, we 
  do not explicitly write the dependence of $F_1$ on its parameters.

  Recall that  $\psi\in L^2(\R^6)$ is a bound state of $H_U$ with energy $E$, if it is a weak solution of the 
  eigenvalue equation $H_U \psi=E\psi$. That is, $\psi\in H^1(\R^6)$ and  
  \begin{align}\label{eq:weak eigenfunction helium}
  	\la \varphi, H_U\psi \ra = E\la \varphi,\psi \ra
  \end{align}
 as quadratic forms for all $\varphi\in H^1$. 
 
\begin{theorem}[Isotropic $L^2$ upper bound near critical coupling] \label{thm:isotropic upper bound}
  If $\psi_U$ is a bound state of $H_U$ with energy $E_U\le -\frac{1}{4}$
  then for any $K>0$ and $1/6<\kappa<1/2$ 
  \begin{align}
  	e^{F_1}\psi_U\in L^2(\R^6)
  \end{align}   
 Moreover, for normalized $\psi_U$, the function $e^{F_1}\psi_U$ is $L^2$ uniformly in the 
 parameter range $1+\mu\le U\le U_c$ for any small fixed 
 $0<\mu<U_c-1$. 
\end{theorem}

Before we give the proof, we want to explain where the usual 
strategy fails. 
Let $\psi$ be a bound state of a Schr\"odinger operator $H_U$ 
with energy $E$. 
Take any bounded, real valued function $\xi\in H^1$. 
Since $E$ is real we can use $\varphi=\xi^2\psi$ 
in \eqref{eq:weak eigenfunction helium} together with the 
IMS localization formula \eqref{eq:IMS} and 
the local energy bound from Proposition \ref{prop:loc energy bound} 
to see 
\begin{align*}
	E\|\xi\psi\|^2 
		&= 
		  	\re (E\la \xi^2\psi,\psi\ra) 
		  	= \re\la \xi^2\psi, H_U\psi\ra = \la \xi\psi, H_U\xi\psi\ra -\la \psi,|\nabla\xi|^2\psi\ra \\
		&\ge 
			\sum_{j=0}^2 \la\chi_j\xi\psi, W_j \chi_j\xi\psi \ra -\la \psi,|\nabla\xi|^2\psi\ra\, , 
\end{align*}
where $\chi_j$, defined in \eqref{eq:localizing functions}, are 
the functions localizing in the regions $A_j$, $j=0,1,2$.  

Now choose $\chi=\chi_R$ to be another smooth  cutoff function outside 
of a centered ball of radius  $R> 2R_0$ and make the ansatz 
$\xi=\chi e^F$ for some bounded function $F\in H^1$. 
Then 
\begin{align*}
	|\nabla \xi|^2
		=  \xi^2|\nabla F|^2 +2 \chi e^{2F}\nabla\chi\cdot\nabla F + e^{2F}|\nabla\chi|^2\, ,
\end{align*}
hence 
\begin{align}\label{eq:refined Agmon}
  \sum_{j=1}^2 \la\chi_j\chi e^F\psi, (W_j-E-|\nabla F|^2) \chi_j\chi e^F \psi \ra
  	\le 
  		\la \psi, e^{2F}(2\chi \nabla\chi\nabla F+|\nabla\chi|^2)\psi\ra \, ,
\end{align}
since $\chi_0$ and $\chi$ have disjoint supports, the missing $j=0$ term in the sum on the left 
hand side above is zero. 

The usual strategy for using \eqref{eq:refined Agmon} is to assume  
$W_j-E\ge c>0$. That is, one needs a spectral gap for the operator $H_U-E$ \emph{near infinity} 
(outside large enough balls), to have a safety distance to the essential spectrum. 
Under such an assumption, one gets from \eqref{eq:refined Agmon} 
\begin{align*}
  \delta c\|\chi e^F\psi\|^2 \le \la \psi, e^{2F}(2\chi \nabla\chi\nabla F+|\nabla\chi|^2)\psi\ra 
  \le C_{F,\chi}\|\psi\|^2\, ,
\end{align*}
for the exponentially weighted bound state $e^F\psi$ on the support of $\chi$, 
as long as $|\nabla F|^2\le (1-\delta)c$ for some small $\delta>0$. This 
condition allows $F$ to grow like $\sqrt{(1-\delta)c}|x|$.  
Of course, one has 
to remove the requirement that $F$ is bounded. This is easy, since the 
constant $C_{F,\chi}$ depends on $F$ and $\chi$ only on the support of 
$\nabla\chi$, which is compact. 
See the argument in the proof of Theorem \ref{thm:isotropic upper bound} below, in particular \eqref{eq:regularized}. 

However, the local energy bound $W_1-E_{U_c}= W_1+1/4$ goes to zero at infinity in the 
\emph{tricky region} $A_1$. Thus $c=0$, i.e., there is no safety distance to the essential spectrum anymore, 
when  $E_U=-1/4$ or when $E_U$ approaches $-1/4$ from below as $U\nearrow U_c$, 
and the above argument does not allow to control $e^F \psi$ anymore.  
So we    have to be more careful. 
\begin{proof}[Proof of Theorem \ref{thm:isotropic upper bound}]
  Let $a=(U-1)_+^{1/2}$, $1/6<\kappa<1/2$, $K>0$, and put 
  \begin{align}\label{eq:F isotropic modified} 
  	G(r)\coloneqq 2ar^{1/2} - Kr^\kappa/2 \, . 
  \end{align}
 We use $K/2$ instead of $K$ in the definition of $G$ to have some wiggle room which allows us to 
 absorb error terms later, see \eqref{eq:isotropic punchline 1} and \eqref{eq:isotropic punchline 2} 
 below. Note that  $G(r)$ is positive when $r$ is large enough, so 
 its regularized version
  \begin{align}\label{eq:regularized}
  	G_\delta(r) = \frac{G(r)}{1+\delta  G(r) } 
  \end{align}
  is well defined and bounded for all large enough $r$ and all $\delta>0$. 
  We also denote $G(x)= G(|x|_\infty)$ and the same for $G_\delta$,  
  with a slight abuse of notation.  
  Note that $G$ and $G_\delta$ are continuous on $\R^6$ and continuously differentiable on $\{|x_1|\not=|x_2|, |x|>R\}$ for large enough $R$.  
  
  Furthermore, let $0\le \chi\le 1$ be a smooth function on $[0,\infty)$ with $\chi(r)=0$ for $0\le r\le 1$, 
  $\chi(r)=1$ for $r\ge 2$, $|\chi'|\le 2$, and  put $\chi_R(x)=\chi(x/R)$. 
  Lemma \ref{lem:IMS domains ok} shows that $\xi=\chi_R e^{G_\delta}$ is in the Sobolev space $W^{1,\infty}(\R^6)$ and 
  multiplication with $\xi$ and $\xi^2$ leaves  the quadratic form domain of $H_U$ invariant.  Setting $r=|x|_\infty$ 
  one calculates  
  \begin{align}
 	\nabla G(x) = (ar^{-1/2} -K\kappa r^{\kappa-1}/2) 
 					\begin{pmatrix}
						\frac{x_1}{|x_1|} \ind_{\{|x_1|>|x_2|\}}\\
						\frac{x_2}{|x_2|} \ind_{\{|x_2|>|x_1|\}}
					\end{pmatrix}\, ,
  \end{align}
  \begin{align*}
  	\nabla G_\delta =\frac{\nabla G}{(1+\delta G)^{2}} \, .
  \end{align*}
  Thus 
  \begin{align*}
  	|\nabla G_\delta|^2 \le |\nabla G|^2 = \big(a|x|_\infty^{-1/2} - K\kappa |x|_\infty^{\kappa-1}/2 \big)^2\, ,
  \end{align*}
  for large enough $|x|_\infty$. 
  Using \eqref{eq:refined Agmon} with $F$ replaced by $G_\delta$ and taking $R$ such that 
  $G(x)>0$ for all $x$ in the support of $\chi_R$ one gets 
  \begin{equation}\label{eq:refined Agmon in G}
  	\begin{split}  
  	  \sum_{j=1}^2 &\la\chi_j\chi_R e^{G_\delta}\psi, (W_j-E-|\nabla G|^2) \chi_j\chi_R e^{G_\delta} \psi \ra \\
  	&\le 
  		\la \psi, e^{2G}(2|\nabla\chi_R||\nabla G|+|\nabla\chi_R|^2)\psi\ra 
  		\le 
  			C_R \|\psi\|^2
    \end{split}
  \end{equation}
  where we also used that $G_\delta(x)\le  G(x)$ and 
  $|\nabla G_\delta(x)|\le |\nabla G(x)|$ once $G(x)>0$. 
  The constant $C_R$ is given by 
  \begin{align*}
  C_R=4\sup_{R\le s\le 2R} e^{2G(s)}\big(|G'(s)|/R+1/R^2 \big)<\infty,
  \end{align*}  
  since $\nabla\chi_R$ is 
  supported on the annulus $R\le |x|\le 2R$ and $|\nabla\chi_R|\le \|\chi'\|_\infty/R\le 2/R$.  
  
  In the following, we will use $C$ for a generic constant, which may change from line to line. Recall that $0\le \veps_U=-1/4 -E_U$ is the ionization energy. On $A_2$ we have   
 \begin{align*}
 	W_2(x)-E_U-|\nabla G(x)|^2 
 		\ge  \frac{1}{4} +\veps_U - \frac{1}{|x|_\infty} - \frac{1}{|x|_\infty^{\gamma}} -\frac{C}{|x|_\infty^{2\gamma}} - |\nabla G(x)|^2 
 		\ge \frac{1}{8}
 \end{align*}
 for large enough $|x|_\infty$. On $A_1$ we get 
 \begin{equation}
 	\begin{split}\label{eq:why kappa > 1/6}
 	W_1(x)&-E_U-|\nabla G(x)|^2 
 			\ge  \veps_U+ \frac{U-1}{|x|_\infty} - \frac{2U}{|x|_\infty^{2-\gamma}} - \frac{C}{|x|_\infty^{2\gamma}} -|\nabla G(x)|^2 \\
 		&\ge \veps_U+ aK\kappa |x|_\infty^{\kappa-3/2} - (K\kappa/2)^2 |x|_\infty^{2(\kappa-1)} 
 				-2U|x|_\infty^{\gamma-2}  - C|x|_\infty^{-2\gamma} \\
 		&\ge \veps_U+  \frac{3}{4}aK\kappa |x|_\infty^{\kappa-3/2} 
				-2U|x|_\infty^{\gamma-2}  - C|x|_\infty^{-2\gamma}
	\end{split}
 \end{equation}
 for all large enough $|x|_\infty$, since $\kappa<1/2$, which implies $\kappa-3/2>2(\kappa-1)$. 
 The minimum of $\max(\gamma-2,-2\gamma)$ is attained at 
 $\gamma=2/3$. The choice $\gamma=2/3$ gives $\gamma-2=-2\gamma= -4/3$, which leads to the lower bound  
 \begin{align*}
 	W_1(x)-E_U-|\nabla G(x)|^2  
 		\ge  \veps_U+ \frac{3}{4}aK\kappa |x|_\infty^{\kappa-3/2} 
 				-(2U+C) |x|_\infty^{-4/3}
		\ge \veps_U+ \frac{aK\kappa}{2} |x|_\infty^{\kappa-3/2} 
 \end{align*}
 uniformly in $1+\mu\le U\le U_c\le 2$ for large enough $|x|_\infty$, since $\kappa >1/6$. 
 Putting everything together, we see that 
 \begin{align}\label{eq:isotropic upper bound 1} 
 	\sum_{j=1}^2 \chi_R\chi_j (W_j -E_U-|\nabla G|^2 ) \chi_j \chi_R 
 		\ge \big(\veps_U+ \frac{aK\kappa}{2} |x|_\infty^{\kappa-3/2}\big) \chi_R^2 
 		\ge \frac{\veps_U+ aK}{12} |x|_\infty^{-4/3} \chi_R^2 
 \end{align}
 for large enough $R$ and all bound state energies $E_U\le -1/4$, i.e., $\veps_U\ge 0$. Using \eqref{eq:isotropic upper bound 1} 
 in \eqref{eq:refined Agmon in G} we get  
 \begin{align}\label{eq:isotropic upper bound 2}
 	\la \chi_Re^{G_\delta}\psi, |x|_\infty^{-4/3}\chi_Re^{G_\delta}\psi \ra 
 	\le \frac{12 C_R}{\veps_U+ (U-1)_+^{1/2}K} \|\psi\|^2 \, 
 \end{align}
 for all $\delta>0$. The right hand side of \eqref{eq:isotropic upper bound 2} is independent of $\delta>0$, so 
 \begin{align*} 
 	\la \chi_R e^{G}\psi, |x|_\infty^{-4/3}\chi_R e^{G}\psi \ra 
 	=  	\lim_{\veps\to 0}\la \chi_Re^{G_\delta}\psi, |x|_\infty^{-4/3}\chi_Re^{G_\delta}\psi \ra 
 	\le \frac{12 C_R}{\veps_U+(U-1)_+^{1/2}K} \|\psi\|^2 
 \end{align*}
 by monotone convergence. Since arbitrary positive multiples of $r^{\kappa}$ control any logarithmic term 
 $\ln r$ for large $r$, we have 
 \begin{equation}\label{eq:isotropic punchline 1}
 	F_1(r)= 2ar^{1/2} - Kr^\kappa \le  2ar^{1/2} - Kr^\kappa/2 - \frac{2}{3}\ln r = G(r) - \frac{2}{3}\ln r
 \end{equation}
 for $r$ large. 
 This shows 
 \begin{align}\label{eq:isotropic punchline 2}
 	\|\chi_R e^{F_1}\psi\|^2\le \la \chi_R e^{G}\psi, |x|_\infty^{-4/3}\chi_R e^{G}\psi \ra 
\le \tfrac{12 C_R}{\veps_U+(U-1)_+^{1/2}K} \|\psi\|^2
 \end{align}
 for any $0\le U\le U_c$. It is easy to see that $U\mapsto \veps_U$ is decreasing and the discussion in the introduction shows that $\veps_1>0$. So $\veps_U+(U-1)_+^{1/2}K\gtrsim 1$ uniformly in $0\le U\le U_c$, hence 
   \eqref{eq:isotropic punchline 2} proves the theorem.
\end{proof}
Our isotropic upper bound also allows for a simple proof for the  existence of a ground state at critical coupling, see Proposition \ref{prop:basic properties} in the appendix. 
In the infinite mass approximation, the existence of a bound state at critical coupling had been proven first in \cite{HofOstHofOstSim83} with PDE methods and in \cite{FraLieSei12} 
with variational methods.

\begin{corollary}\label{cor:existence}
  At critical coupling the operator $H_{U_c}$ has a simple eigenvalue at the edge of its 
  essential spectrum. That is, there exists a bound state $\psi_c\in L^2(\R^6)$ with energy $-1/4$ 
  of $H_{U_c}$ which is unique up to a phase.  
\end{corollary}
The proof is given in the appendix, where we also gather more information 
about the ground states $\psi_U$ for $0\le U\le U_c$, see Proposition \ref{prop:basic properties}.  

%%%%%%%%%%%%%%%%%%%%%%%%%%%%%%%%%%%%%%%%%%%%%%%%%%%%%%%%%%%%%%%%%%%%%
%%%%%%%%%%%%%%%%%%%%%%%%%%%%%%%%%%%%%%%%%%%%%%%%%%%%%%%%%%%%%%%%%%%%%
\section{First anisotropic upper bound} \label{sec:first anisotropic upper bound}
%%%%%%%%%%%%%%%%%%%%%%%%%%%%%%%%%%%%%%%%%%%%%%%%%%%%%%%%%%%%%%%%%%%%%
%%%%%%%%%%%%%%%%%%%%%%%%%%%%%%%%%%%%%%%%%%%%%%%%%%%%%%%%%%%%%%%%%%%%%
In this section we prove a preliminary version of the anisotropic upper bound from Theorem \ref{thm:global pointwise anisotropic upper bound}. 
This will be done in two steps: First we prove a  $L^2$ version with the help of energy methods and convert this into a pointwise bound in a second step. 

Unfortunately, our first anisotropic upper bound given in Proposition 
\ref{prop:global anisotropic L2 upper bound} is 
\emph{suboptimal in a transition region} where 
$|x|_\infty^\gamma\le |x|_0 \le 2|x|_\infty^\gamma $.  
So it does not give the sharp anisotropic upper bound we are aiming for. 
However, it has the \emph{correct asymptotics} well within the 
\emph{tricky region}, more precisely, when $|x|_0\le |x|_\infty^\alpha$ for 
$\alpha<1/2$,  and in the region \emph{near the diagonal}, i.e., 
where $|x|_0\sim |x|_\infty$ are large. 
This provides essential a-priori information for the decay of the ground state well within the tricky region and near the diagonal, which is a crucial for the proof of the sharp global anisotropic upper bound in the next section. 

Recall that we set $a=a_U\coloneqq (U-1)_+^{1/2}$ and 
$\veps= \veps_U\coloneqq -\frac{1}{4} - E_U\ge 0$, where $E_U$ is the ground state energy of $H_U$. 
Moreover recall that 
\begin{equation}
  \begin{split}
	F_{\veps,a} 
	  = \Big( \veps +\frac{a^2}{r}\Big)^{1/2}r 
	    + \frac{a^2}{\sqrt{\veps}} 
        \big(\ln( \sqrt{\veps r+a^2} +\sqrt{\veps r} ) -\ln a   \big)\, .
  \end{split}
\end{equation}
For convenience, we abbreviate 
\begin{equation} 
    F_U=F_{\veps_U,a_U}
\end{equation}
and define 
\begin{align}\label{eq:F anisotropic}
		F_{2,U}(r_1,r_2)= F_U(r_1)  - K_1 r_1^{\kappa_1} 
						 	+\frac{1}{2}\big( 
						 					r_2 -K _2 r_2^{\kappa_2} - 2r_1^{\gamma}
						 					\big)_+\, .
\end{align}
With a slight abuse of notation, we also use $F_{2,U}(x)=F_{2,U}(|x|_\infty,|x|_0)$ for  $(x_1,x_2)=x\in\R^6$. 
\begin{remark} 
  For $U=U_c$ we have 
  \begin{equation*}		
  	F_{2,U_c}(r_1,r_2)= 2(U_c-1)^{1/2}r_1^{1/2}  - K_1 r_1^{\kappa_1} 
						 	+\frac{1}{2}\big( 
						 					r_2 -K _2 r_2^{\kappa_2} - 2r_1^{\gamma}
						 					\big)_+\, ,
  \end{equation*}
  see Remark \ref{rem:interpolating}.
\end{remark}
\begin{remark} 
  Let us motive the slightly odd looking dependence of  $F_{2,U}(r_1,r_2)$ on $r_2$: At first sight, a more natural choice would be 
  \begin{align}\label{eq:wrong ansatz}
    F_U(r_1,r_2) = F_U(r_1) + \frac{1}{2} r_2 - K_1 r_1^{\kappa_1} - K_2 r_2^{\kappa_2}
  \end{align}
  and to use $F_U(x)= F_U(|x|_\infty,|x|_0)$ as an exponential weight in the energy estimates. 
  For simplicity, let us look at the critical case where 
  $U=U_c$. In this case we have 
  $F_{U_c}(r_1)= 2(U_c-1)^{1/2} r^{1/2}$ 
  and one calculates 
  \begin{align*}
      |\nabla F_{U_c}(x)|^2 
        & = \big( (U_c-1)^{1/2}r_1^{-1/2} - \kappa_1 K_1 r_1^{\kappa_1-1} \big)^2 
        		+ \big( 1/2 - \kappa_2 K_2 r_2^{\kappa_2-1} \big)^2\\
        & = \frac{(U_c-1)_+}{r_1} +\frac{1}{4}  
            - 2(U_c-1)_+^{1/2}K_1 \kappa_1 r_1^{\kappa_1-3/2} - K_2 \kappa_1  r_2^{\kappa_2-1} \\
        & \phantom{==}  + \kappa_1^2K_1^2 r_1^{2(\kappa_1-1)} + \kappa_2^2 K_2^2 r_2^{2(\kappa_2-1)}\, .
  \end{align*}
  The cross terms will allow to control errors, in particular, the localization error and all terms involving $r_1^{2(\kappa_1-1)}$ and $r_2^{2(\kappa_2-1)}$. However, leading order term in  $ |\nabla F_{U_c}(x)|^2$ now includes the additional constant term $1/4$, which needs to be compensated by the ionization energies. In the tricky region, where one particles escapes to infinity, the ionization energy is zero and  we only have an additional local  boost of the form 
  $\frac{U_c-1}{r_1}$ in the energy estimates from Lemma \ref{lem: Coulomb lower bound} and Proposition \ref{prop:loc energy bound} and \emph{no additional term coming from the second ionization energy}, which is the energy to remove the second particle.  
  The additional constant term $1/4$ in $|\nabla F_{U_c}(x)|^2$ cannot be compensated by the \emph{local energy boost in the tricky region}. 
  Thus for the seemingly natural and simpler 
  ansatz \eqref{eq:wrong ansatz}  
  the gradient of $F_U$ is too big. 
  We need to modify the ansatz to ensure that $F$, hence also its gradient, only depends on $r_1$ inside the tricky region. This is the reason for choosing 
  $\frac{1}{2}\big( r_2 -K _2 r_2^{\kappa_2} - 2r_1^{\gamma} \big)_+$ as the additional term in \eqref{eq:F anisotropic} capturing the decay of the ground state in $r_2=|x|_0$, since it vanished by construction in the tricky region $A_1$ and only becomes positive inside the green shaded region in Figure \ref{fig:regions}, where both particles are far from the nucleus and the additional ionization energy of the second particle can compensate the constant term $1/4$ in the gradient of the exponential weight. 
  
  Eventually we have to choose $1/2<\gamma<1$ in order to be able to control the localization error.  
  The somewhat surprising fact is that the additional term $-2r_1^{\gamma}$ does not mess up the asymptotic behavior of 
  $F_{2,U}(r_1,r_2)$ for large $r_1$ even at criticality, where the leading order term in $F_{2,U}$ only grows proportional to $\sqrt{r_1}$. 
\end{remark}

\begin{proposition}[First global anisotropic $L^2$ upper bound at critical coupling]\label{prop:global anisotropic L2 upper bound}
	Choose parameters $K_1,K_2>0$, $1/6<\kappa_1<1/2$ and $1/2-\kappa_1<\kappa_2<1$,   as well as $(3-2\kappa_1)/4<\gamma< \min((\kappa_2+1)/2,\kappa_1+1/2)$. Then  the 
	ground state $\psi_U$ of helium-type atoms at coupling $U$ has the $L^2$ upper bound  
	\begin{align}\label{eq:global upper bound L2}
		e^{F_{2,U}}\psi_U \in L^2(\R^6)
	\end{align}
	for all  $0\le U\le U_c$ and with $F_{2,U}$ defined in \eqref{eq:F anisotropic}.
	Moreover, in the range  $0\le U\le U_c-\mu$ for some small $\mu>0$, the bound \eqref{eq:global upper bound L2} holds with any choice of parameters $0<\kappa_1, \kappa_2<1$ and $1/2<\gamma<(\kappa_2+1)/2$.  
\end{proposition}
\begin{remark}\label{rem:allowed range of gamma}
  The most important part of Proposition \ref{prop:global anisotropic L2 upper bound} is of course the statement which holds uniformly in $0\le U\le U_c$. By this we mean 
  that for any choice of ground state $\psi_U$, which is normalized, i.e., $\|\psi_U\|=1$, we have
  \begin{align*}
  	\sup_{0\le U\le U_c} \|e^{F_{2,U}}\psi_U\| <\infty \, .
  \end{align*}   
  A simple calculation shows that  
  $(3-2\kappa_1)/4 < \min((\kappa_2+1)/2,\kappa_1+1/2)$ is equivalent to $\kappa_1>1/6$ and $\kappa_2>1/2-\kappa_1$. So the range of allowed  values for $\gamma$ in Proposition   \ref{prop:global anisotropic L2 upper bound} is not empty. 
  Moreover, for any such choice we have $\gamma> (3-2\kappa_1)/4> 1/2$, since 
  $\kappa_1<1/2$. The bound is uniform in $0\le U\le U_c$ and the implicit constant depends only on the parameters  $K_1,K_2>0$, $1/6<\kappa_1<1/2$ and $1/2-\kappa_1<\kappa_2<1$. 
\end{remark}
\begin{proof}
  As in the proof of the isotropic bound, we use a modified version of $F_{2,U}$, 
  \begin{align}
  	G_2(r_1,r_2)\coloneqq F_U(r_1) - K_1 r_1^{\kappa_1}/2
						 	+\frac{1}{2}\big( 
						 					r_2 -K _2 r_2^{\kappa_2} - 2r_1^{\gamma}
						 					\big)_+
  \end{align}
  replacing $K_1$ by $K_1/2$.   In the remainder, we will use $\partial_1=\partial_{r_1}$, respectively, $\partial_2=\partial_{r_2}$ and also freely abbreviate $a=a_U$. 
  
  We again use the smooth cut--off functions $\chi_R$, which projects outside of large balls of radius 
  $R$ centered at zero and  whose gradient  
  $\nabla\chi_R$ is supported on the annulus $R\le |x|\le 2R$ and bounded by $|\nabla\chi_R|\lesssim R^{-1}$. 
  
  We also put  $G_2(x)=G_2(|x|_\infty,|x|_0)$ as a function on $\R^6$, with a slight abuse of notation. 
  Note that $G_2(x)\ge 0$ for all $x$ in the support of $\chi_R$ as long as $R$ is large enough,  
  so we can regularize $G_2$ by using 
  \begin{align}
  	G_{2,\delta} = \frac{G_2}{1+\delta G_2}
  \end{align}
  which is then well defined on the support of $\chi_R$ and bounded for all $\veps>0$.  
  Clearly $G_{2,\delta}$ is continuous on  and  differentiable on 
  $\{|x|_0 -K _2 |x|_0^{\kappa_2} - 2|x|_\infty^{\gamma} > 0\}\cap\{ |x|>R_0\}$ and  
  $\{|x|_0 -K _2 |x|_0^{\kappa_2} - 2|x|_\infty^{\gamma} < 0\}\cap\{ |x|>R_0\}$, for all large enough $R_0$. 
  Up to the smooth zero Lebesgue measure surface 
    $\{|x|_0 -K _2 |x|_0^{\kappa_2} - 2|x|_\infty^{\gamma} = 0, |x|>R_0\}$ 
  these two sets cover $\{|x|>R_0\}$. 
  Hence Lemma \ref{lem:IMS domains ok} shows that the exponential weight 
  $\xi=\chi_R e^{G_{2,\delta}}$ has, for all large enough $R$, a bounded weak derivative on $\R^6$, which 
  is almost everywhere given  by its classical gradient and we can use $\xi$ in the IMS localization formula.

  As in the proof of the isotropic upper bound, see \eqref{eq:refined Agmon}, one deduces from  
  the IMS localization formula and the local energy bound from Proposition \ref{prop:loc energy bound} that 
  \begin{align}\label{eq:start anisptropic}
  \sum_{j=1}^2 \la\chi_j\chi_R e^{G_{2,\delta}}\psi, (W_j - E_U-|\nabla G_2|^2) \chi_j\chi_R e^{G_{2,\delta}} \psi \ra
  	\le 
  		\la \psi, e^{2G_2}(2|\nabla\chi_R||\nabla G_2|+|\nabla\chi_R|^2)\psi\ra \, ,
  \end{align}
  by choosing $\varphi= (\chi_R e^{G_{2,\delta}})^2\psi$ as a test function in the quadratic form version  
  of the eigenvalue equation.  
  $W_1$ and $W_2$ are given in \eqref{eq:W1} and 
  \eqref{eq:W2}. 
  Similar to the derivation of \eqref{eq:refined Agmon in G}, 
  we used  
  $|\nabla G_{2,\delta}(x)|\le |\nabla G_2(x)|$ and $G_{2,\delta}(x)\le G_2(x)$ when 
  $|x|_\infty$ is large and there is no $j=0$ term in \eqref{eq:start anisptropic} since $\chi_0\chi_R=0$ 
  for all $R>2R_0$.

  Recall that $\chi_1$ localizes into the tricky region  $A_1=\{|x|_0<  2|x|_\infty^\gamma, |x|>R_0\}$. 
  Clearly $G_2(x)= 2a |x|_\infty^{1/2} - K_1 |x|_\infty^{\kappa_1}/2$ for  $x\in A_1$. Thus, setting 
  $r_1=|x|_\infty$, 
   \begin{align}
 	\nabla G_2(x) = 	\begin{pmatrix}
 	 				  \partial_1 G_2 \frac{x_1}{|x_1|}\ind_{\{ |x_1|>|x_2|\}}\\
 	 				  \partial_1 G_2 \frac{x_2}{|x_2|}\ind_{\{ |x_2|>|x_1|\}}
 	 				\end{pmatrix}
 	 			=  
 	 				((\veps +a^2/r_1)^{1/2} -K_1\kappa_1 r_1^{\kappa_1-1}/2 )
 	 				\begin{pmatrix}
 	 				\frac{x_1}{|x_1|}\ind_{\{ |x_1|>|x_2|\}}\\
 	 				\frac{x_2}{|x_2|}\ind_{\{ |x_2|>|x_1|\}}
 	 				\end{pmatrix}
 \end{align}
 for all $x\in A_1$. 
  Using $-E_U= \frac{1}{4}+\veps$ we have on the support of $\chi_1$ 
  \begin{align*}
  	W_1(x)-& E_U-|\nabla G_2|^2  \\
  		&\ge \veps+ a^2 r_1^{-1} - 2U r_1^{\gamma-2} - ((\veps +a^2/r_1)^{1/2}- K_1\kappa_1r_1^{\kappa_1-1}/2)^2 - C r_1^{-2\gamma} \\
  		&= K_1\kappa_1 r_1^{\kappa_1-1}\big(2(\veps +a^2/r_1)^{1/2}- K_1\kappa_1r_1^{\kappa_1-1}/2)\big)/2  - 2U r_1^{\gamma-2} - C r_1^{-2\gamma} \, .
  \end{align*}
 If $0\le U\le U_c-\mu$ then $\veps=\veps_U\ge c>0$ for some constant depending on $\mu>0$. 
 In this case we get 
    \begin{align*}
  	W_1(x)-& E_U-|\nabla G_2|^2  \\
  		&\ge K_1\kappa_1 r_1^{\kappa_1-1}\big(2 c^{1/2}- K_1\kappa_1r_1^{\kappa_1-1}/2)\big)/2  - 2U r_1^{\gamma-2} - C r_1^{-2\gamma} \\
  		& \gtrsim  r_1^{\kappa_1-1}  - r_1^{\gamma-2} - r_1^{-2\gamma} 
  		  \gtrsim  r_1^{\kappa_1-1} \ge r_1^{-1} 
  \end{align*}
  for large enough $r_1$, i.e., large enough $R_0$,  as long as $\kappa_1-1 > \max(\gamma-2,-2\gamma)$, which is 
  equivalent to $(1-\kappa_1)/2<\gamma<1+\kappa_1$.  
  
  In the range $0\le U\le U_c$, we note that when $r_1\ge 1$ we have 
  $\veps+ a^2/r_1= \veps_U+ a_U^2/r_1\ge (\veps_U+(U-1)_+)/r_1\ge c/r_1$, with 
  $c=\inf_{0\le U\le U_c} \big( \veps_U+ (U-1)_+\big)>0$. So uniformly in 
  $0\le U\le U_c$ we have 
    \begin{align*}
      W_1(x)- E_U-|\nabla G_2|^2 
  		&\ge   K_1\kappa_1 r_1^{\kappa_1-1}\big(2(c/r_1)^{1/2}- K_1\kappa_1r_1^{\kappa_1-1}/2)\big)/2  - 2U r_1^{\gamma-2} - C r_1^{-2\gamma} \\
  		& \gtrsim  r_1^{\kappa_1-3/2} - r_1^{\gamma-2} - r_1^{-2\gamma} 
  		  \gtrsim  r_1^{\kappa_1-3/2} \ge r_1^{-4/3}
  \end{align*}
  for large enough $r_1$, i.e., large enough $R_0$,  as long as $\kappa_1-3/2 > \max(\gamma-2,-2\gamma)$, which is equivalent to 
  $(3-2\kappa_1)/4<\gamma<\kappa_1+1/2$. We also used that 
  $\kappa_1>1/6$ in the last bound. 
  
  For the energy bound on the support of $\chi_2\subset A_2$, we split  $A_2$ into the two regions 
  $A_2^- =\{|x_1|^\gamma<|x_2|< |x_1|, |x_1|>R_0 \}$ and 
  $A_2^+= \{|x_2|^\gamma<|x_1|< |x_2|, |x_2|>R_0 \}$ which cover $A_2$ up to a null set within the 
  diagonal  $\{ |x_1|=|x_2| \}$.  
  It is enough to provide a lower bound for  
  $ W_2-E_U- |\nabla G_2|^2$ on $A_2^-$, the same bound 
  will then also hold on $A_2^+$,  by symmetry,  
  Since null sets are irrelevant such a bound will then hold on the support of $\chi_2$.  
  
  For the same reason, we can also disregard the null set 
  $\{ |x_2|-K_2 |x_2|^{\kappa_2}- 2|x_1|^\gamma= 0\}\cap A_2^- $, on which the classical gradient of 
  $G_{2}$ does not exist.

  On $\{ |x_2|-K_2 |x_2|^{\kappa_2}- 2|x_1|^\gamma< 0\}\cap A_2^- $ we again have  
  $G_2(x)= 2a|x_1|^{1/2}-K_1|x_1|^{\kappa_1}$. Thus  
  \begin{align*}
  	   W_2(x) -&E_U - |\nabla G_2(x)|^2\\
  	  			&\ge \frac{1}{4} +\veps - r_1^{-1} -r_1^{-\gamma} - ((\veps+a^2/r_1)^{1/2}- K_1\kappa_1r_1^{\kappa_1-1}/2)^2 - C r_1^{-2\gamma} \\
  	  			&\ge \frac{1}{4} - (1+a^2)r_1^{-1} - r_1^{-\gamma} 
  	  			  		-(K_1\kappa_1 r_1^{\kappa_1-1}/2)^2 \ge \frac{1}{8}
  \end{align*} 
  on this set and all large enough $R>0$. 
  
  For $x\in \{ |x_2|-K_2 |x_2|^{\kappa_2}- 2|x_1|^\gamma> 0\} \cap A_2^-$, we have 
  \begin{align}
  	G_2(x) =  F_U(r_1) - K_1 r_1^{\kappa_1}/2 
						 	+\frac{1}{2} \big(	r_2 -K _2 r_2^{\kappa_2} - 2r_1^{\gamma} \big)
  \end{align}
 with $r_1=|x_1|$ and $r_2=|x_2|$. Hence 
 \begin{align}
 	\nabla G_2(x) = 	\begin{pmatrix}
 	 				  \partial_1 G_2 \frac{x_1}{|x_1|}\\
 	 				  \partial_2 G_2 \frac{x_2}{|x_2|}
 	 				\end{pmatrix}
 	 			=  
 	 				\begin{pmatrix}
 	 				  \big((\veps_U+a^2/r_1)^{1/2} -K_1\kappa_1 r_1^{\kappa_1-1}/2 - \gamma r_1^{\gamma-1}\big) \frac{x_1}{|x_1|}\\
 	 				  \frac{1}{2}\big(1 -K_2\kappa_2 r_2^{\kappa_2-1}\big) \frac{x_2}{|x_2|}
 	 				\end{pmatrix}
 \end{align}
 which implies  
  \begin{align*}
  	  	W_2&(x) -E_U - |\nabla G_2(x)|^2 
  	  	   \ge 	W_2(x) +\frac{1}{4}+ \veps - |\nabla G_2(x)|^2 
 \\
  	  			&= \frac{1}{4} -\frac{1}{4}\left(1-K_2\kappa_2r_2^{\kappa_2-1}\right)^2- r_1^{-1} -r_2^{-1} \\
  	  			&\phantom{=}~~ 
  	  				+\veps - \left((\veps+a^2/r_1)^{1/2}- K_1\kappa_1r_1^{\kappa_1-1}/2 -\gamma r_1^{\gamma-1}\right)^2 - C r_1^{-2\gamma} \\
  	  			&\ge   \frac{1}{4}K_2\kappa_2 r_2^{\kappa_2-1}\left( 2-K_2\kappa_2 r_2^{\kappa_2-1} \right) - r_1^{-1} - r_2^{-1}
  	  				- \frac{a^2}{r_1} \\
  	  			&\phantom{=~~} + \left( \frac{\veps+a^2}{r_1} \right)^{1/2} \left( K_1\kappa_1 r_1^{\kappa_1-1} +2\gamma r_1^{\gamma-1}\right) 
  	  				 - \left( K_1\kappa_1 r_1^{\kappa_1-1}/2 +\gamma r_1^{\gamma-1}\right)^2   	
					- C r_1^{-2\gamma} \, .
  \end{align*}
  Using again $\veps+a^2= \veps_U+(U-1)_+\gtrsim 1$ uniformly in 
  $0\le U\le U_c$ and also that $\gamma>1/2> \kappa_1$, so terms such as 
  $r_1^{\gamma-1}$ control $r_1^{\kappa_1-1}$ for large $r_1$,  
  we arrive at the lower bound 
    \begin{align*}
  	  	W_2&(x) -E_U - |\nabla G_2(x)|^2 \\
  	  		  &\ge  \frac{1}{4}K_2\kappa_2 r_2^{\kappa_2-1}\left( 2-K_2\kappa_2 r_2^{\kappa_2-1} \right) - r_1^{-1} - r_2^{-1}
  	  				- \frac{a^2}{r_1} \\
  	  			&\phantom{=~~} + \left( \frac{\veps+a^2}{r_1} \right)^{1/2} \left( K_1\kappa_1 r_1^{\kappa_1-1} +2\gamma r_1^{\gamma-1}\right) 
  	  				 - \left( K_1\kappa_1 r_1^{\kappa_1-1/2} +\gamma r_1^{\gamma-1}\right)^2   	
					- C r_1^{-2\gamma} \\
				&\gtrsim r_2^{\kappa_2-1} - r_2^{-1} + r_1^{\gamma-3/2} -r_1^{2(\gamma-1)}- r_1^{-2\gamma} 
					\ge   r_2^{\kappa_2-1} - r_2^{-1} + r_1^{\gamma-3/2} -r_2^{2(\gamma-1)}- r_1^{-2\gamma}  \\
				&\ge r_2^{\kappa_2-1} \ge r_1^{\kappa_2-1}\ge r_1^{-1}
  \end{align*}

  for all $r_1^\gamma < r_2\le r_1$, and large enough $r_1$,  
  as long as $0<\kappa_2<1$, 
  $\kappa_2-1 >  2(\gamma-1)$, and $\gamma-3/2>-2\gamma$, which is equivalent to $1/2<\gamma<(\kappa_2+1)/2$.   
  
  It is straightforward to check that 
  $1/2<\gamma<(\kappa_2+1)/2$ and 
  $(3-2\kappa_1)/4<\gamma<\kappa_1+1/2$ is equivalent to 
  $(3-2\kappa_1)/4<\gamma< \min((\kappa_2+1)/2,\kappa_1+1/2)$,
  since $(3-2\kappa_1)/4>1/2$ when $\kappa_1<1/2$. 
  This is the condition on the parameter $\gamma$ for the range $0\le U\le U_c$. 
  
  One also easily checks that $(1-\kappa_1)/2<\gamma<1+\kappa_1$ and $1/2<\gamma<(\kappa_2+1)/2$ is 
  equivalent to $1/2<\gamma<(\kappa_2+1)/2$ for $0<\kappa_1,\kappa_2<1$. This is the  condition on the parameter $\gamma$ when $U$ stays away from the critical coupling.  
  
  Collecting the lower bounds on $\frac{1}{4}+ W_j- |\nabla G_2|^2$ on the supports of $\chi_1$ and $\chi_2$   
  and plugging this into \eqref{eq:start anisptropic}, one arrives at
  \begin{align*}
  	\la \chi_R e^{G_{2,\delta}}\psi, |x|_\infty^{-4/3} \chi_R e^{G_{2,\delta}}\psi  \ra 
  	\le C\|\psi\|^2 \, .
  \end{align*} 
  The constant $C$ is uniform in $\veps>0$, since 
  $G_2$ is bounded on the support of $\nabla\chi_R$ for any fixed $R>0$.  
  Hence we can again use 
  the monotone convergence theorem to conclude   
  \begin{align*}
  	\la \chi_R e^{G_2}\psi, |x|_\infty^{-4/3 } \chi_R e^{G_2}\psi  \ra 
  	=\lim_{\veps\to 0} \la \chi_R e^{G_{2,\delta}}\psi, |x|_\infty^{-4/3} \chi_R e^{G_{2,\delta}}\psi  \ra
  	\le C\|\psi\|^2 \, 
  \end{align*}
  which proves the theorem since $G_2(x)- \tfrac{2}{3}\ln |x|_\infty \ge F_2(x) $ for all large 
  $|x|_\infty$. 
\end{proof}

\begin{corollary}[Pointwise version of Proposition \ref{prop:global anisotropic L2 upper bound}]\label{cor:first anisotropic pointwise upper bound} 
  	Given parameters   $K_1,K_2>0$, $1/6<\kappa_1<1/2$, and $1/2-\kappa_1<\kappa_2<1$, as well as 
  	$(3-2\kappa_1)/4<\gamma< \min((\kappa_2+1)/2,\kappa_1+1/2)$, 
	there exists a constant $C>0$, depending only on the above parameters such that for any  
	normalized ground state $\psi_U$ of the helium--type system 
	Hamiltonian $H_U$ we have  
	\begin{align}\label{eq:global upper bound pointwise}
		\psi_U(x) \le C\exp[-F_{2,U}(|x|_\infty,|x|_0)]
		\quad \text{for all } x\in\R^6\, ,
	\end{align}
	uniformly in $0\le U\le U_c$, with $F_2$ defined in \eqref{eq:F anisotropic}.
	
	Moreover, in the subcritical range  $0\le U\le U_c-\mu$ for 
	some small 
	$\mu>0$, the bound \eqref{eq:global upper bound pointwise} 
	holds with any choice of parameters $0<\kappa_1, \kappa_2<1$ 
	and $1/2<\gamma<(\kappa_2+1)/2$.  
\end{corollary}
\begin{remark}\label{rem:suboptimal in transition region}
  As discussed in the introduction, we expect the ground state to have an asymptotic anisotropic decay given by  
  \begin{equation}
      \psi_c(x)
       \le C \exp\big(-2(U_c-1)|x|_\infty^{1/2} - \frac{1}{2}|x|_0\big )\, .
  \end{equation}
  up to some lower order corrections. The bound provided by Corollary 
  \ref{cor:first anisotropic pointwise upper bound} does not quite achieve to prove this. If $|x|_\infty^\gamma \le |x|_0 \le 2 |x|_\infty^\gamma$ we expect the bound state to decay at least as fast as  
    \begin{equation}
      \psi_c(x)
       \le C \exp\big(-2(U_c-1)|x|_\infty^{1/2} - \frac{1}{2}|x|_\infty^\gamma \big )\, .
  \end{equation}
  and since $1/2<\gamma<1$, the second term now dominates. However, the upper bound from Corollary \ref{cor:first anisotropic pointwise upper bound} only shows that 
  \begin{equation}
      \psi(x)\le C\exp\big(- 2(U_c-1)_+^{1/2}|x|_\infty^{1/2} +K|x|_\infty^{\kappa_1}\big)
  \end{equation}
  in the region $|x|_0\le 2|x|_\infty^\gamma$. This is the reason why we have to refine the upper bound from Corollary \ref{cor:first anisotropic pointwise upper bound} in the next section. Nevertheless, the bound from Corollary \ref{cor:first anisotropic pointwise upper bound} provides essential a--priori information for this refinement, see the proof of Lemma \ref{lem:upper bound on boundary layer}.  
\end{remark}
\begin{proof}
  Proposition \ref{prop:basic properties} shows that $H_{U}$ has a unique ground state $\psi_U$ for any $0\le U\le U_c$,  which up to a phase can be chosen positive. 
  The subsolution estimate of Trudinger \cite{Tru1973} in 
  the version of  Aizenman and Simon \cite{AizSim82}, see also \cite[Theorem C.1.3]{Simon-semigroups}  shows that for any $r>0$  and  $x\in\R^6$ 
  \begin{equation}\label{eq:subsolution bound Trudinger Aizenman Simon}
  	|\psi_U(x)|\le C_1 \int_{|x-y|\le r} |\psi_U(y)|\, d y 
  \end{equation}
  since the Coulomb potential is in the Kato class. The constant $C_1$ 
  depends only on $r$ and the Kato norm of the potential, 
  see, e.g.,  \cite[Section C.1]{Simon-semigroups}. 
  Thus, with $|B^d_1|$ the volume of the unit ball in $\R^d$,   
  \begin{equation}\label{eq:pointwise mother}
  \begin{split}
  	e^{F_{2,U}(x)} |\psi_U(x)| 
  		&\le 
  		  C_1|B^d_1|^{1/2}\left( \int_{|x-y|\le 1} e^{2F_{2,U}(x)} |\psi_U(y)|^2\, d y  \right)^{1/2} \\
  		&\le 
  		  C_1|B^d_1|^{1/2} \sup_{|x-y|\le 1} e^{F_{2,U}(x)- F_{2,U}(y)} \|e^{F_{2,U}}\psi_U\|
  \end{split}
  \end{equation}
  for all $x\in\R^6$. 
  One easily checks that for all $|r_1| \gg 1$ and $|r_1-s_1|, |r_2-s_2| \ge 1$
  \begin{align*}
  	& F_{2,U}(r_1,r_2) - F_{2,U}(s_1,s_2) \\
  		&\lesssim 
  			|r_1^{1/2}-s_1^{1/2}| + |r_1^{\kappa_1}-s_1^{\kappa_1}| 
  			+ |r_1-s_1| + |r_2^{\kappa_2}-s_2^{\kappa_2}| + |r_1^\gamma- s_1^\gamma| \\
  		&\lesssim 
  			t^{-1/2} + t^{\kappa_1-1} +1 + \min\{2,u^{\kappa_2-1}\} + t^{\gamma-1}
  \end{align*}
  with $t=\max(r_1,s_1)$, $u=\max(r_2,s_2)$ %\MLi{where we used that if $|r_i - s_i| \le 1$ one has that $s_i = r_i + e$ with $e \in [0,1]$ and $i\in\{1,2\}$ (or vice versa). Thus 
  %\begin{align*}
   %   r_i^\alpha - s_i^\alpha 
%      = r_i^\alpha - (r_i + e)^\alpha
 %     = r_i^\alpha - (r_i^2 + e^2 + 2\langle e,r_i\rangle)^{\alpha/2}
  %    = r_i^\alpha \Big(1 - (1+\frac{e^2 + 2\langle e,r_i\rangle}{|x|^2}\Big)^{\alpha/2}
  %    = ...
 % \end{align*}
 % }
  Hence 
  \begin{align}
  	C_2\coloneqq \sup_{x\in\R^6} \sup_{|x-y|\le 1}|F_{2,U}(x) -F_{2,U}(y)| <\infty\, .
  \end{align}
  So \eqref{eq:pointwise mother} gives 
  the pointwise exponential upper bound 
  \begin{equation}
  	|\psi_U(x)| \le C C_1 C_2 |B^d_1|^{1/2}e^{-F_{2,U}(x)} \quad \text{for all } x\in \R^6
  \end{equation}
  for the ground state $\psi_U$,  where $C$ is the constant from the 
  $L^2$ upper bound of Proposition \ref{prop:global anisotropic L2 upper bound}. 
  This proves Corollary \ref{cor:first anisotropic pointwise upper bound}.
\end{proof}

%%%%%%%%%%%%%%%%%%%%%%%%%%%%%%%%%%%%%%%%%%%%%%%%%%%%%%%%%%%%%%%%%%%%%
%%%%%%%%%%%%%%%%%%%%%%%%%%%%%%%%%%%%%%%%%%%%%%%%%%%%%%%%%%%%%%%%%%%%%
\section{Global anisotropic upper bound} \label{sec:global anisotropic upper bound}
%%%%%%%%%%%%%%%%%%%%%%%%%%%%%%%%%%%%%%%%%%%%%%%%%%%%%%%%%%%%%%%%%%%%%
%%%%%%%%%%%%%%%%%%%%%%%%%%%%%%%%%%%%%%%%%%%%%%%%%%%%%%%%%%%%%%%%%%%%%
In this section we give the proof of Theorem \ref{thm:global pointwise anisotropic upper bound} and the proof of the upper bound 
from Theorem \ref{thm:sharp upper and lower bounds all coupling}. 
Recall that the function 
$F_+$ is given by 
	\begin{align}\label{eq:F+ again}
		F_+(r_1,r_2)= 2(U_c-1)^{1/2}r_1^{1/2} - K_1 r_1^{\kappa_1} 
						 	+\frac{1}{2} r_2- K_2 r_2^{\kappa_2}
	\end{align}
and the function $F^U_+$ is given by 
	\begin{align}\label{eq:F_U_+ again}
		F^U_+(r_1,r_2)= F_U(r_1) - K_1 r_1^{\kappa_1} 
						 	+\frac{1}{2} r_2- K_2 r_2^{\kappa_2}
	\end{align}
for $r_1,r_2\ge 0$ with $F_U$ defined in \eqref{eq:F_U}. We also set $F_+(x)=F_+(|x|_\infty,|x|_0)$ and the same for $F^U_+$. 
We choose $\psi_U$ to be the unique 
positive ground state of $H_U$. We want to use the comparison principle from Theorem \ref{thm:Agmon subsolution bound} 
to show that there exist a constant 
$0<C<\infty$, depending only on the parameters  $\kappa_1,\kappa_2, K_1, K_2$ in the definition of $F^U_+$ 
such that 
\begin{align}
	\psi_U(x) \le C\exp(-F^U_+(|x|_\infty,|x|_0))
\end{align}
for all $x\in\R^6$ and all $0\le U\le U_c$.  
Since 
$\psi_U$ is bounded by Proposition \ref{prop:basic properties} 
and $\exp(-F^U_+)$ is bounded away from zero on compact sets uniformly in $U$, 
it is enough to assume that $|x|_\infty> R$ for some large $R>0$. In this section, we abbreviate
\begin{align}
	f_U\coloneqq \exp(-F^U_+)
\end{align}
and note that, similarly as for $F^U_+$,  we will not explicitly write the dependence of $f_U$ on the other parameters except $U$, for simplicity of notation. 
\begin{remark}
  Note that $F^U_+$, hence also $f$, is twice continuously differentiable for all $0<|x|_0<|x|_\infty$. 
  However, it is not twice differentiable on any neighborhood of the diagonal $|x_2|=|x_1|$, 
  since the gradient of $F^U_+$ has a jump discontinuity across the diagonal.  
  Nevertheless, we could use Agmon's quadratic form version of the comparison principle, 
  see Theorem \ref{thm:Agmon subsolution bound}, as long as one could control certain boundary terms on the 
  diagonal $|x_1|=|x_2|$, which appear from integration by parts. However, as the proof of the 
  \emph{global lower bound} in Section \ref{sec:global lower bound}, in particular, the proof 
  of Lemma \ref{lem: if classical subsolution then ok}, will show, the boundary terms 
  have the \emph{wrong sign and cannot be discarded}.    
  This is why we cannot apply the comparison theorem directly. Instead,  
  one has  to apply the comparison theorem separately on $\{R^\alpha<|x_2|<|x_1|\}$ and 
  $\{R^\alpha<|x_1|<|x_2|\}$, for large enough  $R$. 
  This forces us to have \emph{very precise a-priori information} about the asymptotic behavior 
  of $\psi_U$ at infinity near the diagonal $|x|_0=|x|_\infty$, since we need the a--priori information 
  that  $\psi_U(x)\lesssim f_U(x)$ near the diagonal $|x_1|=|x_2|$. Fortunately this is exactly what 
  Corollary \ref{cor:first anisotropic pointwise upper bound} provides. We summarize the necessary a-priori 
  information about the asymptotic decay of $\psi_c$ near the diagonal and within the tricky region in the 
  following  
\end{remark} 
\begin{lemma}\label{lem:upper bound on boundary layer}
  Given any $K_1,K_2>0$,  $1/6<\kappa_1<1/2$, $0<\alpha<\kappa_1$,  and $3/4-\kappa_1/2 < \kappa_2<1$, there exist a constant $0<C<\infty$ such that   
  \begin{align}\label{eq:upper bound on boundary layer}
  	  	\psi_U(x) &\le C f_U(x) 
  \end{align}
  on the sets   
  $\{|x|_0\le |x|_\infty^\alpha,  |x|_\infty\ge R\}$ and $\{|x|_\infty-1\le |x|_0\le |x|_\infty, |x|_\infty\ge R\}$ 
  for all large enough $R>0$ and all $0\le U\le U_c$. 
  
  	Moreover, in the subcritical range  $0\le U\le U_c-\mu$ for 
	some small $\mu>0$, for any choice of parameters  
	$0<\kappa_1, \kappa_2<1$, $\max(\kappa_1,\kappa_2)>1/2$, 
    $0<\alpha<\min(\kappa_1,1/2)$, and   $K_1,K_2>0$, 
	there exist $R>0$ 
	such that  the bound \eqref{eq:upper bound on boundary layer}  
	holds on the above sets and the constant $C$ is again 
	independent of $0\le U\le U_c-\mu$.    
\end{lemma}
\begin{remark}
  The first part of Lemma \ref{lem:upper bound on boundary layer} provides an upper bound on the ground state 
  $\psi_U$ in the green shaded region near the diagonal 
  $r_1=r_2$ and deep within the tricky region near 
  $r_2= r_1^\alpha$ 
  in Figure \ref{fig:boundary layer global upper bound} up to the critical case when $U=U_c$. 
  
  The second part allows for a wider range of parameters 
  $\kappa_1,\kappa_2$ but works only for subcritical couplings $U$ which stay away from $U_c$. 
\end{remark}

\begin{proof}
The conditions  $1/6<\kappa_1$ and $3/4-\kappa_1/2 < \kappa_2$ are equivalent to 
\begin{align*}
	(3-2\kappa_1)/4<\min(\kappa_2,\kappa_1+1/2)\, .
\end{align*}
Since $\kappa_2<1$, we also  have 
$\kappa_2<(\kappa_2+1)/2$, so any $\gamma$ with 
$(3-2\kappa_1)/4<\gamma<\min(\kappa_2,\kappa_1+1/2)$ fulfills the condition of 
Corollary \ref{cor:first anisotropic pointwise upper bound} in the whole range $0\le U\le U_c$. 

In the subcritical range where $U$ stays away from $U_c$, 
we choose $1/2<\gamma<\max(\kappa_1,\kappa_2)$. Note that for 
this choice the conditions of 
Corollary \ref{cor:first anisotropic pointwise upper bound} 
for the subcritical case are also satisfied since 
$\kappa_2<(\kappa_2+1)/2$.

  Given the  parameters $\kappa_1,\kappa_2,\gamma$ and $K_1,K_2>0$ we use the exponential weight $F_{2,U}$ from 
  Corollary \ref{cor:first anisotropic pointwise upper bound}, but with $K_1$ replaced by $K_1/2$, that is, with a slight abuse of notation, we use
  \begin{align*}
  	F_{2,U}(r_1,r_2) 
  		= F_U(r_1) -K_1r_1^{\kappa_1}/2 
  		+\frac{1}{2}
  		 	\left( 
  		 		r_2-K_2r_2^{\kappa_2} - 2r_1^\gamma 
  		 	\right)_+ \, .
  \end{align*} 
  Due to  $\alpha<1/2<\gamma$, we have 
  $F_{2,U}(r_1,r_2) = F_U(r_1) - \frac{K_1}{2} r_1^{\kappa_1}$ when $r_2\le r_1^\alpha$ and $r_1$ is large. 
  In particular, 
  \begin{align*}
  	F_{2,U}(r_1,r_2)&= F_U(r_1) - \frac{K_1}{2} r_1^{\kappa_1}
  	  \ge 
  	  	F_U(r_1) - K_1 r_1^{\kappa_1} +\frac{1}{2}r_2 - K_2 r_2^{\kappa_2} 
  	  		+ \frac{K_1}{2} r_1^{\kappa_1} -\frac{1}{2} r_2 \\
  	  &\ge 
  	    F^U_+(r_1,r_2) +  \frac{K_1}{2} r_1^{\kappa_1} -\frac{1}{2} r_1^\alpha 
  	    >F^U_+(r_1,r_2)
  \end{align*}
  if $r_2\le r_1^\alpha$ and $r_1$ is large enough since $\alpha<\kappa_1$. Similarly, 
  \begin{align*}
  	F_{2,U}(r_1,r_2)
  	  &= F_U(r_1) - \frac{K_1}{2} r_1^{\kappa_1} + \frac{1}{2}r_2 
  		-\frac{K_2}{2}r_2^{\kappa_2} -r_1^\gamma \\
  	  &\ge 
  		   F^U_+(r_1,r_2) + \frac{K_1}{2}r_1^{\kappa_1}
  		   + \frac{K_2}{2}(r_1-1)^{\kappa_2} 
  		   	-r_1^\gamma 
  		     > F^U_+(r_1,r_2)
  \end{align*}
  when $r_1-1<r_2\le r_1$ and $r_1$ is large, since $\gamma<\kappa_2$ in the critical case and 
  $\gamma<\max(\kappa_1,\kappa_2)$ in the subcritical case. 
  From Corollary \ref{cor:first anisotropic  pointwise upper bound} we get, 
  the pointwise  upper bound 
  \begin{align}
  	\psi_U(x) \le C \exp[-F_{2,U}(x)] \le C \exp[-F^U_+(x)] = Cf_U(x)
  \end{align}
  in the regions $|x|_0\le |x|_\infty^\alpha$, respectively, $|x|_\infty-1<|x|_0\le |x|_\infty$, 
  when $|x|_\infty$ is large enough.  
\end{proof}

The next Lemma shows that $f_U$ is a supersolution on  a set ``sandwiched between'' the sets 
$\{|x|_0\le |x|_\infty^\alpha,  |x|_\infty\ge R\}$ and  $\{|x|_\infty-1\le |x|_0\le |x|_\infty, |x|_\infty\ge R\}$.   
\begin{lemma}\label{lem:classical supersolution}
  Let $0<\kappa_1<1/2<\kappa_2<1$, $K_1,K_2>0$, and $0<\alpha<1$. Then the function 
  $
	f_U= \exp(-F^U_+)
  $, 
 with  $F^U_+$ given in \eqref{eq:F+ again}, is a classical supersolution of $H_U$ 
 at energy $E_U$ on the set 
 \begin{align}
 B_R=\{ |x|_\infty^\alpha -1<|x|_0<|x|_\infty, |x|_\infty>R \}
 \end{align}
 for all large enough $R>0$. 	That is,
 \begin{align}
 	(H_U-E_U)f_U \ge 0 \text{ pointwise in } B_R\, .
 \end{align}
  In particular, $f_U$ is a supersolution in 
  the quadratic form sense a la Agmon: for large enough $R$ and all $0\le \varphi\in \calC^\infty_0(B_R)$ 
  we have the quadratic form inequality  
  \begin{align}\label{eq:supersolution a la Agmon in B_R}
  	\la \varphi, (H_U-E_U) f_U \ra \ge  0 \, .
  \end{align}
  	Moreover, in the subcritical range, where  
  	$0\le U\le U_c-\mu$ for some small $\mu>0$,  
  	for any choice of parameters $0<\kappa_1, \kappa_2<1$,  
  	and $0<\alpha<1$,  
	there exists $R>0$ such that 
	the function $f_U$ is a classical supersolution in the set 
	$B_R$. 
\end{lemma}
\begin{proof} 
  In order to be able to control the errors, it will be important that both particles are far from 
  the nucleus in $B_R$. 
  For large enough $R>0$, the set $B_R$ is the disjoint union of  
  $B_R^-\coloneqq\{ |x_1|^\alpha -1<|x_2|<|x_1|, |x_1|>R  \}$, the part of $B_R$ ``below the diagonal'' $|x_2|=|x_1|$, 
  and the part  $B_R^+$ above the diagonal, defined similarly as $B_R^-$ but with $x_1$ 
  and $x_2$ interchanged.	 
  By symmetry of $f_U$, which is clear from the symmetry of $F^U_+$,  it is enough to show that 
  $f_U$ is a classical supersolution of $H_U$ at energy $E_U$ on $B_R^-$. 
   
  Let $\nabla$ be the gradient and $\Delta$ the Laplacian on $\R^6$. We abbreviate $F=F^U_+$ and $f=f_U$. Since they are twice differentiable on $B_R^-$ we clearly have 
  \begin{align*}
  	-\Delta f= f\big[-|\nabla F|^2 +\Delta F\big]
  \end{align*}
  on $B_R^-$. 
  Also recall that $\veps= -\frac{1}{4}-E_U\ge 0$ is 
  the ionization energy (which is zero when $U=U_c$) and we also abbreviate $a=(U-1)_+^{1/2}$, so 
  $F_U'(r_1) = (\veps+a^2/r_1)^{1/2}$. 
  Moreover, we have $f(x)= \exp(-F(r_1,r_2))$ with  $r_1=|x_1|$ and $r_2=|x_2|$ on   $B_R^-$, 
  so   one easily calculates    
  \begin{align}\label{eq:gradient F_+}
  	\nabla F(x)
  		&= 	\begin{pmatrix}
 				\partial_1 F\frac{x_1}{|x_1|} \\
 				\partial_2 F\frac{x_2}{|x_2|}
 			\end{pmatrix}
 		=
 			\begin{pmatrix}
 				\big( F_U'(r_1) - K_1\kappa_1 r_1^{\kappa_1-1}\big)\frac{x_1}{|x_1|} \\
 				\big(\frac{1}{2} -K_2\kappa_2r_2^{\kappa_2-1}\big)\frac{x_2}{|x_2|}
 			\end{pmatrix} 
 			\quad \text{on } B_R^- \\
    \intertext{and }
    \Delta F(x)
       &=  \partial_1^2F +\partial_1 F \frac{2}{|x_1|} + \partial_2^2F +\partial_2 F \frac{2}{|x_2|} 
       \quad \text{on } B_R^- \, .
  \end{align}
  Thus 
  \begin{align*}
  -\Delta f = f\left[ 
  				-|\partial_1 F|^2 - |\partial_2F|^2 + \partial_1^2F +\partial_1 F \frac{2}{|x_1|} + \partial_2^2F +\partial_2 F \frac{2}{|x_2|} 
  				\right]\, ,
  \end{align*}
  on $B_R^-$, hence also 
  \begin{align*}
  	(H_U-E_U)f 
  	  &= f\Big[ 
  	        \veps -|\partial_1 F|^2 +\frac{1}{4} - |\partial_2F|^2 + \partial_1^2F +\partial_1 F \frac{2}{|x_1|} + \partial_2^2F +\partial_2 F \frac{2}{|x_2|}  \\
  	  &\phantom{=w\big[~~} - \frac{1}{|x_1|} - \frac{1}{|x_2|}+ \frac{U}{|x_1-x_2|} 
  	      \Big]\\
  	  &\ge f\Big[ 
  	          \veps -|\partial_1 F|^2 +\frac{1}{4}- |\partial_2F|^2 + \partial_1^2F   
  	          +\partial_2 F \frac{2}{r_2}  - \frac{1}{r_1} - \frac{1}{r_2}  
  	        \Big]\, ,
  \end{align*}
  where we dropped the positive terms $\partial_2^2F$, $U/|x_1-x_2|$, and also $\partial_1F\frac{2}{r_1}$, which 
  is positive for large enough $r_1$. 
  Furthermore, using 
  $F_U'(r_1)= (\veps+a^2/r_1)^{1/2}\ge (\veps+a^2)^{1/2}r_1^{-1/2} \ge c r_1^{-1/2}$ with the constant $c^2= \inf_{0\le U\le U_c}(\veps_U+a_U^2)>0$  
  we have  
  \begin{align*}
  	\veps -|\partial_1F|^2 
  	  &= 
  	    \veps-\left(F_U'(r_1)-K_1\kappa_1 r_1^{\kappa_1-1}\right)^2 \\
  	  &= -a^2 r_1^{-1} +2K_1\kappa_1 F_U'(r_1) r_1^{\kappa_1-1} -(K_1\kappa_1 r_1^{\kappa_1-1})^2 \\
  	  &\gtrsim - r_1^{-1} + r_1^{\kappa_1-3/2} 
  	   			- r_1^{2(\kappa_1-1)} 
  	   		\gtrsim -r_1^{-1} + r_1^{\kappa_1-3/2} \, , \\
  \intertext{since $\kappa_1-3/2>2(\kappa_1-1)$. Moreover,}
  	\frac{1}{4}- |\partial_2F|^2 
  	  &= 
  	    K_2\kappa_2r_2^{\kappa_2-1}(1 -K_2\kappa_2r_2^{\kappa_2-1}) \sim  r_2^{\kappa_2-1}\, ,\\
  	\partial_1^2 F
  	  &= F_U''(r_1) + K_1\kappa_1(1-\kappa_1) r_1^{\kappa_1-2} 
  	  	= -\frac{a^2}{r_1^2}\left( \veps+a^2/r_1 \right)^{-1/2} 
  	  	+ K_1\kappa_1(1-\kappa_1) r_1^{\kappa_1-2} \\
  	  &\ge -\frac{a^2}{r_1^2}\frac{r_1^{1/2}}{c} 
  	  		+ K_1\kappa_1(1-\kappa_1) r_1^{\kappa_1-2}
  	  		\gtrsim -r_1^{-3/2} \, ,\\
  	\partial_2F\frac{2}{r_2} - r_2^{-1}
  	  &= 
  	    ( \frac{1}{2} - K_2\kappa_2 r_2^{\kappa_2-1})\frac{2}{r_2} -r_2^{-1} \sim - r_2^{\kappa_2-2}
  \end{align*} 
  for all large enough $r_1,r_2$. So we get  
  \begin{align*}
  	(H_U-E_U)f
  	  &\gtrsim 
  	       f\Big[ -r_1^{-1} + r_1^{\kappa_1-3/2} 
  	       		+r_2^{\kappa_2-1} -r_1^{-3/2}  -r_2^{\kappa_2-2}  
  	        \Big]
  	       \gtrsim   f\Big[ 
  	          -r_1^{-1} +r_2^{\kappa_2-1}  
  	        \Big]\\
  	  &\ge f\Big[ 
  	           -r_1^{-1} +r_1^{\kappa_2-1}   
  	        \Big] \gtrsim f \, r_1^{\kappa_2-1} > 0 \, ,
  \end{align*}
  for all $x\in B_R^-$ and large enough $R$. 
  The second inequality holds since $r_1^{\kappa_1-3/2}>r_1^{-3/2}$ for large $r_1$, the third because 
  $r_2^{\kappa_2-1}\ge  r_1^{\kappa_2-1}$ and the fourth, because 
  $r_1^{\kappa_2-1} - r_1^{-1}= r_1^{\kappa_2-1}(1 -r_1^{-\kappa_2})\gtrsim r_1^{\kappa_2-1}$. This proves Lemma \ref{lem:classical supersolution} uniformly in 
  $0\le U\le U_c$. 
  
  In the subcritical case, when $0\le U\le U_c-\mu$ for some fixed small $\mu>0$, we have the bound  
  \begin{align*}
  	  \veps -|\partial_1F|^2 
  	    &\gtrsim - r_1^{-1} + r_1^{\kappa_1-1} 
  	   			- r_1^{2(\kappa_1-1)} 
  	   		\gtrsim  r_1^{\kappa_1-1}\, ,
  \end{align*}
  uniformly in $0\le U\le U_c-\mu$ and  therefore 
  \begin{align*}
  	(H_U-E_U)f
  	  &\gtrsim 
  	       f\Big[ r_1^{\kappa_1-1} 
  	       		+r_2^{\kappa_2-1} -r_1^{-3/2}  -r_2^{\kappa_2-2}  
  	        \Big] > 0 \, ,
  \end{align*}
  since $0<\kappa_1, \kappa_2<1$,  $r_1\ge R$, 
  $r_2\ge R^\alpha -1$ and $R$ is large enough, which proves the second claim. 
\end{proof}

Now we come to the proof of the global upper bound.
\begin{theorem}[Sharp upper bound, arbitrary coupling]\label{thm:sharp upper bound all coupling most general}
  For any choice of parameters 
  $K_1,K_2>0$, $1/6 <\kappa_1<1/2$, and 
	$(3-2\kappa_1)/4<\kappa_2<1$ 
  there exist constants 
  $C_+$ depending only on $\kappa_1,\kappa_2, K_1, $ and $K_2$, 
  such that for the unique positive choice of the ground state of the helium-type operator $H_U$ the pointwise bound 
  \begin{align}\label{eq:sharp upper bound critical range}
    \psi_U(x)\le C_+ \exp\left( -F^U_+(|x|_\infty,|x|_0) \right)
  \end{align}
  holds uniformly in $0\le U\le U_c$. 
  
  For the subcritical case, where for fixed small $\mu>0$ the repulsion parameter $U$ is allowed to vary uniformly 
  in $0\le U\le U_c-\mu$ assume that $0<\kappa_1,\kappa_2<1$, 
  $\max(\kappa_1,\kappa_2)>1/2$, and $K_1,K_2>0$. Then there 
  exist a constant  $\wti{C}_+$, depending only on $\kappa_1,\kappa_2, K_1, K_2$, 
  and also $\mu$, such that the upper bound  
    \begin{align}\label{eq:sharp upper bound subcritical range}
    \psi_U(x)\le \wti{C}_+ \exp\left( -F^U_+(|x|_\infty,|x|_0) \right)
  \end{align}
  holds for all  $0\le U\le U_c -\mu$. 
\end{theorem}
\begin{remark}
  Note that Theorem \ref{thm:global pointwise anisotropic upper bound} is a special 
case of  Theorem \ref{thm:sharp upper bound all coupling most general} for $U=U_c$. 
\end{remark}

\begin{proof}[Proof of Theorem \ref{thm:sharp upper bound all coupling most general}{\rm :}]
  Since $\psi_U$ is bounded uniformly in $0\le U\le U_c$, see Proposition \ref{prop:basic properties}, and $f_U=\exp(-F^U_+)$ is bounded away from zero on compact sets uniformly in $0\le U\le U_c$,  
  we clearly have $\psi_U(x)\le Cf_U(x)=C\exp(-F^U_+(x))$ for all $|x|_\infty\le R$ for 
  some constant $C$, which might depend on $R$ and the 
  parameters but not on $U$.  
  
  So it is enough to show  there exists some constant 
  $C$ such that $\psi_U(x)\le C f_U(x)$ on $\{|x|_\infty>R\}$ for some $R>0$. 
  By symmetry, it is enough to prove that there exists a constant $C$ such that 
  \begin{align}\label{eq:upper bound on A_R}
  	\psi_U(x) \le Cf_U(x)= C\exp(-F^U_+(x))  
  \end{align}
  for all $\{ |x_2|\le |x_1|, |x_1|>R \}$ and large enough $R>0$.
    
  Fix any $0<\alpha<\min(\kappa_1,1/2)$. 
  Due to the assumptions on the parameters 
  $\kappa_1,\kappa_2,K_1,K_2$ in Theorem \ref{thm:sharp upper bound all coupling most general} the assumptions of Lemma \ref{lem:upper bound on boundary layer} are satisfied 
  with this choice of $\alpha$. 
  Hence the bound 
  \eqref{eq:upper bound on A_R} holds for some constant $C$ on the sets   
  $B_{1,R}\coloneqq \{ |x_2|\le |x_1|^\alpha, |x_1|>R \}$ and 
  $B_{2,R}\coloneqq\{  |x_1|-1\le |x_2|\le |x_1|, |x_1|>R \}$, both in the critical and subcritical case. 
  
  So we only have to prove the same bound on the intermediate region 
  $B_R^-=\{ |x_1|^\alpha -1<|x_2|<|x_1|, |x_1|>R  \}$. 
  Consider 
  \begin{align*}
  	\wti{\partial} B_R^1 
  		&\coloneqq 
  			\{ |x_1|-1\le |x_2| <  |x_1|, |x_1|>R+1\} \\
  	\wti{\partial} B_R^2 
  		&\coloneqq 
  			\{ |x_1|^\alpha-1 <  |x_2| \le  |x_1|^\alpha, |x_1|>R+1\} \\
  	\intertext{and}
  	\wti{\partial} B_R^0 
  		&\coloneqq 
  			\{ |x_1|^\alpha-1 <  |x_2| < |x_1|, R< |x_1|\le R+1\} \, .
  \end{align*}
  Then $\wti{\partial} B_R= \cup_{j=0}^2 \wti{\partial} B_R^j$ is a boundary layer of $B_R^-$ in the sense 
  of Definition \ref{def:boundary layer}. 
  \begin{figure}
  \centering
  \begin{tikzpicture}
    \node[anchor=south west,inner sep=0] (image) at (0,0) {\includegraphics[width=0.4\textwidth]{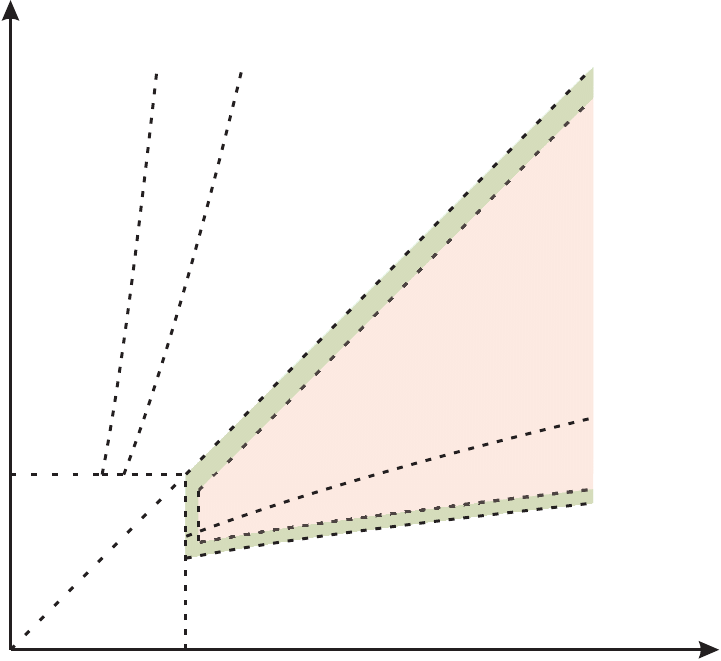}};
    \begin{scope}[x={(image.south east)},y={(image.north west)}]
        \draw (0.97,0) node[below] {$r_1$};
                  \draw (-0.04,.91) node[above] {$r_2$};
                                          \draw (0.93,.85) node[above] {$r_1=r_2$};
                               \draw (0.95,.31) node[above] {$r_2=r_1^{\gamma}$};
                                          \draw (0.935,.18) node[above] {$r_2=r_1^{\alpha}$};
    \end{scope}
  \end{tikzpicture}
  \caption{The intermediate region $B_R^-$ and its boundary layer (colored in green). 
  Lemma \ref{lem:upper bound on boundary layer}, which is heavily based on Corollary \ref{cor:first anisotropic pointwise upper bound},  
  provides sharp a--priori upper bounds on the ground state in the part of the boundary layer of 
  $B_R^-$ near the diagonal $r_2=r_1$ and deep within the tricky region when $r_2=r_1^\alpha$.} 
  \label{fig:boundary layer global upper bound}
  \end{figure}
  Moreover, as already mentioned above, we know from 
  Lemma \ref{lem:upper bound on boundary layer} that 
  \eqref{eq:upper bound on A_R} holds on the parts $\wti{\partial} B_R^1$ and $\wti{\partial} B_R^2$ when 
  $R$ is large enough.  
  Since $f_U= \exp(-F^U_+)$  is bounded away from zero on any compact set and the closure of $\wti{\partial} B_R^0$ 
  is bounded, we see that $f_U$ is bounded away from zero on $\wti{\partial} B_R^0$. Using that $\psi_U$ is bounded, 
  shows that the bound \eqref{eq:upper bound on A_R} also holds on $\wti{\partial} B_R^0$ for 
  some constant $C$.  
  Thus, enlarging the constant if necessary, we see that the upper bound \eqref{eq:upper bound on A_R} holds 
  on the boundary layer $\wti{\partial} B_R$ for some 
  constant $C$. This constant only depends on the constant from Lemma \ref{lem:upper bound on boundary layer}, on uniform bounds on $\psi_U$ from Proposition \ref{prop:basic properties}, and lower bounds on $f_U$ on compact sets, which are independent of $U$.     
  
  To finish the argument, we note that $f_U$ is a classical supersolution, hence also in the quadratic form sense,  
  of $H_U$ at energy $E_U$ on $B_R^-$ by Lemma \ref{lem:classical supersolution} (for the precise notion see the discussion in  Appendix \ref{app:sub-super-solutions}). Moreover,   
  $\psi_U\in H^1(\R^6)$ is a solution, hence also a subsolution, in the quadratic form sense 
  a la Agmon on $B_R^-$. 
  The subharmonic comparison principle given in Theorem \ref{thm:Agmon subsolution bound} then yields the upper bound 
  \eqref{eq:upper bound on A_R} on all of $B_R^-$.  This proves  
  the upper bound from Theorem \ref{thm:sharp upper and lower bounds all coupling} in the critical and subcritical case. 	
\end{proof}
\begin{remark}
  It is very convenient to have a quadratic form version of the subharmonic comparison principle, since it allows 
  us to directly work with weak eigenfunctions, which are only in $H^1(\R^6)$ and not in $H^2(\R^6)$. 
  To the best of our knowledge, the quadratic form version of the subharmonic comparison principle goes back 
  to a beautiful paper by Agmon \cite{Agm85}. Theorem \ref{thm:Agmon subsolution bound} in the appendix 
  is a slight extension of Agmon's original result. Agmon works on open sets which are neighborhoods of 
  infinity, i.e., complements of compacts sets, while we have to work on unbounded sets, which are not 
  necessarily  neighborhoods of infinity.  
\end{remark}

%%%%%%%%%%%%%%%%%%%%%%%%%%%%%%%%%%%%%%%%%%%%%%%%%%%%%%%%%%%%%%%%%%%%%
%%%%%%%%%%%%%%%%%%%%%%%%%%%%%%%%%%%%%%%%%%%%%%%%%%%%%%%%%%%%%%%%%%%%%
\section{Lower bound in the tricky region} \label{sec:lower bound tricky region}
%%%%%%%%%%%%%%%%%%%%%%%%%%%%%%%%%%%%%%%%%%%%%%%%%%%%%%%%%%%%%%%%%%%%%
%%%%%%%%%%%%%%%%%%%%%%%%%%%%%%%%%%%%%%%%%%%%%%%%%%%%%%%%%%%%%%%%%%%%%
\begin{theorem}[Lower pointwise bound in the tricky region]\label{thm:lower bound tricky region}
Let $H_U$ be the helium--type  Schr\"odinger operator given in \eqref{eq:HeliumHamiltonian} and let 
$\psi_U \in L^2(\R^6)$ be the positive ground state of $H_{U}$ for $0\le U\le U_c$. 
Then for any $1/6<\kappa<1/2$ and any 
$0<\gamma<\kappa+1/2 $ and any $K>0$  
\begin{equation}\label{eq:lower bound tricky region}
	\psi_U (x)\gtrsim \exp\Big(-F_U(|x|_\infty)-K|x|_\infty^\kappa-\frac{1}{2}|x|_0\Big)
	\quad \text{for all } |x|_0\le |x|_\infty^\gamma
\end{equation}
with $F_U$ defined in \eqref{eq:F_U}. 
Again we use  $|x|_0=\min\{|x_1|,|x_2|\}$, $|x|_\infty=\max\{|x_1|,|x_2|\}$. The implicit constant in the above bound depends on the parameters $\kappa, \gamma$, and $K$,  
but not on $U$ in the range $0\le U\le U_c$. 

Moreover, in the subcritical case, i.e., when $U$ stays away from $U_c$,  the lower bound 
\eqref{eq:lower bound tricky region} holds for any fixed 
small $\mu$  uniformly in  $0\le U\le U_c-\mu$ and  any 
$0<\kappa<1$ and $0<\gamma<1$.   
\end{theorem}
\begin{remark}   
  Theorem \ref{thm:lower bound tricky region} provides a lower bound for the ground state when one particle escapes while the other stays close to 
  the nucleus. This is an important a-priori bound for 
  our proof of the global lower bound. 
  
  Since the leading order behavior of $F_U(r)$ switches from linear to $r^{1/2}$ at $U=U_c$, we need the 
  restriction $\kappa<1/2$ in the first case, so that $r^{\kappa}$ is a lower order correction compared to 
  the leading order term $F_U(r)$,  uniformly in $0\le U\le U_c$.
  
  When $U$ stays away from $U_c$, the leading order term $F_U(r)$ is always linear in $r$, which allows for a larger range for $\kappa$ and $\gamma$. 
\end{remark}

\begin{proof} 
  For $0<\gamma<1$ define  
  $\wti{A}=\{|x|_0\le |x| _\infty^\gamma \}$. 
  For large enough $R$, the regions 
  $\wti{A}_R^-=\{ |x_2|\le |x_1|^\gamma, |x_1|> R \}$ 
  and $\wti{A}_R^+=\{ |x_1|\le |x_2|^\gamma, |x_2|> R \}$ 
  are disjoint. 
  By symmetry, a lower bound of the form 
  \eqref{eq:lower bound tricky region} on the set $\wti{A}_R^-$, for some $R>0$ implies the same bound on  
  $\wti{A}_R^+$. The set  $\wti{A}\setminus (\wti{A}_R^-\cup \wti{A}_R^+) =\{|x|_0\le |x|_\infty^\gamma\le R^\gamma\}$ is a 
  compact subset of $\R^6$. Clearly the right hand side of \eqref{eq:lower bound tricky region} bounded above on compact sets and, because of 
  Proposition \ref{prop:basic properties}, $\psi_U$ 
  is bounded away from zero on compact sets. Hence there is some constant $C>0$ with 
  \begin{align*}
  	\psi_U(x)\ge C\exp\Big(-F_U(|x|_\infty)-K|x|_\infty^\kappa-\frac{1}{2}|x|_0\Big) 
  	\quad \text{for all } x\in \wti{A}\setminus (\wti{A}_R^-\cup \wti{A}_R^+)\, .
  \end{align*}
  Thus, shrinking the involved constant, if necessary, one sees that  
  the lower bound \eqref{eq:lower bound tricky region} holds on $\wti{A}$ iff it holds on $\wti{A}_R^-$ for some $R>0$.
  
  For technical reasons, we need to work on the slightly larger set 
  \begin{equation*}
  	  A_{R}^-\coloneqq\{ |x_2|< 2|x_1|^\gamma, |x_1|\ge R \}.  
  \end{equation*}
  which is just the (lower) tricky region from Section \ref{sec:local energy bound}.  
  Recall that $a=(U-1)_+^{1/2}$ and $\veps= -\frac{1}{4}-E_U$ is the ionization energy. $F_U$, defined in \eqref{eq:F_U}, depends on these two parameters. 
  Finally we define
  \begin{equation} \label{eq:def F tricky region}
  	\begin{split}
  	F_{3,U}(r_1,r_2)
  		&= F_U(r_1) +K r_1^\kappa -\tfrac{1}{2}r_2\quad \text{ for } r_1,r_2\ge 0\, ,\\
  	\Phi(s) &\coloneqq
  		\begin{cases} 
			1\,, &\quad 0\leq s\leq 1\\
  			\cos\left(\frac\pi 2(s-1)\right)\,, &\quad 1<s<2\\
			0\,, &\quad 2\leq s\\
		\end{cases} \,  \\
	\chi(x)
		&\coloneqq 
			\Phi\Big( \frac{|x_2|}{|x_1|^\gamma} \Big) \quad \text{for } x=(x_1,x_2)\in A_{R}^- 
	\end{split}
  \end{equation}
  and 
\begin{equation}\label{eq:def g tricky region}
  	g_U(x)\coloneqq \chi(x) \exp(-F_{3,U}(|x_1|, |x_2|)) \quad \text{for } 
  	x=(x_1,x_2)\in A_{R}^-\, .
  \end{equation}
  We will suppress the dependence of $F_{3,U}$ and $g_U$ on 
  the other parameters, except on $U$, for simplicity of notation. 
  Note that $\chi(x)=1$ on $\wti{A}_R^-$, so the bound \eqref{eq:lower bound tricky region} will hold on $\wti{A}_R^-$, 
  hence also on $\wti{A}$, once we show that $\psi_U\ge C g_U$ on $A_{R}^-$ for some $R>0$ and some constant $C>0$, which can depend on all the parameters, but not on $U$. 
  
  The boundary $\partial A_{R}^-$ is the union of  the unbounded part 
  $\partial A_{R,1}^-=\{ |x_2|= 2|x_1|^\gamma, |x_1|\ge R \}$ and 
  the compact set $\partial A_{R,2}^-=\{|x_1|=R, |x_2|\le 2 R^\gamma\}$. 
  We clearly have  $\psi_U(x)>0 =g_U(x)$ for all $x\in \partial A_{R,1}^-$. By continuity of $\psi_U$ 
  and $g_U$, there exists an open neighborhood $O_1$  
  of $\partial A_{R,1}^-$ such that $\psi_U(x)> g_U(x)$ for all $x\in O_1$. 
  Hence $\psi_U \ge g_U$ on $O_1\cap A_R^-$. 
  
  Moreover, $\psi_U>0$ is bounded from below by a positive constant on compact subsets of $\R^6$, uniformly in $0\le U\le U_c$, see Proposition \ref{prop:basic properties} and 
  $g_U\ge 0$ is continuous and depends continuously on the parameters. Thus, since $\partial A_{R,2}^-$ is compact, there exists  
  a constant $C>0$ such that $\psi_U\ge 2Cg_U>Cg_U$ on  
  $\partial A_{R,2}^-$ for all $0\le U\le U_c$. 
  By continuity, there exist an open neighborhood $O_2$ of $\partial A_{R,2}^-$ such that 
  $\psi_U>Cg_U$ on $O_2$, hence $\psi_U> Cg_U$ on $O_2\cap A_R^-$. Thus 
  \begin{align*}
  	\psi_U\ge \min(C,1)g_U \text{~~ on the boundary layer } (O_1\cup O_2)\cap A_R^- \, .
  \end{align*}
  
  Now assume that  $1/6<\kappa<1/2$ and 
  $(3-2\kappa)/4<\gamma <\kappa+1/2$.  
  In Lemma \ref{lem:classical subsolution in A_{R}^-} we 
  show that under these conditions there exists $R>0$ 
  such that $g_U$ is a subsolution of $H_U$ at energy $E_U$ in $A_{R}^-$, for all $0\le U \le U_c$, in the quadratic 
  form sense a la  Agmon. Moreover, it is easy to see that $g_U\in L^2(\R^6)$.  
  Hence we can apply Theorem \ref{thm:Agmon subsolution bound} to conclude that 
  $\psi_U\ge C g_U$ on all of $A_{R}^-$, hence on all of 
  $\{|x|_0\le |x|_\infty^\gamma\}$. 
  
  It remains to get rid of the lower bound on $\gamma$. 
  Note that the sets $\{|x|_0\le |x|_\infty^\gamma\}$ are 
  increasing in $0<\gamma<1$, i.e., once the bound 
  \eqref{eq:lower bound tricky region} holds for some $(3-2\kappa)/4<\gamma<\kappa+1/2$  it holds for all $0<\gamma\le \kappa+1/2$. This finishes the proof of 
  Theorem  \ref{thm:lower bound tricky region}. 
\end{proof}

It remains to prove that for suitable choices of the parameters, 
the function $g_U$ defined in \eqref{eq:def g tricky region} is 
a classical subsolution in the tricky region. 
\begin{lemma}\label{lem:classical subsolution in A_{R}^-}
  Let $\kappa>1/6$ and $(3-2\kappa)/4<\gamma<\min(\kappa+1/2,1)$. 
  Then there exists $R>0$ such that  the function 
  $g_U$ defined in \eqref{eq:def g tricky region}   
  is a classical subsolution of $H_U$ at energy $E_U= -1/4-\veps_U\le -1/4$ 
  on the set $A_{R}=\{ |x|_0<2|x|_\infty^\gamma, |x|_\infty>R \}$, that is,  
  \begin{align}\label{eq:classical subsolution tricky region}
  	(H_U-E_U) g_U \le 0 \quad \text{pointwise in } A_{R} \, 
  \end{align}
  for any $0\le U\le U_c$.  
  In particular, it is a subsolution in 
  the quadratic form sense a la  Agmon: for large enough $R$ and all $0\le \varphi\in \calC^\infty_0(A_{R})$ we have the quadratic form inequality  
  \begin{align}\label{eq:subsolution a la Agmon tricky region}
  	\la \varphi, (H_U-E_U) g_U \ra \le 0 \, .
  \end{align}
  
  Moreover, in the subcritical case, i.e., when $U$ stays away from $U_c$,  the same  holds for any fixed 
small $\mu$  uniformly in  $0\le U\le U_c-\mu$ and  any 
$\kappa>0$ and $\max((1-\kappa)/2,0) <\gamma<1$.   
\end{lemma}
\begin{remark}
  It is easy to see that $(3-2\kappa)/4<\kappa+1/2$ is equivalent to 
  $\kappa>1/6$. So the set of allowed values for $\gamma$ is 
  not empty iff $\kappa>1/6$. 
\end{remark}

\begin{proof}
  The bound \eqref{eq:subsolution a la Agmon tricky region} immediately follows from 
  \eqref{eq:classical subsolution tricky region} by integration by parts, since $\varphi$ has 
  compact support inside $A_{R}$ and is non--negative on $A_{R}$. 
  Moreover, since $A_{R}$ splits into the regions 
  $A_{R}^-=\{ |x_2|<2|x_1|^\gamma, |x_1|>R \}$ and $A_{R}^+=\{ |x_1|<2|x_2|^\gamma, |x_2|>R \}$ which are disjoint for large enough $R$, it is enough to prove \eqref{eq:classical subsolution tricky region} on $A_{R}^-$, by symmetry.
    
  In the remainder of the proof, we will work only on 
  $A_{R}^-$,   abbreviate $F=F_{3,U}$ 
  and $g=g_U$, and also identify the function $F$ with $F(|x_1|,|x_2|)$, for simplicity of notation. 
  
  Clearly,  $F$ is twice continuously differentiable on $A_{R}^-$. Moreover, $\Phi$ is twice 
  differentiable on $[0,2)$ and one easily checks that 
  $g\in H^1_0(A_{R}^-)\cap H^2(A_{R}^-)$.  
  Since $\nabla g = \nabla\chi e^{-F}-\chi \nabla F e^{-F}$, we have 
  \begin{equation*}
  	-\Delta g
  		= e^{-F}\big[ \chi \Delta F -\chi |\nabla F|^2 -\Delta\chi +2\nabla\chi \nabla F\big] \, .
  \end{equation*}
  From the definition of $F$ and $\chi$, setting $r_1=|x_1|$, $r_2=|x_2|$, 
  and recalling $\partial_1=\partial_{r_1}$, $\partial_2=\partial_{r_2}$,   one calculates 
  \begin{align*}
  	\nabla F(x)
  		&= 	\begin{pmatrix}
 				\partial_1 F\frac{x_1}{|x_1|} \\
 				\partial_2 F\frac{x_2}{|x_2|}
 			\end{pmatrix}
 		=
 			\begin{pmatrix}
 				\big( F_U'(r_1)+K\kappa r_1^{\kappa-1}\big)\frac{x_1}{|x_1|} \\
 				\frac{1}{2}\frac{x_2}{|x_2|}
 			\end{pmatrix}\, \\
 	\intertext{with $F_U'(r_1) =(\veps+a^2/r_1)^{1/2}$ and $\veps=1/4-E_U\ge0$ and $a=(U-1)_+^{1/2}$. Moreover,}
 	\nabla \chi(x) 
 		&= \Phi'\Big(\frac{r_2}{r_1^\gamma} \Big) \begin{pmatrix}
 					-\gamma\frac{r_2}{r_1^{\gamma+1}}\frac{x_1}{|x_1|} \\
 					\frac{1}{r_1^\gamma}\frac{x_2}{|x_2|}
 				\end{pmatrix}\, .
  \end{align*}
 Thus 
  \begin{align*}
  	\nabla\chi(x) \nabla F(x) 
  		&= \Phi'\Big(\frac{r_2}{r_1^{\gamma}}\Big) \, 
  				\left[\partial_2 F r_1^{-\gamma} - \gamma\partial_1 F \,r_2 r_1^{-\gamma-1} 
  			\right]
  \end{align*}
  and 
  \begin{align*}
  	\Delta\chi(x) 
  		&= 
  			 \Phi''\Big(\frac{r_2}{r_1^{\gamma}}\Big)\, 
  				\left( 
  					\gamma^2 r_2^2 r_1^{-2\gamma-2} + r_1^{-2\gamma}
  				\right)
  			+\Phi'\Big(\frac{r_2}{r_1^{\gamma}}\Big) 
  				\left(
  					2r_1^{-\gamma} r_2^{-1}- \gamma(1-\gamma) r_2r_1^{-\gamma-2}
  				\right) \, ,
  \end{align*}
  so 
  \begin{equation} \label{eq:drop}
  	\begin{split}
  	2\nabla\chi(x) \nabla F(x) & -\Delta\chi(x) \\
  		&= - \Phi''\Big(\frac{r_2}{r_1^{\gamma}}\Big)\, 
  				\left( 
  					\gamma^2 r_2^2 r_1^{-2\gamma-2} + r_1^{-2\gamma}
  				\right) \\
  		&\phantom{=~~}	+ \Phi'\Big(\frac{r_2}{r_1^{\gamma}}\Big) \, 
  				r_1^{-\gamma}\left[2\partial_2 F - 2\gamma\partial_1 F\, r_2 r_1^{-1} 
  					+\gamma(1-\gamma) r_2r_1^{-2} - 2 r_2^{-1}
  			\right] \, .
    \end{split}
  \end{equation}
  Note that $\Phi'\le 0$ and since the support of  $\Phi'$ is contained in the interval $[1,2]$, we have $r_1^\gamma\le r_2\le 2r_1^\gamma$ for the second term on the right hand side above. Thus we get the lower bound 
  \begin{align*}
  	2\partial_2 F - 2\gamma\partial_1 F\, r_2 r_1^{-1} 
  					+\gamma(1-\gamma) r_2r_1^{-2} - 2 r_2^{-1}
  		\ge 2\partial_2 F - 4\gamma\partial_1 F\, r_1^{\gamma-1} 
  					 - 2 r_2^{-\gamma}
  		\gtrsim 1
  \end{align*}
  since $2\partial_2F = 1$ and $\partial_1F$ is bounded at infinity uniformly in $0\le U\le U_c$. 
  
  So we can again use $\Phi'\le 0$ to drop 
  the second term on the right hand side of \eqref{eq:drop} to arrive at  
  \begin{align*}
  	2\nabla\chi(x) \nabla F(x) -\Delta\chi(x) 
  		&\le - \Phi''\Big(\frac{r_2}{r_1^{\gamma}}\Big)\, 
  				\left( 
  					\gamma^2 r_2^2 r_1^{-2\gamma-2} + r_1^{-2\gamma}
  				\right) 
  		\le \frac{\pi^2}{4}\chi(x) \, \left( 
  					4\gamma^2 r_1^{-2} + r_1^{-2\gamma}
  				\right)
  \end{align*}
  for all large enough $r_1$, where we also used that $-\Phi''\le \tfrac{\pi^2}{4} \Phi$ on $[0,2)$ by definition of $\Phi$.  
  Hence  
   \begin{equation}\label{eq:subsolution tricky region key}
  	-\Delta g
  		\le  g \big[ \Delta F - |\nabla F|^2 
  						+  \frac{\pi^2}{4}( 
  							4\gamma^2 r_1^{-2} + r_1^{-2\gamma}) \big] \quad \text{on } A_{R}^- \, .
  \end{equation}
  for large enough $R$. 
  
  Now the rest of the argument is easy: One checks 
  \begin{align*}
  	\Delta F = \partial_1^2 F +\partial_1 F \frac{2}{r_1} + \partial_2^2 F + \partial_2 F\frac{2}{r_2}
  		= \partial_1^2 F +\partial_1 F \frac{2}{r_1}  + \frac{1}{r_2}
  \end{align*}
  since $\partial_2 F=1/2$. 
  Using this, $1/4-(\partial_2F)^2=0$, and  \eqref{eq:subsolution tricky region key},  we get 
  \begin{align}\label{eq:subsolution tricky region key-2}
  	(H_U-E_U) g 
  		&\le g\left[ 
  				\veps-(\partial_1 F)^2 +\frac{U}{|x_1-x_2|} -\frac{1}{r_1}  
  				+ \partial_1^2 F +\partial_1 F \frac{2}{r_1}   
  				+ \frac{\pi^2}{4}( 
  							4\gamma^2 r_1^{-2} + r_1^{-2\gamma})
  			\right]
  \end{align}
  On $A_{R}^-$ we have $|x_2|\le 2 |x_1|^\gamma$, so $|x_1-x_2|\ge |x_1|-|x_2| \ge r_1-2r_1^{\gamma}$. Thus 
  \begin{align*}
  	\frac{U}{|x_1-x_2|} - \frac{1}{r_1} 
  	  \le \frac{U}{r_1-2r_1^\gamma}-\frac{1}{r_1} 
  		=  \frac{U-1}{r_1} + \frac{2U r_1^{\gamma-2}}{1-2r_1^{\gamma-1}} \quad \text{on } A_{R}^-
  \end{align*}
  for large enough $R$. Since $\partial_1F = F_U'(r_1)+K\kappa r_1^{\kappa-1}$ and $\gamma<1$, we have   
  \begin{align*}
  	\veps-(\partial_1 F)^2 +\frac{U}{|x_1-x_2|} -\frac{1}{r_1}
  		&\le -2K\kappa_1 F_U'(r_1)r_1^{\kappa-1} 
  				+ 4U r_1^{\gamma-2} 
  \end{align*}
  for all $r_1>0$ such that $1-2r_1^{\gamma-1}\ge 1/2$.  
   The bound  $F_U''\le 0$ implies 
  \begin{align*}
  	\partial_1^2 F +\partial_1 F\frac{2}{r_1} 
  		&=
  			F_U''(r_1) + K\kappa(\kappa+1) r_1^{\kappa-2} 
  			+ 2F_U'(r_1)r_1^{-1} \\
  		&\le 2F_U'(r_1)r_1^{-1} 
  			+ K\kappa(\kappa+1) r_1^{\kappa-2}
  \end{align*}
  which leads to   
  \begin{align*}
  	  	\veps-(\partial_1 F)^2 &+\frac{U}{|x_1-x_2|} -\frac{1}{r_1}
  	  	+   	\partial_1^2 F +\partial_1 F\frac{2}{r_1} \\
  	  	&\lesssim - F_U'(r_1) r_1^{\kappa-1} + r_1^{\gamma-2} + r_1^{\kappa-2}
  	  	  \lesssim -r_1^{\kappa-3/2} + r_1^{\gamma-2} 
  \end{align*}
  for large $r_1$, using again the bound $F_U'(r_1)\ge c r_1^{-1/2}$ with some constant $c>0$, uniformly in 
  $0\le U\le U_c$. 
  Thus from \eqref{eq:subsolution tricky region key-2} we get 
  \begin{align*}
  	 	(H_U-E_U) g 
  		&\lesssim  g\left[ 
  				-r_1^{\kappa-3/2} +r_1^{\gamma-2} +r_1^{-2}  
  				+r_1^{-2\gamma} 
  				\right] \le 0
  			\quad \text{in } A_{R}^-
  \end{align*}
  for all large enough $R$, under the condition that 
  $\kappa-3/2>\max(\gamma-2, -2\gamma)$, which is equivalent 
  to $(3-2\kappa)/4<\gamma<\kappa+1/2$.  
  Since we have the constraint $\gamma<1$, this 
  is equivalent to the condition on the parameters in the 
  range $0\le U\le U_c$. 
  
  In the subcritical case we use that 
  $F_U'(r_1)\gtrsim c_\mu>0$, for any $0\le U\le U_c-\mu $ 
  so $ F_U'(r_1) r_1^{\kappa-1}\gtrsim r_1^{\kappa-1}$
   uniformly in  $0\le U\le U_c-\mu$.  Using this as before, we see that $(H_U-E_U) g \le 0$ on $A_{R}^-$ for all large enough $R$, under the condition that 
   $\kappa-1>\max(\gamma-2, -2\gamma)$. 
   This is equivalent to $(1-\kappa)/2 <\gamma< \kappa +1$. 
   Since we have the constraint $0<\gamma<1$, this 
  is equivalent to the condition on the parameters in the 
  subcritical case. 
\end{proof}

%%%%%%%%%%%%%%%%%%%%%%%%%%%%%%%%%%%%%%%%%%%%%%%%%%%%%%%%%%%%%%%%%%%%%
%%%%%%%%%%%%%%%%%%%%%%%%%%%%%%%%%%%%%%%%%%%%%%%%%%%%%%%%%%%%%%%%%%%%%
\section{Global lower bound}\label{sec:global lower bound}
%%%%%%%%%%%%%%%%%%%%%%%%%%%%%%%%%%%%%%%%%%%%%%%%%%%%%%%%%%%%%%%%%%%%%
%%%%%%%%%%%%%%%%%%%%%%%%%%%%%%%%%%%%%%%%%%%%%%%%%%%%%%%%%%%%%%%%%%%%%
In this section we prove the global pointwise lower bound for the ground state of helium type atoms, including the  critical coupling. 
The Coulomb repulsion of the particles, which can become \emph{arbitrary large} when the particles 
are close to each other, 
requires a considerably more complicated ansatz for the subsolution  compared to the proof of the 
lower bound in the \emph{tricky region} in Section \ref{sec:lower bound tricky region}. 
On the other hand, since we already proved a lower bound in Theorem \ref{thm:lower bound tricky region} 
in the tricky region, 
we can now assume that both $|x|_0$ and $|x|_\infty$ are large, which helps to control the errors. 

Recall that for non-negative constants $\kappa_1,\kappa_2, K_1, K_2$, and $s,r_1,r_2\ge 0$ 
the weight function $F^U_-$ is given by  
\begin{align}\label{eq:F^U_-}
	F^U_-(r_1,r_2) 
		&\coloneqq F_U(r_1) +\frac{1}{2} r_2 +K_1 r_1^{\kappa_1} +K_2 r_2^{\kappa_2} 
\end{align}
with $F_U$ defined in \eqref{eq:F_U} and we consider the 
couplings in the range $\mu\le U\le U_c$ for some positive $\mu$. 
Moreover,  we will use 
\begin{align}
	\theta(s)
		&\coloneqq \frac{1}{1+s} \label{eq:def u}\\
	h(s,r_2) 
		&\coloneqq s\, \theta(r_2^{-\nu} s)  \label{eq:def h}
\end{align}
for some $0<\nu<1$. 
Recall also $1< U_c\le 2$. Furthermore, we will use 
\begin{align}\label{eq:def L lower bound}
	L_U(x)= L_U(x_1,x_2)  \coloneqq F^U_-(|x|_\infty,|x|_0) - h(|x_1-x_2|,|x|_0)
\end{align}
and abbreviate 
\begin{align}\label{eq:def g lower bound}
	g_U\coloneqq \exp(-L_U)
\end{align}
in the remainder of this section. Again, for simplicity of notation, we do not indicate the dependence of $L_U$ and $g_U$ on the other parameters except for $U$, in our notation. 
Note that $g_U\in H^1(\R^6)$, but its gradient has a jump discontinuity along the diagonal $|x_1|=|x_2|$, so it is 
not even locally in $H^2(U)$ for any open set $U\cap\{|x_1|=|x_2|\}\neq \emptyset$. Fortunately, 
the subharmonic comparison principle which we will use does not require local $H^2$ regularity. 

\begin{proposition}\label{prop:subsolution upper region} 
	For all $0<\beta,\nu<1$, $0<\kappa_1<\kappa_2< 1$, $\max(\nu,1-\nu)<\kappa_2$,  and $K_1,K_2>0$, the function $g_U=\exp(-L_U)$ is a subsolution of $H_U$ at energy $E_U$ in the sense of Agmon in the region 
	\begin{align}
		C_{R,\beta}\coloneqq\big\{ |x|_0> |x|_\infty^\beta, |x|_\infty>R \big\}
	\end{align}
	for all large enough $R$ uniformly in $0\le U\le U_c$. 
	That is, for all $0\le \varphi\in H^1(C_{R,\beta})$ one has, as quadratic forms,  
	\begin{align}\label{eq:subsolution everywhere}
		\la \varphi, (H_U-E_U)g_U \ra \le 0 \, 
	\end{align}
	and $R$ depends on the parameters $\kappa_1,\kappa_2$, and 
	$\nu$,  but not on $U$ in the range $0\le U\le U_c$. 
	
	Moreover, if $0<\beta,\nu, \kappa_1,\kappa_2< 1$, 
    $\max(\nu,1-\nu)<\kappa_2$, and $0<\mu<U_c$ 
    then there exists $R>0$ independent of $U$ in the 
    range  $\mu\le U\le U_c$ 
 such that \eqref{eq:subsolution everywhere} again holds. 
\end{proposition} 
\begin{proof}
  The proof of Proposition \ref{prop:subsolution upper region} follows directly from 
  Lemma \ref{lem: if classical subsolution then ok} and \ref{lem:classical subsolutions} at the end of this section. \ref{lem:classical supersolution}
\end{proof}
\begin{remark} 
  Since $\max(\nu,1-\nu)\ge 1/2$ we always have $1/2<\kappa_2<1$ in Proposition \ref{prop:subsolution upper region}. 
  
  The lower bound $U\ge \mu>0$ is needed in the proofs of Lemma 
  \ref{lem: if classical subsolution then ok} and 
  \ref{lem:classical subsolutions} when 
  $0<\kappa_2\le \kappa_1<1$. 
  It is not needed if $0<\kappa_1<\kappa_2<1$.

  The additional term $h$ in the comparison function 
  $g_U= \exp(-L_U)$ allows us to control the 
  Coulomb repulsion between the particles. We believe 
  that our choice of subharmonic comparison function 
  $g_U= \exp(-L_U)$ is not only considerably simpler than 
  the one used by Thomas Hoffmann--Ostenhof \cite{HofOst80-PhysLett} in 
  his derivation of exponential lower bounds for the ground state 
  of subcritical helium-type systems but it also allows us to 
  get the \emph{sharp coefficients} of the leading order terms 
  in the lower bound! 
     
  Moreover, the proof shows that one has considerable flexibility in the choice of $\theta$.  
  The only properties of $\theta$ which we need are
  \begin{smallenum}
      \item $\theta\ge 0$ is continuous on $[0,\infty)$ and 
            twice differentiable on $(0,\infty)$, 
      \item $\theta(s)\ge 1/2$ for $0\le s\le 1$, 
      \item $s\mapsto s\theta(s) $ is increasing and bounded on $[0,\infty)$,  
      \item $s\mapsto (1+s+s^2)|\theta'(s)|$ and  
            $s\mapsto (s+s^2+s^3)|\theta''(s)|$ are bounded on $(0,\infty)$
  \end{smallenum}
\end{remark}
 
Assuming Proposition \ref{prop:subsolution upper region} and given the quadratic form version of the 
sub-harmonic comparison principle, the proof of the following lower bound is relatively straightforward. 

\begin{theorem}\label{thm:global pointwise anisotropic lower bound with h}
  Let $\psi_U$ be the positive ground state of helium--type operator $H_{U}$, 
  i.e, $H_U\psi_U\!=E_U\psi_U $, in the quadratic form sense. Then for 
  $1/6< \kappa_1<\frac{1}{2}$, $0<\nu<1$,  $\max(\nu,1-\nu)<\kappa_2<1$,  and $K_1,K_2>0$, 
  we have the lower bound 
  \begin{align}\label{eq:global lower bound with h}
  	\psi_U(x)\ge C \exp\big( -L_U(x) \big)
  \end{align}
  for all $x\in \R^6$ where $L_U$ is defined in \eqref{eq:def L lower bound}, the constant $C$ depends on the parameters $\kappa_1,\kappa_2,\nu$, and $K_1,K_2$ but not on $U$ in the range $0\le U\le U_c$. 
  
  Moreover, if $0< \nu, \kappa_1< 1$, $\max(\nu,1-\nu)<\kappa_2<1$,  $K_1,K_2>0$, 
  and $0<\mu<U_c/2$, then we again have the bound \eqref{eq:global lower bound with h} for some constant $C$, depending on the parameters $\kappa_1,\kappa_2,\nu$, $K_1,K_2$, and $\mu$ 
  but not on $U$ in the range $\mu\le U\le U_c-\mu$. 
\end{theorem}
\begin{proof}\label{eq: a priori lower bound}
  The lower bound given in Theorem \ref{thm:lower bound tricky region} shows that uniformly in $0\le U\le U_c$ and for all fixed $1/6<\kappa_1<1/2$, $0<\gamma<\kappa_1+1/2 $,  
  and $K_1>0$ 
  \begin{align}
  	\psi_U(x)\gtrsim \exp\big(-F_U(|x|_\infty) -\frac{K_1}{2} |x|_\infty^{\kappa_1} -\frac{1}{2}|x|_0\big)\, .
  \end{align}
  for all $|x|_0\le |x|_\infty^\gamma$. 
  Since $s\theta(s)\le 1$, one sees  $0< h(s,r_2)= s \theta(r_2^{-\nu}s)\le r_2^\nu $ uniformly 
  in $s\ge 0$. Thus, since  $0<\nu<\kappa_2$, 
  \begin{align*}
  	F_U(r_1)  + K_1 r_1^{\kappa_1} +\frac{1}{2}r_2 
  		 \le F^U_-(r_1,r_2) -h(s,r_2) + r_2^\nu -K_2r_2^{\kappa_2} 
  		   \le F^U_-(r_1,r_2) - h(s,r_2) +C 
  \end{align*}
  for all $s,r_1,r_2 \ge 0$ and some constant $0\le C<\infty$. 
  This proves the lower bound \eqref{eq:global lower bound with h} 
  in the tricky region $\wti{A}=\{|x|_0\le |x|_\infty^\gamma\}$ for any $0<\gamma<\kappa_1+1/2$.    
  
  The same argument applies, for subcritical $U$ uniformly in 
  the range $0\le U\le U_c-\mu$, for small fixed $\mu>0$, 
  for all fixed $0<\gamma,\kappa_1<1$ and $0<\nu<\kappa_2$.  
  \smallskip
  
  Let $0<\beta<\gamma$ and abbreviate $ C_R=C_{R,\beta}=\{ |x|_0 > |x|_\infty^\beta, |x|_\infty>R \}$ for the remainder of this proof. Its boundary is given by 
  $\partial C_R= \partial C_R^1\cup \partial C_R^2$ with 
  $\partial C_R^1=\{ R^\beta\le |x|_0\le R, |x|_\infty=R\}$, which is compact, and 
  $\partial C_R^2=\{ |x|_0=|x|_\infty^\beta, |x|_\infty>R \}$, which is \emph{unbounded}. Since the bound \eqref{eq:global lower bound with h} 
  holds on the tricky region $\wti{A}$ it clearly holds on $\partial C_R^2$, that is, there exist a constant $C_1>0$ such that 
  \begin{align*}
  	\psi_U(x)\ge C_1 g_U(x) \quad \text{for all } x\in \partial C_R^2\,  
  \end{align*}
  where the constant $C_1$ does not depend on $U$. 
  Moreover, since $\psi_U$ is bounded away from zero uniformly in $0\le U\le U_c$ on compact sets, see Proposition \ref{prop:basic properties}, 
  and $g_U$ is continuous and $\partial C_R^1$ is compact, there exist a 
  constant $C_2>0$ such that 
  \begin{align*}
  	\psi_U(x)\ge C_2 g_U(x) \quad \text{for all } x\in \partial C_R^1\,, 
  \end{align*}
  with $C_2$ independent of $U$. 
  Hence, for some constant $C>0$ we have 
  $ \psi_U(x)\ge 2C g_U(x)$ for all $ x\in \partial C_R $. 
  By continuity of $\psi_U$ and $g_U$, there exists a boundary layer $\wti{\partial  C}_R$ 
  such that   
  \begin{align}\label{eq:lower bound on boundary layer helium}
  	\psi_U(x)\ge C g_U(x) \quad \text{for all } x\in \wti{\partial C}_R \,  .
  \end{align}
  Such a boundary layer is an open subset of 
  $C_R$ near the boundary $\partial C_R$ with $\dist(x,\partial C_R)>0$ for all $x\in C_R\setminus \wti{\partial C}_R $, see Appendix \ref{app:sub-super-solutions}. 
  
  Due to Proposition \ref{prop:subsolution upper region} we can use  
  Theorem \ref{thm:Agmon subsolution bound}, to extend the bound \eqref{eq:lower bound on boundary layer helium} 
  to almost all $x\in C_R$.  
  Since $\psi_U$ and $g_U=\exp(-L_U)$ are continuous, this lower bound clearly holds for all $x\in C_R$ and since we already 
  showed that the same type of lower bound holds in the region $\wti{A}=\big\{|x|_0\le |x|_\infty^\gamma \big\}$, 
  it holds on $C_R\cup \wti{A}$. 
  
  The complement of $C_R\cup \wti{A}$ in $\R^6$ is bounded, hence its closure is compact. Using Proposition \ref{prop:basic properties}
  one sees that $\psi_U$ is bounded away from zero on compact sets and since $g_U$ is bounded uniformly in $0\le U\le U_c$, a bound of the form 
  $\psi_U\ge C g_U $ also holds on 
  $\R^6\setminus(C_R\cup \wti{A})$. 
  Thus, at the expense of decreasing the constant $C>0$, if necessary,  
  the lower bound 
  \eqref{eq:global lower bound with h} holds globally uniformly in in the critical range $0\le U\le U_c$.
  
  \smallskip 
  
  In the subcritical case, where we allow for a wider range of parameters $\kappa_1,\kappa_2$,  the same arguments apply, but now $U$ has to stay 
  away from both $U_c$ and zero. That is, we 
  have to restrict the range of allowed parameters $U$ to $\mu\le U\le U_c-\mu$
  for arbitrary but fixed small $\mu$, due to the additional 
  lower bound on $U$ in Proposition 
  \ref{prop:subsolution upper region} and the upper bound on $U$ from 
  Theorem \ref{thm:lower bound tricky region}. 
  \end{proof}
  
Now we come to the proof of the global lower bound.
\begin{theorem}[Sharp lower bound, arbitrary coupling]\label{thm:sharp lower bound most general}
  For any choice of parameters 
  $K_1,K_2>0$, $1/6 <\kappa_1<1/2$, and 
	$1/2<\kappa_2<1$ 
  there exist a constant 
  $C_-$ depending only on $\kappa_1,\kappa_2, K_1, $ and $K_2$, 
  such that for the unique positive choice of the ground state of the helium-type operator $H_U$ the pointwise bound 
  \begin{align}\label{eq:sharp lower bound critical range most general}
    \psi_U(x)\geq C_- \exp\left( -F^U_-(|x|_\infty,|x|_0) \right)
  \end{align}
  holds uniformly in $\mu \le U\le U_c$. 
  
  For the subcritical case let $0<\kappa_1<1$, 
  $1/2<\kappa_2<1$, $K_1,K_2>0$, and $0<\mu <U_c/2$ be arbitrary.  Then there 
  exist a constant  $\wti{C}_-$, depending only on 
  $\kappa_1,\kappa_2, K_1, K_2$, 
  and also on $\mu$, such that the lower bound  
    \begin{align}\label{eq:sharp lower bound subcritical range most general}
    \psi_U(x)\ge  \wti{C}_- \exp\left( -F^U_-(|x|_\infty,|x|_0) \right)
  \end{align}
  holds for all  $\mu \le U\le U_c -\mu$. 
\end{theorem}
\begin{remark}
  Note that Theorem \ref{thm:global pointwise anisotropic lower bound} is a special 
case of  Theorem \ref{thm:sharp lower bound most general} 
for $U=U_c$. 
\end{remark}
 
\begin{proof}[Proof of Theorem \ref{thm:sharp lower bound most general}{\rm:}]
  Given $1/6<\kappa_1<1/2<\kappa_2<1$ we choose any  $0<\nu<1$ with $\max(\nu,1-\nu)<\kappa_2$. 
  Theorem \ref{thm:global pointwise anisotropic lower bound with h} gives the global lower bound 
  $\psi_U\ge C \exp(-L_U) =C\exp(-F^U_- +h)\ge C \exp(-F^U_-)$, since $h\ge 0$ with the constant $C$ from Theorem \ref{thm:global pointwise anisotropic lower bound with h}  which does not depend on $U$ in the range 
  $0\le U\le U_c$.  The same argument applies for the 
  subcritical range $\mu\le U\le U_c-\mu$. 
\end{proof}

We still have to give the proof of Proposition \ref{prop:subsolution upper region}. 
This will be done in two steps. We split the region $C_{R,\beta}$ in two subregions 
\begin{align}
	C_{R,\beta}^-&\coloneqq \big\{ |x_1|^\beta< |x_2|<|x_1|,\, |x_1|>R \big\}\, , \label{eq:C_R-}\\
	C_{R,\beta}^+&\coloneqq \big\{ |x_2|^\beta< |x_1|<|x_2|,\, |x_2|>R \big\}\, , \label{eq:C_R+}\\
	\intertext{and the diagonal}
	D_R&\coloneqq \big\{ |x_1|= |x_2|>R \big\}\,  .
\end{align}
While $e^{-L_U}$ with $L_U$ defined in \eqref{eq:def L lower bound} is not $H^2(C_{R,\beta})$, due to the jump of the gradient of $L_U$ 
along the diagonal $D_R$, it is twice differentiable in $C_{R,\beta}^\pm$. 
The next lemma shows that $e^{-L_U}$ is a subsolution on $C_{R,\beta}$ in the quadratic form sense 
a la  Agmon , see \cite{Agm85} and Appendix \ref{app:sub-super-solutions}, 
as soon as it is a classical subsolution locally on $C_{R,\beta}^\pm$, 
that is, $(H_U -E_U) e^{-L_U}\le 0$ on $C_{R,\beta}^+$ and on $C_{R,\beta}^-$   separately. 
\begin{lemma}\label{lem: if classical subsolution then ok} 
  Let $0<\beta<1$ and assume that $g_U\coloneqq e^{-L_U}$ is a classical   subsolution of $H_U$ at energy $E_U$, i.e.,   
  $(H_U-E_U)g_U\le 0$ pointwise, both in $C_{R,\beta}^+$ and $C_{R,\beta}^-$.

  If $0<\nu,\kappa_1<\kappa_2<1$, then
  $g_U$ is a subsolution in the quadratic form sense a la  Agmon on $C_{R,\beta}$ for any large enough $R>0$ independent of 
  $0\le U\le U_c$.  That is 
  \begin{align}\label{eq:subsolution quadratic form}
  	\la \varphi, (H_U-E_U) g_U \ra \le 0
  \end{align}
 for all $0\le \varphi\in \calC^\infty_0(C_{R,\beta})$. 
 
 Moreover, if $0<\nu,\kappa_1,\kappa_2<1$ and $0<\mu<U_c$ 
 then there exists $R>0$ independent of $U$ in the range  
 $\mu\le U\le U_c$ 
 such that \eqref{eq:subsolution quadratic form} again holds. 
\end{lemma}
\begin{proof}
  We will write $g$, $L$, and $F$ for $g_U$, $L_U$, respectively, $F^U_-$ in the remainder of this proof, for simplicity of notation. 
  The contribution of the Coulomb potential is local. So we only have to show that for $R$ large enough,  
  \begin{align}\label{eq:subsolution breakup}
  	\la \nabla \varphi, \nabla g\ra_{L^2(C_{R,\beta})} 
  	  \le
  	     \la \varphi, -\Delta g\ra_{L^2(C_{R,\beta}^+)} + \la \varphi, -\Delta g\ra_{L^2(C_{R,\beta}^-)}     
  \end{align}
  for all $\varphi \in\calC^\infty_0(C_{R,\beta})$, since then for all $\varphi\ge 0$ the 
  pointwise bounds $(H_U-E_U)g\le 0$ in $C_{R,\beta}^\pm$ together with 
  \eqref{eq:subsolution breakup} will imply \eqref{eq:subsolution quadratic form}. 
  The bound \eqref{eq:subsolution breakup} follows from splitting the domain $C_{R,\beta}$ into two pieces along 
  the diagonal $D_R=\{|x_2|=|x_1|>R\}$ and integration by parts, since the boundary term on $D_R$ has the right sign when 
  $R$ is large enough, as we will show in a moment.  Thus we can drop the boundary term to get  
  \eqref{eq:subsolution breakup}. 
  
  Given $x=(x_1,x_2)\in D_R$, the vectors 
  \begin{align}\label{eq:outer normals}
  	n_-(x)\coloneqq \frac{1}{\sqrt{2}}\begin{pmatrix}
  								\frac{-x_1}{|x_1|} \\
  								 \frac{x_2}{|x_2|}
 							\end{pmatrix} 
 							\quad \text{and }
  	n_+(x)\coloneqq \frac{1}{\sqrt{2}}\begin{pmatrix}
  								\frac{x_1}{|x_1|} \\
  								 \frac{-x_2}{|x_2|}
 							\end{pmatrix}  							
  \end{align}
  are the outer normals of $C_{R,\beta}^-$, respectively $C_{R,\beta}^+$, on the diagonal $D_R$. To see this  simply note that 
  level sets $\{N(x)=\lambda\}\cap C_{R,\beta}^-$ with $\lambda<1$ of the function $N(x)= |x_2|/|x_1|$ converge to the diagonal $D_R$ from within $C_{R,\beta}^-$ as $\lambda\nearrow 1$. 
  The gradient of $N$ points into the direction of the largest increase, hence  the vector $n_-(x)$ is a positive, and  
  $n_+(x)$ a negative, multiple  of $\nabla N(x)$ when $x=(x_1,x_2)\in D_R$. This proves  \eqref{eq:outer normals}, 
  because  
  \begin{align*}
  	\nabla N(x) = \begin{pmatrix}
  					\frac{-|x_2|}{|x_1|^2} \frac{x_1}{|x_1|} \\
  					\frac{1}{|x_1|}\frac{x_2}{|x_2|}
 				  \end{pmatrix} 
 				 =
 				  \frac{1}{|x_1|}\begin{pmatrix}
  					\frac{-x_1}{|x_1|} \\
  					\frac{x_2}{|x_2|}
 				  \end{pmatrix} 
  \end{align*}
  when $|x_1|=|x_2|$. 
  Since 
  \begin{align}\label{eq:splitting the domain along the diagonal}
  	\la \nabla \varphi, \nabla g\ra_{L^2(C_{R,\beta})} 
  		= \la \nabla \varphi, \nabla g\ra_{L^2(C_{R,\beta}^-)}+ \la \nabla \varphi, \nabla g\ra_{L^2(C_{R,\beta}^+)}\, , 
  \end{align}
  Gau\ss{}' formula, i.e., integration by parts, gives 
  \begin{align}\label{eq:Gauss 1}
  	\la \nabla \varphi, \nabla g\ra_{L^2(C_{R,\beta}^-)}
  	= \la \varphi, -\Delta g\ra_{L^2(C_{R,\beta}^-)} +\int_{D_R}\varphi \nabla g \cdot n_-\, dS
  \end{align} 
  where $S$ is $5$-dimensional Hausdorff (surface) measure on $D_R$. 
  There are no contributions from the other parts of the boundary of $C_{R,\beta}$, since $\varphi$ vanishes there. 
  Similarly, 
  \begin{align}\label{eq:Gauss 2}
  	\la \nabla \varphi, \nabla g\ra_{L^2(C_{R,\beta}^+)}
  	= \la \varphi, -\Delta g\ra_{L^2(C_{R,\beta}^+)} +\int_{D_R}\varphi \nabla g \cdot n_+\, dS \, .
  \end{align} 
  Thus \eqref{eq:subsolution breakup} will follow once the boundary integrals 
  in \eqref{eq:Gauss 1} and \eqref{eq:Gauss 2} are non-positive for all $\varphi\ge 0$ 
  and  large enough $R$. By symmetry, it is enough to show this for the  
  boundary integral in \eqref{eq:Gauss 1}. 
  
  Clearly  $\nabla g = -e^{-L}\nabla L$. Using $\varphi\ge 0$ and $e^{-L}\ge 0$, one sees that 
  the second term in \eqref{eq:Gauss 1} is non-positive as soon as    
  \begin{align}
  	\nabla L \cdot n_- \ge 0 \quad\text{on } D_R\, .
  \end{align}
  On $C_{R,\beta}^-$ the function $L$ is given by 
  \begin{align}
  	L(x)= F(|x_1|, |x_2|) - h(|x_1-x_2|, |x_2|)
  \end{align}
  with $h(s,r_2)= s \theta(s^{-\nu}s)$. So 
  \begin{align}\label{eq:gradient L}
  	\nabla L = 
  		\begin{pmatrix}
  			\nabla_{x_1}(F -h) \\
  			\nabla_{x_2}(F -h)
  		\end{pmatrix}
  		=
  		\begin{pmatrix}
 	      \partial_1 F \frac{x_1}{|x_1|} - \partial_s h\frac{x_1-x_2}{|x_1-x_2|} \\
 	      \partial_2 F \frac{x_2}{|x_2|} - \partial_2h \frac{x_2}{|x_2|} -\partial_s h\frac{x_2-x_1}{|x_1-x_2|}
 		\end{pmatrix}\, ,
  \end{align}
  hence on $D_R$ we have 
  \begin{equation}
  \begin{split} \label{eq:gradient L-0}
  	  	\sqrt{2}\nabla L\cdot n_- 
  	&= 
  		- \partial_1 F + \partial_s h\frac{x_1\cdot(x_1-x_2)}{|x_1||x_1-x_2|} 
  		+ \partial_2 F - \partial_2h -\partial_s h\frac{x_2\cdot(x_2-x_1)}{|x_2||x_1-x_2|} \\
  	&= 
  	    - \partial_1 F + \partial_2 F   
  		- \partial_2 h +\partial_s h\frac{x_1\cdot(x_1-x_2)-x_2\cdot(x_2-x_1)}{|x_1||x_1-x_2|} 
  \end{split}
  \end{equation}
  using also $|x_1|=|x_2|$ on $D_R$. 
  Since  $x_1\cdot(x_1-x_2)- x_2\cdot(x_2-x_1)= |x_1|^2-|x_2|^2= 0$ on $D_R$ 
  this simplifies to 
  \begin{equation} 
  	 \sqrt{2}\nabla L\cdot n_- 
  	    	=- \partial_1 F + \partial_2 F   
  				- \partial_2 h \, .
  \end{equation}
  Also 
  \begin{equation}\label{eq:partial_2 h}
  	0\le \partial_2 h = \partial_2(s\theta(r_2^{-\nu}s)) 
  		= -\nu r_2^{\nu-1} (r_2^{-\nu}s)^2\theta'(r_2^{-\nu}s) \lesssim  r_2^{\nu-1}
  \end{equation}
  by our choice of $\theta$. Thus, with $r_1=|x_1|=|x_2|=r_2$, we get for all large enough $r_1$    
  \begin{align*}
  	\sqrt{2}\nabla & L\cdot n_- 
  	   \ge  
  	    - \partial_1 F + \partial_2 F -\partial_2 h
  	   \ge -F_U'(r_1) -K_1\kappa_1r_1^{\kappa_1-1} +\frac{1}{2} +K_2\kappa_2 r_2^{\kappa_2-1} - Cr_1^{\nu-1} \\
  	  &= -\left(\veps+\frac{a^2}{r_1}\right)^{1/2} +\frac{1}{2}  -K_1\kappa_1r_1^{\kappa_1-1} +K_2\kappa_2 r_1^{\kappa_2-1} - Cr_1^{\nu-1} 
  	   > 0 
  \end{align*}
  where we used the bound \eqref{eq:important} below,  
  $r_2= r_1$ is large,   and $0<\nu,\kappa_1<\kappa_2<1$ 
  uniformly in $0\le U\le U_c$. 
  If $U$ stays away from zero, we also get  
  $\sqrt{2}\nabla L\cdot n_- \ge 0 $ 
  uniformly in $U$ in the range  $\mu\le U\le U_c$ for 
  large $r_1$, any small 
  $\mu>0$, and $0<\nu,\kappa_1,\kappa_2<1$, using \eqref{eq:important-2}.  
  
  Thus the second term in 
  \eqref{eq:Gauss 1}, hence by symmetry also the second term in \eqref{eq:Gauss 2}, is non-positive  when $R$ is large enough. 
\end{proof}
\begin{remark}\label{rem:ionization energy}
  In the proof above, and also in the proof of Lemma 
  \ref{lem:classical subsolutions} below the inequality 
    \begin{align}\label{eq:important}
  	\sup_{0\le U\le U_c}\Big(\veps_U+ \frac{a_U^2}{r_1}\Big)  \le \frac{1}{4}\, 
  \end{align}
  for all large enough $r_1$ plays an important role. 
  Also the refined bound 
  \begin{align}\label{eq:important-2}
  	\limsup_{r_1\to\infty}\sup_{\mu\le U\le U_c}
  	\Big(\veps_U+ \frac{a_U^2}{r_1}\Big)  < \frac{1}{4}
  \end{align}
  for any small $\mu>0$.  The second bound clearly implies 
  \begin{align}\label{eq:important-3}
  	c_\mu
  	= \liminf_{r_1\to\infty}\inf_{\mu\le U\le U_c}
  	\Big(\frac{1}{2}- \Big(\veps_U+ \frac{a_U^2}{r_1}\Big)^{1/2} \Big)  
  	> 0 
  \end{align}  
  for any $0<\mu<U_c$ which is needed in the proof of Lemma \ref{lem:classical subsolutions} below. 
  
  These bounds can be seen as follows: 
  Since the Coulomb repulsion is positive and the ground state energy of $H_0$ is $-1/2$ (twice the energy of hydrogen), we clearly have $E_U\ge E_0= -1/2$, so 
  $\veps_U = -1/4-E_U\le 1/4$, always. 
  Since $E_U$ is strictly increasing in $0\le U\le U_c$, by the Hellman-Feynman formula,  we also have  
  $\veps_U= -1/4-E_U<-1/4-E_{\mu}=\veps_{\mu}<1/4$ for any $0\le \mu<U\le U_c$.   
  Since $a_U=(U-1)_+^{1/2}$ we get 
  \begin{align*}
    \sup_{\mu\le U\le U_c} (\veps_U +\frac{a_U^2}{r_1})\le \veps_\mu + \frac{U_c-1}{r_1} <\frac{1}{4}
  \end{align*}
  for all large enough $r_1>0$,  which immediately implies 
  \eqref{eq:important-2}. 
  Moreover, for $0\le U\le 1$ we have 
  \begin{align*}
  	 \veps_U+ \frac{a_U^2}{r_1} = \veps_U\le \veps_0= \frac{1}{4}
  \end{align*}
  for all $r_1>0$. Since $U_c\le 2$, we get for $1\le U\le U_c$, 
  \begin{align*}
  	\veps_U+ \frac{a_U^2}{r_1} \le \veps_1+ \frac{1}{r_1}
  		< \frac{1}{4}
  \end{align*}
  for all large enough $r_1$, independently of $1\le U\le U_c$, since $\veps_U\le \veps_1<1/4$ for $U\ge 1$. The last two bounds prove \eqref{eq:important}.    
\end{remark}

\begin{lemma}\label{lem:classical subsolutions}
  Assume $0<\beta,\nu<1$, $0<\kappa_1<\kappa_2< 1$ and $\max(\nu,1-\nu)<\kappa_2$  
  and $L_U$ given by \eqref{eq:def L lower bound}. Then  $g_U= e^{-L_U}$ 
  is a classical subsolution of $H_U$ at energy $E_U$ on $C_{R,\beta}^\pm$  
  for all large $R$ independently of $0\le U\le U_c$.  
  
  Moreover, if $0<\beta,\nu, \kappa_1,\kappa_2< 1$, 
  $\max(\nu,1-\nu)<\kappa_2$, and $0<\mu<U_c$ 
  then there exists $R>0$ independent of $U$ in the range  
  $\mu\le U\le U_c$ such that   $g_U$ 
  is again a classical subsolution of $H_U$ at energy $E_U$ on $C_{R,\beta}^\pm$.
\end{lemma}
\begin{proof} Again, we abbreviate $g=g_U$, $L=L_U$, and $F=F^U_-$. 
  Clearly, $L$ and hence $g= e^{-L}$ is $\calC^2$ on $C_{R,\beta}^\pm$ and 
  the derivatives of $g$ up to order 
  two are in $L^2(C_{R,\beta}^\pm)$ for large enough $R$.  
  Moreover, we only have to show the pointwise bound $(H_U-E_U)g\le 0$ 
  in $C_{R,\beta}^-$, since by symmetry the same bound then also holds on $C_{R,\beta}^+$. 
  
  From   $\nabla g= -e^{-L}\nabla L$, one gets   
  \begin{align}
  	-\Delta g = g\left[ -|\nabla L|^2 +\Delta L \right]\, .
  \end{align}
  On $C_{R,\beta}^-$ we have  $L(x) = F(|x_1|,|x_2|) - h(|x_1-x_2|,|x_2|)$. 
  Using formula \eqref{eq:gradient L} for $\nabla L$ 
  we have  
  \begin{equation}
	\begin{split}
  	  \Delta  L 
  	  	&= \nabla_{x_1} \left( \partial_1 F \frac{x_1}{|x_1|} - \partial_s h\frac{x_1-x_2}{|x_1-x_2|}  \right) \\
  	  	&\phantom{=~}	+ \nabla_{x_2} \left( 
  	  		\partial_2 F \frac{x_2}{|x_2|} - \partial_2h \frac{x_2}{|x_2|} -\partial_s h\frac{x_2-x_1}{|x_1-x_2|} 
  	  	  \right) \\
  	  &= \partial_1^2 F + \partial_1 F \frac{2}{|x_1|}  
  	  		+ \partial_2^2 F + \partial_2 F \frac{2}{|x_2|} -\partial_2^2 h - \partial_2 h \frac{2}{|x_2|} 
  	  		- 2\partial_s^2 h \\
  	  &\phantom{=~~} 
  	  	 -2\partial_2\partial_s h \frac{x_2\cdot(x_2-x_1)}{|x_2||x_1-x_2|} - \partial_s h \frac{4}{|x_1-x_2|} \, .
   \end{split}
  \end{equation}
  The term $\partial_s h \tfrac{4}{|x_1-x_2|}$ will allow us to control the Coulomb repulsion between the electrons. 
  Dropping the positive term $(\partial_s h)^2$ we also get from  \eqref{eq:gradient L}   
  \begin{align*}
   |\nabla L|^2 &= |\nabla_{x_1}L|^2+ |\nabla_{x_2}L|^2 \\
   	&\ge (\partial_1 F)^2 -2\partial_s h\partial_1 F \frac{x_1\cdot(x_1-x_2)}{|x_1||x_1-x_2|} 
   	+ (\partial_2 F -\partial_2 h)^2 -2(\partial_2 F -\partial_2 h)\partial_s h \frac{x_2\cdot(x_2-x_1)}{|x_2||x_1-x_2|} \, 
  \end{align*} 
  hence
  \begin{equation} \label{eq:subharmonic details}
  	  \begin{split}
  	  	\big( H_U-E_U\big) g 
  	  	&= \big( -\Delta -\frac{1}{|x_1|} -\frac{1}{|x_2|} +\frac{U}{|x_1-x_2|} +\frac{1}{4} +\veps \big) g\\ 
  		&\le  
  			g\Big[ 
  				\underbrace{\veps-(\partial_1 F)^2}_{{\rm I}}  
  				+ \underbrace{ \frac{1}{4}-(\partial_2 F -\partial_2 h)^2}_{{\rm II}} \\
  		&\phantom{= ~v\Big[}		
  				+ \underbrace{2\partial_s h
  				  (\partial_1 F \frac{x_1\cdot(x_1-x_2)}{|x_1||x_1-x_2|}
  				    +(\partial_2 F-\partial_2 h)\frac{x_2\cdot(x_2-x_1)}{|x_2||x_1-x_2|} }_{{\rm III}} \\
  		&\phantom{= ~v\Big[} 
  				+\underbrace{\partial_1^2 F +\partial_1 F\frac{2}{|x_1|} -\frac{1}{|x_1|}}_{{\rm IV}} 
  	  		+ \underbrace{\partial_2^2 F + \partial_2 F \frac{2}{|x_2|} -\frac{1}{|x_2|}}_{{\rm V}} \\
  	  	&\phantom{= ~v\Big[}
  	  		-\underbrace{\partial_2^2 h - \partial_2 h \frac{2}{|x_2|} - 2\partial_s^2 h}_{{\rm VI}} 
  	  		 - \underbrace{2\partial_2\partial_s h \frac{x_2\cdot(x_2-x_1)}{|x_2||x_1-x_2|}}_{{\rm VII}} \\
  	  	&\phantom{= ~v\Big[} 
  	  		+ \underbrace{ \frac{U}{|x_1-x_2|} - \partial_s h \frac{4}{|x_1-x_2|} }_{{\rm VIII}}
  	  		\Big] \, .
  	  \end{split}
  \end{equation}
  Abbreviating $r_1=|x_1|, r_2=|x_2|, s=|x_1-x_2|$, we have 
  \begin{align*}
  	\partial_1 F = F_U'(r_1) +K_1\kappa_1 r_1^{\kappa_1-1}, 
  \end{align*}
  and since $F_U'(r_1)=\left(\veps+a^2/r_1 \right)^{1/2}\ge 0$ 
  we have $\veps-(F_U'(r_1))^2\le 0$. Hence   
  \begin{align*}
  	{\rm I}\le 0 \, .
  \end{align*}
  Moreover,   
  \begin{align*}
  	\partial_2 F 
  		&= \frac{1}{2} + K_2\kappa_1 r_2^{\kappa_2-1}\, , \\
  	\partial_2 h 
  		&= -\nu r_2^{\nu-1}(r_2^{-\nu}s)^2 \theta'(r_2^{-\nu}s) 
  		\le \nu r_2^{\nu-1}\, ,
  \end{align*}
  where the last bound holds since $0\le t\mapsto  t^2 |\theta'(t)|\le 1$. 
  Since $\nu<\kappa_2<1$, we get  
  \begin{align*}
  	{\rm II} &= \big(\frac{1}{2} -\partial_2F+\partial_2h\big)\big(\frac{1}{2}+\partial_2F -\partial_2 h\big) \\
  	  &\le -\big(K_2\kappa_2 r_2^{\kappa_2-1} -\nu r_2^{\nu-1}\big)\big( 1+K_2\kappa_2 r_2^{\kappa_2-1} -\nu r_2^{\nu-1} \big)	
  			\lesssim -r_2^{\kappa_2-1}
  \end{align*}
  for $r_2 \gg 1$. 
  
  The bound for the third term ${\rm III}$ is tricky, since neither 
  $\partial_1 F$, $\partial_2F$ nor $\partial_s h$ go to zero as 
  $r_1,r_2\to \infty$. We will do this one last and bound the other terms first. 
 Since  $\partial_1 F 
  		= F_U'(r_1) + K_1\kappa_1 r_1^{\kappa_1-1} $ we get 
  \begin{align*}
  	\partial_1^2 F
  		&= F_U''(r_1) -K_1\kappa_1(1-\kappa_1)r_1^{\kappa_1-2}
  				\le 0\, ,
  \end{align*} 
  since $\kappa_1<1$ and $F_U''\le 0$. Thus   
  \begin{align*}
  	{\rm IV}\le (2F_U'(r_1)-1) r_1^{-1} = \left( 2\big(\veps+\frac{a^2}{r_1}\big)^{1/2}-1\right)r_1^{-1}\le 0
  \end{align*}
  for $r_1\gg 1$ because of \eqref{eq:important}.  
  Similarly, 
  \begin{align*}
  	{\rm V} = K_2\kappa_2(\kappa_2-1)r_2^{\kappa_2-2} +\big(\frac{1}{2} 
  			+K_2\kappa_2 r_2^{\kappa_2-1}\big) \frac{2}{r_2} - r_2^{-1}
  		= K_2\kappa_2(\kappa_2+1) r_2^{\kappa_2-2}\, .
  \end{align*}
  Moreover, with  $t=r_2^{-\nu}s$, we have   
  \begin{align*}
  	|\partial_2^2 h| 
  		&\lesssim r_2^{\nu-2}\big(|t^2|\theta'(t)| + t^3|\theta''(t)| \big) 
  		  \lesssim r_2^{\nu-2}\, ,\\
  	|\partial_s^2 h| 
  		&= r_2^{-\nu}\big| 2\theta'(t) + t\theta''(t) \big|
  		  \lesssim r_2^{-\nu}
  \end{align*}
  since $\theta'(t)$ and $t\theta''(t)$ are bounded for $t\ge 0$. Thus we get   
  \begin{align*}
  	{\rm VI} \lesssim r_2^{\nu-2} + r_2^{-\nu} 
  \end{align*}
  for all $r_2\gg 1$ and since 
  \begin{align*}
  	|\partial_2\partial_s h|
  		=  \nu r_2^{-1}\big| 2 t\theta'(t) +t^2\theta''(t) \big| 
  		\lesssim r_2^{-1}
  \end{align*}
  we have 
  \begin{align*}
  	{\rm VII} \lesssim r_2^{-1}\, .
  \end{align*}
 For the term ${\rm VIII}$ which contains the Coulomb repulsion, we have 
 \begin{align*}
 	{\rm VIII} = \frac{1}{s}\big( U- 4\theta(r_2^{-\nu}s)\big) 
 			- 4r_2^{-\nu} \theta'(r_2^{-\nu}s) 
 		\le  \frac{1}{s}\big( 2- 4\theta(r_2^{-\nu}s)\big) 
 			  + 4r_2^{-\nu} \, .
 \end{align*}
 Recall that $U\le U_c\le 2$. 
 If $r_2^{-\nu}s\le 1$, we have $\theta(r_2^{-\nu}s)\ge 1/2$, hence also  
 $2-4\theta((r_2^{-\nu}s))\le 0$.  
 If $r_2^{-\nu}s\ge 1$ we have   $s^{-1}(2-4\theta(r_2^{-\nu}s))\le 2r_2^{-\nu}$. Altogether, 
 \begin{align*}
 	{\rm VIII}\le 2 r_2^{-\nu} + 4r_2^{-\nu} = 6 r_2^{-\nu} 
 \end{align*}
  for all $r_2>0$. 
  
  Now we come to the term ${\rm III}$, which is the hardest term to bound. Recall that  
  $|\partial_2 h|\lesssim r_2^{\nu-1}$. 
  Moreover,  $\partial_sh=   
    \theta(r_2^\nu s)+r_2^{-\nu}s \theta'(r_2^{-\nu}s)\ge 0$, so 
    $|\partial_sh|\lesssim 1$. Thus   
  \begin{align*}
  	{\rm III} &= 2\partial_sh \left( \partial_1 F \frac{x_1\cdot(x_1-x_2)}{|x_1||x_1-x_2|}
  				    +\partial_2 F\frac{x_2\cdot(x_2-x_1)}{|x_2||x_1-x_2|} \right) -\partial_sh \partial_2 h\frac{x_2(x_2-x_1)}{|x_2||x_1-x_2|} \\
  		&\le 2\partial_sh \left( \partial_1 F \frac{x_1\cdot(x_1-x_2)}{|x_1||x_1-x_2|}
  				    +\partial_2 F\frac{x_2\cdot(x_2-x_1)}{|x_2||x_1-x_2|} \right) 
  				    +C r_2^{\nu-1}\,.
  \end{align*}
  Note also  
  \begin{align*}
  	x_1\cdot(x_1-x_2) &= \frac{1}{2}\left( x_1^2-x_2^2 +|x_1-x_2|^2\right)\\
  	\intertext{and }
  	x_2\cdot(x_2-x_1) &= \frac{1}{2}\left( x_2^2-x_1^2 +|x_1-x_2|^2\right)\, ,
  \end{align*}
  so with $r_1^2=x_1^2 \ge r_2^2= x_2^2$ we have 
  \begin{align*}
  	 \partial_1 F \frac{x_1\cdot(x_1-x_2)}{|x_1||x_1-x_2|}
  				    +\partial_2 F\frac{x_2\cdot(x_2-x_1)}{|x_2||x_1-x_2|} 
  		&\!= \!\!
  			\left( \frac{\partial_1F}{2r_1} - \frac{\partial_2 F}{2r_2}\right)\!
  			\frac{r_1^2-r_2^2}{|x_1-x_2|} 
  			+   			\left( \frac{\partial_1F}{2r_1} + \frac{\partial_2 F}{2r_2}\right)\!|x_1-x_2| \, .
  \end{align*}
  Due to $r_2\le r_1$ and \eqref{eq:important}, we have uniformly in  
  $0\le U\le U_c$ 
  \begin{align*}
  	 \frac{\partial_1F}{r_1} - \frac{\partial_2 F}{r_2}
  	   &= \frac{(\veps+\frac{a^2}{r_1})^{1/2}}{r_1} + K_1\kappa_1 r_1^{\kappa_1-2}
  	   	- \frac{1}{2r_2} -K_2\kappa_2 r_2^{\kappa_2-2}\\
  	   &\le  r_1^{-1}\left[ \Big(\veps+\frac{a^2}{r_1} \Big)^{1/2} -\frac{1}{2}
  	   			 + K_1\kappa_1 r_1^{\kappa_1-1}  - K_2\kappa_2 r_1^{\kappa_2-1} 
  	   		 \right] 
  	   		 	\lesssim - r_1^{\kappa_2-1} <  0
  \end{align*}
  for all large $r_1$ when $0<\kappa_1<\kappa_2<1$. 
  
  When $0<\kappa_1,\kappa_2<1$ and $U$ stays away from zero, we again get  
  $\frac{\partial_1F}{r_1} - \frac{\partial_2 F}{r_2}\le -r_1^{-1}\, c_\mu/2 < 0$ 
  uniformly in $U$ in the range  $\mu\le U\le U_c$ for large $r_1$  and 
  small $\mu>0$  
  using now \eqref{eq:important-3}.  
    
 Thus with $s=|x_1-x_2|$ we see that  
 \begin{align*}
 	  	{\rm III} &\le s\partial_s h \left(  \frac{\partial_1F}{r_1} + \frac{\partial_2 F}{r_2}\right) 
  				    +C r_2^{\nu-1}
 \end{align*}
 for large $r_1$.   Since 
    $\partial_s h = \theta(r_2^{-\nu} s)+r_2^{-\nu}s \theta'(r_2^{-\nu}s)\ge 0$
  we have 
  \begin{align*}
  	s\partial_s h = r^\nu\big( (r_2^{-\nu}s) \theta(r_2^{-\nu} s)+(r_2^{-\nu}s)^2 \theta'(r_2^{-\nu}s)\big)\lesssim r_2^\nu
  \end{align*}
  uniformly in $s\ge 0$. Moreover, $\partial_1 F r_1^{-1}\lesssim r_1^{-1}+r_1^{\kappa_1-2}$ and 
  $\partial_2 F r_2^{-1}\lesssim r_2^{-1}+r_2^{\kappa_2-2}$, so we arrive at 
  \begin{align*}
  	{\rm III} \lesssim r_2^{\nu}\left[ r_1^{-1}+r_1^{\kappa_1-2}+  r_2^{-1}+r_2^{\kappa_2-2} \right] 
  		\lesssim r_2^{\nu-1} + r_2^{\nu+\kappa_1-2} + r_2^{\nu+\kappa_2-2} 
  \end{align*}
  since $\kappa_1, \kappa_2<1$ and  $r_2\le r_1$. 

  Collecting the above bounds in \eqref{eq:subharmonic details} we 
  finally  arrive at 
  \begin{align*}
  	\big( H_U +E_U\big) g
  		& \lesssim 
  			g \big[ 
  				-r_2^{\kappa_2-1} + r_2^{\nu-1} + r_2^{\nu+\kappa_1-2} + r_2^{\nu+\kappa_2-2} + r_2^{\kappa_2-2} 
  				+ r_2^{\nu-2} + r_2^{-\nu} + r_2^{-1}
  			\big] \\
  		&\le g r_2^{\kappa_2-1}\!\!
  					\left[ 
  						-1 \!+ r_2^{\nu-\kappa_2} + r_2^{\nu+\kappa_1-\kappa_2-1}+ r_2^{\nu-1} +r_2^{-1}
  						+ r_2^{\nu-1-\kappa_2} + r_2^{1-\nu-\kappa_2} + r_2^{-\kappa_2}
  					\right] \\
  					&< 0
  \end{align*}
  for all large enough $r_2$, since $0<\nu,\kappa_1,\kappa_2<1$ and  
  $\max(\nu,1-\nu)<\kappa_2<1$. This proves the lemma. 
\end{proof}

\backmatter

\bmhead{Acknowledgements}

It is a pleasure to thank  Ioannis Anapolitanos, Andreas Bitter, and Semjon Wugalter for discussions and comments 
and the anonymous referee for comments and constructive 
criticism, which significantly helped to improve the 
presentation of our results. 
We also thank the  Mathematisches Forschungs\-institut 
Oberwolfach (MFO) for their Research in Pairs programs, where part of this work was done. 

Dirk Hundertmark thanks the Deutsche Forschungsgemeinschaft (DFG,  German Research Foundation) for financial support under -- Project-ID 258734477 -- SFB 1173. 
Michal Jex received postdoctoral funding 
from the European Research Council (ERC) under the 
European Union's Horizon 2020 research and 
innovation programme (grant agreement MDFT No.\ 725528) and received financial support from the 
Ministry of Education, Youth and Sport of the Czech Republic under the Grant No.\ RVO 14000. 
Markus Lange was supported by NSERC of Canada and also acknowledges financial support from the European Research Council (ERC) under the European Union’s Horizon 2020 research and innovation programme (ERC StG MaMBoQ, grant agreement No.\ 802901). 

This version of the article has been accepted for publication, after peer review but is 
not the Version of Record and does not reflect post-acceptance improvements, or any corrections. The Version of Record is 
available online at: https://doi.org/10.1007/s00205-026-02167-7

\section*{Declarations}

\noindent
\textbf{Data availability statement:} All data generated or analyzed during this study are included in this article.

\noindent
\textbf{Conflict of interest:} None of the authors have, to the best of our knowledge, any conflict of interest.

%\bigskip

%\bigskip\noindent
%Springer journals and proceedings: \url{https://www.springer.com/gp/editorial-policies}

%\bigskip\noindent
%Nature Portfolio journals: \url{https://www.nature.com/nature-research/editorial-policies}

%\bigskip\noindent
%\textit{Scientific Reports}: \url{https://www.nature.com/srep/journal-policies/editorial-policies}

%\bigskip\noindent
%BMC journals: \url{https://www.biomedcentral.com/getpublished/editorial-policies}
%\end{flushleft}

\begin{appendices}
%%%%%%%%%%%%%%%%%%%%%%%%%%%%%%%%%%%%%%%%%%%%%%%%%%%%%%%%%%%%%%%%%%%%%
%%%%%%%%%%%%%%%%%%%%%%%%%%%%%%%%%%%%%%%%%%%%%%%%%%%%%%%%%%%%%%%%%%%%%
\section{A quadratic form version of the IMS localization formula}
\label{app:IMS formula}
%%%%%%%%%%%%%%%%%%%%%%%%%%%%%%%%%%%%%%%%%%%%%%%%%%%%%%%%%%%%%%%%%%%%%
%%%%%%%%%%%%%%%%%%%%%%%%%%%%%%%%%%%%%%%%%%%%%%%%%%%%%%%%%%%%%%%%%%%%%
In this section we derive  the well-known IMS localization formula 
under rather weak assumptions on the localization functions. 
Our derivation is motivated by \cite{Griesemer2004} 
who used quadratic form methods to derive an IMS localization formula 
for pseudo--relativistic Schr\"odinger operators under weak assumptions on the localization functions. 
While we believe  that the result for non-relativistic 
Schr\"odinger operators is known, at least to the experts, 
we could not find any reference in the literature for the version we need. 

In the following, we consider the momentum operator 
$P=-i\nabla$ on $\R^d$ and acting in $L^2(\R^d)$ 
for general $d\in \N$. The domain of $P$ is the Sobolev space $\calD(P)=H^1(\R^d)=H^1$. The kinetic energy is given by the 
the quadratic form  $\la P\varphi, P\varphi\ra$, which is a closed quadratic form on $\calD(P)$. In order to be able to even formulate the IMS localization formula, on needs to have weight functions 
$\xi:\R^d\to\R$ such that $\xi^2\varphi, \xi\varphi\in\calD(P)$ for any 
$\varphi\in\calD(P)$. 
 
\begin{lemma}\label{lem:IMS domains ok}
  Assume that $\xi:\R^d\to \R$ is bounded and continuous  and there exist $k\in\N$ and pairwise disjoint 
  open sets $U_j$, $j=1,\ldots, k$ with Lipshitz boundary, such that $\R^d=\cup_{j} \ol{U_j}$ and 
  \begin{align}  
     \xi\big{|}_{U_j} \in W^{1,\infty}(U_j) \quad \text{for all } j=1,\ldots,k 
  \end{align}
  that is, the weak derivative $\nabla\xi$ is bounded on each $U_j$, $j=1,\ldots,k$. 
  Then for all $\varphi\in\calD(P)$ we have 
  $
  	\xi^2\varphi, \xi\varphi \in \calD(P)
  $, 
  that is, as multiplication operators $\xi^2$ and $\xi$ map $\calD(P)$ into itself.
\end{lemma}
\begin{proof}
  First, we show that the weak derivative of $\xi$ exists and is given by the function $\wti{ \nabla\xi}$ defined by $\wti{ \nabla\xi}\coloneqq\nabla\xi$ on $U_j$ and $\wti{ \nabla\xi}(x)\coloneqq 0$ if $x\not\in \cup_{j}U_j $.
  Of course, we can set $\wti{ \nabla\xi}(x)$ equal to an arbitrary fixed vector in $\R^d$ for $x\not\in \cup_{j}U_j $. 
  
  Since the boundary $\partial U_j$ is Lipshitz regular it has zero $d$-dimensional Lebesgue measure. 
  By definition of the distributional derivative, we get 
  \begin{align}
  	\la \nabla\xi,\varphi \ra = -\la \xi,\nabla\varphi\ra 
  	= - \sum_{j=1}^k \int_{U_j} \ol{\xi} \nabla\varphi \, \d x 
  	=  \sum_{j=1}^k\left( 
  			\int_{U_j} \ol{{\nabla\xi}} \varphi\, d x 
  			- \int_{\partial U_j} \ol{\xi} \varphi \nu_j\d H^{d-1} 
  		\right)\, 
  \end{align}
  for any test function $\varphi\in\calC^\infty_0(\R^d)$, 
  where the last equality follows from the  weak form of 
  Gau\ss's theorem, see  \cite[Theorem A6.8]{alt92} or 
  \cite[Theorem 1 in Chapter 5.8]{EvaGar92}, $H^{d-1}$ 
  is $d-1$ dimensional Hausdorff measure, and $\nu_j$ 
  is the outer normal, which is well defined $H^{d-1}$ 
  almost everywhere on 
  $\partial U_j$, see \cite[Section 6.5]{alt92} or 
  \cite[Section 5.1]{EvaGar92}. 
  
  Since in the sum over $j$ each part of the boundaries $\partial U_j$ 
  comes in pairs with \emph{opposite} direction of their 
  corresponding normals and the integrand $\ol{\xi}\varphi$ is the same on each of these pairs, we have 
  \begin{align}
  	\sum_{j=1}^k \int_{\partial U_j} \ol{\xi} \varphi \nu_j\d H^{d-1} 
  	=0\, .
  \end{align}
  Thus 
  \begin{align}\label{eq:weak derivative punshline}
  	- \la \xi,\nabla\varphi \ra 
  	= \sum_{j=1}^k \int_{U_j} \ol{{\nabla\xi}} \varphi\, d x 
  	= \la \wti{\nabla\xi},\varphi \ra
  \end{align}
  for any test function $\varphi\in\calC^\infty_0(\R^d)$, 
  so $\xi\in W^{1,\infty}(\R^d)$.

  Now we check that $\xi:\calD(P)\to\calD(P)$:  Clearly, since $\xi$ is bounded, $\|\xi\varphi\|\le \|\xi\|_\infty\|\varphi\|$.
 Since $\xi$ has weak derivative in $W^{1,\infty}$ we have 
  \begin{align}
  	\nabla(\xi\varphi) = (\nabla\xi)\varphi + \xi\nabla\varphi
  \end{align}
  for all $\varphi\in\calD(P)$. Thus 
  \begin{align*}
  	\|P(\xi\varphi)\| &\le  \|(\nabla\xi)\varphi\| + \|\xi P\varphi\|
  		\le \|\nabla\xi\|_\infty\|\varphi\|+\|\xi\|_\infty\|P\varphi\| \\
  		&\le (\|\nabla\xi\|_\infty+\|\xi\|_\infty)\|\varphi\|_{H^1}
  \end{align*}
 which shows that the multiplication operator $\xi:\calD(P)\to\calD(P)$ 
 is bounded. 
 
 This argument shows also that $\xi^2: \calD(P)\to\calD(P)$ 
 is bounded, since  
 replacing $\xi$ by $\xi^2$ in the argument leading to \eqref{eq:weak derivative punshline} shows $\xi^2\in W^{1,\infty}(\R^d)$ as soon as  $\xi$ is bounded and continuous,  
 with bounded weak derivatives on $U_j$, since on each component $U_j$ we 
 clearly have $\nabla(\xi^2) = 2\xi\nabla\xi \in W^{1,\infty}(U_j)$. Thus, by what we already proved, $\xi^2\in W^{1,\infty}(\R^d)$  and 
 $\xi^2: \calD(P)\to\calD(P)$  is bounded.
\end{proof}
\begin{remark} 
  As the proof shows, the continuity assumption on $\xi$ can be 
  weakened to $\xi$ is continuous in a neighborhood of the boundary 
  set $\cup_j\partial U_j$ and this can also be further weakened 
  using the appropriate notion of traces, see, e.g., 
  \cite[Section A.6]{alt92}. Furthermore, the natural regularity assumption on the boundary of $U_j$ 
  is that $U_j$ has locally finite perimeter, see \cite[Chapter 5]{EvaGar92}, so that a suitable  
  integration--by--parts formula holds.   
\end{remark}
 
Our version of the famous IMS localization formula is 
\begin{lemma}[IMS localization formula, quadratic form version]
\label{lem:IMS}
  Let $\xi:\R^d\to\R$ be bounded and continuous and assume 
  that there exist $k\in\N$ and pairwise disjoint open sets $U_j$, $j=1,\ldots, k$ with a Lipshitz boundary, 
	 such that $\R^d=\cup_{j} \ol{U_j}$ and 
  \begin{align}  
     \xi\big{|}_{U_j} \in W^{1,\infty}(U_j) \quad \text{for all } j=1,\ldots,k\,.
  \end{align}
  Then for all $\varphi\in\calD(P)=H^1(\R^d)$ we have 
  \begin{align}\label{eq:IMS}
  	\re\la  \nabla(\xi^2\varphi), \nabla\varphi\ra 
  		= \la  \nabla(\xi\varphi), \nabla(\xi\varphi)\ra - \la  \varphi, |\nabla\xi|^2 \varphi\ra\, .
  \end{align}
  In particular, for a Schr\"odinger operator $H= P^2 +V$, defined in the sense of quadratic forms with a 
  local potential $V$,  we have 
  \begin{align}\label{eq:IMS2}
  	\re\la  \xi^2\varphi, H\varphi\ra 
  		= \la  \xi\varphi, H \xi\varphi\ra - \la  \varphi, |\nabla\xi|^2 \varphi\ra \, .
  \end{align}

\end{lemma}
\begin{remark}
  A natural choice for $\xi$ is $\xi^2=1=  \sum_j\chi_j^2$ for suitable 
  ``partition of unity" given by cut--off functions $\chi_j$. A second choice, which is useful to derive weighted 
  $L^2$ bounds, is given by an exponentially weighted function $\xi= \chi e^F$ for some (bounded) 
  function $F$. 
  Our version of the IMS localization formula allows us to have rather minimal, weak regularity properties on the involved functions 
  $\chi_j$, $\chi$, and $F$, which is very convenient for the proof of our upper bounds. 
\end{remark}

\begin{proof}
  Lemma \ref{lem:IMS domains ok}  shows that $\xi^2\varphi$ and $\xi\varphi$ 
  are in $H^1(\R^d)$ as soon as $\varphi$ 
  is. So both sides of 	\eqref{eq:IMS} are well defined. 
  Given $\varphi\in H^1$ and using that $\xi$ is real-valued, a short calculation reveals
  \begin{align*}
  	\re \la \nabla(\xi^2\varphi), \nabla\varphi \ra 
  	  &= \re \la \xi\nabla(\xi\varphi)+ (\nabla\xi)\xi\varphi,\nabla\varphi \ra 
  	    = \re \la \nabla(\xi\varphi),\xi\nabla\varphi \ra 
  	    	+ \re \la (\nabla\xi)\varphi,\xi\nabla\varphi \ra \\
  	  &=  \re \la \nabla(\xi\varphi),\nabla(\xi\varphi)-(\nabla\xi)\varphi \ra +  \re \la (\nabla\xi)\varphi,\xi\nabla\varphi \ra \\
  	  &=  \la \nabla(\xi\varphi),\nabla(\xi\varphi)\ra 
  	  		- \re \la (\nabla\xi)\varphi + \xi\nabla\varphi, (\nabla\xi)\varphi \ra
  	  		+ \re \la (\nabla\xi)\varphi,\xi\nabla\varphi \ra 
  	  		\\
   	  &=  \la \nabla(\xi\varphi),\nabla(\xi\varphi)\ra 
  	  		- \la(\nabla\xi)\varphi,(\nabla\xi)\varphi \ra
  \end{align*}
 since also $\re \la (\nabla\xi)\varphi,\xi\nabla\varphi \ra 
 = \re \la \xi\nabla\varphi, (\nabla\xi)\varphi \ra$. Thus \eqref{eq:IMS} holds. 
 
 The formula \eqref{eq:IMS2} follows immediately from 
 \eqref{eq:IMS} since the potential $V$ is a local 
 multiplication operator, hence we have  
 $\la \xi^2\varphi, V\varphi\ra = \la \xi\varphi, V\xi\varphi\ra $ for the quadratic form.  
\end{proof}

%%%%%%%%%%%%%%%%%%%%%%%%%%%%%%%%%%%%%%%%%%%%%%%%%%%%%%%%%%%%%%%%%%%%%
%%%%%%%%%%%%%%%%%%%%%%%%%%%%%%%%%%%%%%%%%%%%%%%%%%%%%%%%%%%%%%%%%%%%%	
\section{Sub-- and supersolution bounds a la Agmon} 
\label{app:sub-super-solutions}
%%%%%%%%%%%%%%%%%%%%%%%%%%%%%%%%%%%%%%%%%%%%%%%%%%%%%%%%%%%%%%%%%%%%%
%%%%%%%%%%%%%%%%%%%%%%%%%%%%%%%%%%%%%%%%%%%%%%%%%%%%%%%%%%%%%%%%%%%%%
Let $H=-\Delta +V$ be a Schr\"odinger operator defined by the quadratic form 
\begin{align}
	\la \varphi, H\psi \ra \coloneqq \la \nabla\varphi, \nabla\psi\ra + \la \varphi, V\psi\ra 
\end{align}
for $\varphi, \psi\in H^1(\R^d)$ and an infinitesimally form small perturbation by a real-valued function $V$. 
Here we define $\la \varphi, V\psi\ra = \la \sgn(V) |V|^{1/2}\varphi, |V|^{1/2}\psi\ra $ where 
$\sgn(t)= t/|t|$ when $t\not = 0$ and $\sgn(0)\coloneqq 0$.  

\begin{definition}[Sub-- and supersolutions in the quadratic form sense a la Agmon]
\label{def:sub-super-solutions}
Let $\Omega$ be an open subset of $\R^d$. 
Then a real-valued function 
$\psi\in H^1_{\text{loc}}(\Omega)$ is a \emph{supersolution} of $H$ at energy $E\in\R$ in $\Omega$ if 
\begin{align}\label{eq:Agmon supersolution}
  \la \varphi, (H-E)\psi\ra \ge  0 
\end{align}
and $\psi$ is a \emph{subsolution} if 
\begin{align}\label{eq:Agmon subsolution}
  \la \varphi, (H-E)\psi\ra \le 0 
\end{align}
for all non-negative $\varphi\in \calC^\infty_0(\Omega)$. 
\end{definition}

For our slight extension of a beautiful result of Agmon \cite{Agm85} we need one more notation.
\begin{definition}\label{def:boundary layer}
  Let $\Omega\subset \R^d$ be an open set with boundary $\partial \Omega$. We call a closed set 
  $\wti{\partial \Omega}$ an 
  (inner) boundary layer of $\Omega$ if it is a subset of the closure of $\Omega$ which contains the boundary     
  $\partial \Omega$ of $\Omega$ in such a way that $\Omega\setminus \wti{\partial \Omega}$ has locally positive 
  distance from the boundary $\partial \Omega$.  More precisely, for any compact set $K\subset \R^d$ we have 
  \begin{align}
  	\dist( \partial \Omega, (K\cap \Omega)\setminus \wti{\partial \Omega}) >0 \, .
  \end{align}
\end{definition}
In other words, a boundary layer $\wti{\partial \Omega}$ provides locally a safety  distance to the boundary:  
For any $R>0$ there exist $\delta>0$ such that 
\begin{align*}
	\dist(x,\partial \Omega) \ge \delta 
\end{align*}
 \emph{uniformly} in  $x\in \Omega\setminus \wti{\partial\Omega}$ with $|x|\le R$. 

\begin{theorem}\label{thm:Agmon subsolution bound}
  Let $\Omega$ be an open subset of $\R^d$, $f$ a positive supersolution  and 
  $g$ a subsolution of $H$ at energy $E$ in $\Omega$. Assume that for some $\lambda>1$ 
  \begin{align}\label{eq:subsolution bound assumption 1}
  	\liminf_{L\to\infty} L^{-2}\int_{\Omega\cap\{ L\le |x|\le \lambda L \}} g_+^2 \, dx =0\, 
  \end{align}
  where $g_+=\sup(g,0)$ is the positive part of $g$,  and 
  \begin{align}
  	f\ge g \quad \text{ for {\rm(}almost{\rm)} all } x\in \wti{\partial \Omega}  
  \end{align}
  where $\wti{\partial \Omega} $ is an  {\rm(}inner{\rm)} boundary layer of $\Omega$.  
  Then $ f\ge g$ almost everywhere in $\Omega$. 
\end{theorem}
\begin{remark}
  We would like to stress that this is basically Theorem 2.7 in \cite{Agm85}. It is a 
  slight extension of Agmon's result, since he considers neighborhoods of infinity, i.e., complements of 
  compact subsets of $\R^d$. We need to work with more general unbounded domains which contain infinity but are not necessarily neighborhoods of infinity. 
\end{remark}
\begin{proof}
  We give a sketch of the proof, for the convenience of the reader. 
  Since $f$ is a positive supersolution and $g$ is a subsolution in $\Omega$, the function 
  \begin{align}
  	u\coloneqq (g-f)_+
  \end{align}
  is a subsolution of $H$ at energy $E$ in $\Omega$, see \cite[2.9 Lemma]{Agm85} whose proof also shows that one only needs 
  $V\in L^1_{\text{loc}}(\R^d)$ for this. So for any real-valued $\xi\in \calC^\infty_0(\Omega)$ we can choose 
  $\varphi= \xi^2u$ as a testfunction in the IMS localization formula  \eqref{eq:IMS}  to get 
  \begin{align*}
  	\la \xi u, (H-E)\xi u\ra -\la u, |\nabla\xi|^2 u\ra= \re \la \xi^2 u, (H-E)u\ra =\la \xi^2 u, (H-E)u\ra \le 0\, . 
  \end{align*}
  That is, we have 
  \begin{align}\label{eq:subsolution 1}
  	\la \xi u, (H-E)\xi u\ra \le \la u, |\nabla\xi|^2 u\ra  \quad \text{for all } \xi\in\calC^\infty_0(\Omega)\, .
  \end{align} 
  Extend $u$ by zero to $\Omega^c$, by a slight abuse of notation, we continue to use $u$ for this extension. 
  Since $u$ vanishes on the boundary layer $\wti{\partial\Omega}$, i.e, within a safety--distance to the 
  boundary  $\partial\Omega$, this extension of $u$ is in the Sobolev space $H^1(\R^d)$. Moreover, thanks to 
  $u$ being zero on the boundary layer $\wti{\partial\Omega}$, it is easy to see that 
  \eqref{eq:subsolution 1} continues to  hold for all real-valued $\xi\in \calC^\infty_0(\R^d)$, 
  since the both sides of the inequality \eqref{eq:subsolution 1}  depend only on the values of $\xi$ on 
  the support of $u$. So 
  \begin{align}\label{eq:subsolution 2} 
  	\la \xi u, (H-E)\xi u\ra \le \la u, |\nabla \xi|^2 u\ra \quad \text{for all } \xi\in\calC_0^\infty(\R^d)\, .
  \end{align}
  By assumption, $f>0$ is a supersolution of $H$ at energy $E$ in $\Omega$.  Thus for any real-valued 
  $\xi\in\calC^\infty_0(\R^d)$ we can choose $\varphi= \rho^2 f$, with $\rho=\xi u/f$,  as a positive test 
  function with support in  
  $\Omega$ to get, from the definition of a supersolution,   
  \begin{align}
  	0\le \la \varphi, (H-E) f \ra = \la \rho^2 f, (H-E)f\ra= \la \xi u, (H-E)\xi u \ra - \la f, \big| \nabla\Big( \frac{\xi u}{f} \Big) \big|^2 f\ra\, 
  \end{align}
  where we again used the IMS localization formula. 
  Together with \eqref{eq:subsolution 2} this yields 
  \begin{align}\label{eq:subsolution 3}
  	\la f, \big| \nabla\Big( \frac{\xi u}{f} \Big) \big|^2 f\ra
	  \le \la u, |\nabla\xi|^2 u\ra \quad \text{for all } \xi\in\calC_0^\infty(\R^d)\, .
  \end{align}
  Now pick $\chi\in \calC^\infty_0(\R^d)$ with $0\le \chi\le 1$, $\chi(x)=1$ for $|x|\le 1$, $\chi(x)=0$ for $|x|\ge 2$, 
  and use the scaled version $\xi(x)=\xi_L(x)=\chi(x/L)$ in \eqref{eq:subsolution 3}. Since $\xi_L\to 1$ pointwise as $L\to\infty$, Fatou's lemma and  \eqref{eq:subsolution 3} imply 
  \begin{equation}
  	\begin{split}\label{eq:subsolution 4}
  	\la f, \big| \nabla\Big( \frac{u}{f} \Big) \Big|^2 f\ra
	  &\le \liminf_{L\to\infty} \la f, \big| \nabla\Big( \frac{\xi_L u}{f} \Big) \big|^2 f\ra 
	     \le \liminf_{L\to\infty}\la u, |\nabla\xi_L|^2 u\ra \\
	  &\le  \liminf_{L\to\infty}\int_\Omega |\nabla\xi_L|^2 g_+^2\, dx =0 
	\end{split}
  \end{equation}
 where in the second inequality we used  $u = (g-f)_+\le  g_+$, which follows from $f> 0$ in $\Omega$. 
 The last equality in \eqref{eq:subsolution 4} is due to the bound $\nabla\xi_L\le \|\nabla\chi\|_\infty^2 L^{-2} \id_{\{L\le |x|\le \lambda L\}}$ and assumption \eqref{eq:subsolution bound assumption 1}. 
 
 Clearly, \eqref{eq:subsolution 4} implies that $u=cf$ in $\Omega$ for some constant $c$. 
 Since $f>0$ in $\Omega$ and  $u=0$ on the boundary layer $\wti{\partial\Omega}$, we must have $c=0$.  
 Thus $u=(g-f)_+=0$ almost everywhere in $\Omega$, which proves the lemma.
\end{proof}

%%%%%%%%%%%%%%%%%%%%%%%%%%%%%%%%%%%%%%%%%%%%%%%%%%%%%%%%%%%%%%%%%%%%%
%%%%%%%%%%%%%%%%%%%%%%%%%%%%%%%%%%%%%%%%%%%%%%%%%%%%%%%%%%%%%%%%%%%%%
\section{A tightness argument}
\label{app:tight}
%%%%%%%%%%%%%%%%%%%%%%%%%%%%%%%%%%%%%%%%%%%%%%%%%%%%%%%%%%%%%%%%%%%%%
%%%%%%%%%%%%%%%%%%%%%%%%%%%%%%%%%%%%%%%%%%%%%%%%%%%%%%%%%%%%%%%%%%%%%
In this section we show how one can prove a simple tightness argument for weakly convergent sequences in 
$L^2(\R^d)$ from  \cite{HunLee12}. Of course, such results are well-known using, for example, the locally 
compact embedding of the Sobolev space $H^1$ into suitable $L^p$ spaces. Our point is that one can work with 
a pure $L^2$--space version. The tightness argument we use is based on the following equivalence.  

\begin{theorem}\textup{(\cite{HunLee12})}\label{thm:tightness}
\label{thm:tight}
Let $(\psi_n)_{n\in\mathbb N}$ be a sequence in $L^2(\mathbb R^d)$ and 
$(\hatt{\psi}_n)_{n\in\N}$ the sequence of Fourier transforms. Then the following are equivalent. 
\begin{smallenum}
\item\label{tightness 1} The sequence $(\psi_n)_{n\in\mathbb N}$ is converging strongly,
\item\label{tightness 2} The sequence $(\psi_n)_{n\in\mathbb N}$ is converging weakly and satisfies
\begin{align}
\lim_{R\rightarrow\infty}\limsup_{n\rightarrow\infty}\int_{|x|>R}|\psi_n(x)|^2\textrm{d}x&=0\,,\label{eqtightA1}\\
\lim_{L\rightarrow\infty}\limsup_{n\rightarrow\infty}\int_{|k|>L}|\hat\psi_n(k)|^2\textrm{d}k&=0\label{eqtightA2}\, ,
\end{align}
\item\label{tightness 3} The sequence $(\psi_n)_{n\in\mathbb N}$ is converging weakly and there exist functions $H,F\ge 0$ with 
  $\lim_{|x|\rightarrow\infty}H(x)=\infty=\lim_{|k|\rightarrow\infty}F(k)$ such that 
\begin{align}
\limsup_{n\rightarrow\infty}\int_{\mathbb R^d}H(x)|\psi_n(x)|^2\textrm{d}x&<\infty\,,\label{eqtightB1}\\
\limsup_{n\rightarrow\infty}\int_{\mathbb R^d}F(k)|\hat\psi_n(k)|^2\textrm{d}k&<\infty\,.\label{eqtightB2}
\end{align}
\end{smallenum}
\end{theorem}
\begin{remark}
	It is straightforward to see that conditions \eqref{eqtightA1} and 
	\eqref{eqtightB1}, respectively, conditions \eqref{eqtightA2} and \eqref{eqtightB2}, are  equivalent.   
\end{remark}

\begin{proof}
  For the convenience of the reader, we give a sketch of the proof. The equivalence of part \ref{tightness 1} and 
  \ref{tightness 2} of  Theorem \ref{thm:tightness} was already shown in \cite{HunLee12}. 
  The easy argument is as follows: 
  It is clear that part \ref{tightness 1} implies part \ref{tightness 2}. For the converse assume that $\psi_n$ 
  converges weakly to zero in $L^2$, without loss of generality. Let $P_L=\ind_{\{|P|\le L\}}$  be a Fourier 
  cutoff onto momenta $|\eta|\le L$ in Fourier space and $Q_R=\ind_{\{|x|\le R\}}$ be a cutoff in $x$-space. 
  Then it is easy to see that $P_L Q_R$ has a square integrable kernel, i.e., it is a Hilbert--Schmidt operator on $L^2$, hence compact, \cite{Sim05}. 
  Since 
  \begin{align*}
  	\psi_n = P_L\psi_n + (1-P_L)\psi_n = P_L Q_R\psi_n + P_L(1-Q_R)\psi_n + (1-P_L)\psi_n  
  \end{align*}
  and $P_L$ is an orthogonal projection  we get 
  \begin{align}\label{eq:tightness intermediate}
  	\limsup_{n\to\infty}\|\psi_n\| \le \limsup_{n\to\infty} \|(1-Q_R)\psi_n\| + \limsup_{n\to\infty}\|(1-P_L)\psi_n\| 
  \end{align}
  using that $\lim_{n\to\infty} \|P_L Q_R \psi_n\|=0$, by compactness of $P_L Q_R$ for all fixed $L,R$. 
  Now simply note that the two terms 
  on the  right hand side of \eqref{eq:tightness intermediate} go to zero as $R,L\to\infty$ due to  
  \eqref{eqtightA1}, respectively \eqref{eqtightA2}.  
  
  Clearly part \ref{tightness 3} implies part \ref{tightness 2} using the  Chebycheff--Markov inequality. 
  For the converse assume that \eqref{eqtightB1} holds and take an increasing  sequence of radii $R_k\to\infty$ with 
  \begin{align}
  	\limsup_{n\rightarrow\infty}\int_{|x|>R_k}|\psi_n(x)|^2\textrm{d}x\le 2^{-2k} \
  \end{align}
  for all $k\in\N$. Setting 
  \begin{equation}
  	H(x)\:= \sum_{k: R_k< |x|}^\infty 2^k 
  \end{equation}
  we get 
  \begin{align*}
  	\limsup_{n\to\infty} \int H(x) |\psi_n(x)|^2\, dx 
  		\le  
  			\sum_{k=0}^\infty 2^k \limsup_{n\to\infty} \int_{|x|>R_k} |\psi_n(x)|^2\, dx 
  		\le \sum_{k=0}^\infty 2^{-k} =1. 
  \end{align*}
  i.e., \eqref{eqtightB1} holds. Similarly one shows that \eqref{eqtightA2} implies \eqref{eqtightB2}. 
\end{proof}

%%%%%%%%%%%%%%%%%%%%%%%%%%%%%%%%%%%%%%%%%%%%%%%%%%%%%%%%%%%%%%%%%%%%%
%%%%%%%%%%%%%%%%%%%%%%%%%%%%%%%%%%%%%%%%%%%%%%%%%%%%%%%%%%%%%%%%%%%%%
\section{Basic properties of the ground state}\label{app:basic properties}
%%%%%%%%%%%%%%%%%%%%%%%%%%%%%%%%%%%%%%%%%%%%%%%%%%%%%%%%%%%%%%%%%%%%%
%%%%%%%%%%%%%%%%%%%%%%%%%%%%%%%%%%%%%%%%%%%%%%%%%%%%%%%%%%%%%%%%%%%%%
In this appendix we gather some basic properties of the ground state 
and the ground state energy. 
We call $E_U=\inf\sigma(H_U)$ the ground state energy, although, technically, it is only the ground state energy when a ground state exists. 

A result of Lieb \cite{Lie84}, based on  a clever idea of Benguria, but now applied for the case 
that the electron--electron repulsion term contains a coupling 
constant $U$, shows that a single atom with nuclear charge $Z$ can only bind 
\begin{align}\label{eq:modified Benguria Lieb bound}
	N< \frac{2Z}{U} +1	
\end{align}
particles, independent of the statistics of the particles.  
That is, if the ground state energy is strictly below the essential spectrum, then  \eqref{eq:modified Benguria Lieb bound} holds. For $N=2$ and $Z=1$ this shows that $E_U<-1/4= \sigmaess(H_U)$ implies $ U<2$. Thus, for the operator given by \eqref{eq:HeliumHamiltonian}, the critical coupling $U_c$ is bounded by   
\begin{align}\label{eq:a priori bound on critical coupling}
	1<U_c\le 2 \, .
\end{align}
These simple rigorous bounds give are rough, but not bad, estimate 
for the critical coupling $U_c$. As discussed in the introduction, calculations by quantum chemists predict 
$U_c\sim 1.097 66$, see \cite{BakFreDavHil}, and even pushed the calculation to $U_c\simeq 1,097\, 660\, 833\, 738\, 56$ in \cite{BusDraEstMoi14}.   
Bethe showed in \cite{Bet29} with the help of a variational calculation that hydrogen can bind two electrons, hence clearly $U_c>1$.  
Moreover, Hill showed that even for finite 
nuclear mass $M$ the  $H^-$ quantum system has a ground state in the center--of--mass frame for some $U>1$ as long 
as the mass ration $m/M\le  0.2101$. 

\begin{proposition}\label{prop:basic properties}
  Given a helium-type operator $H_U$, with infinite nuclear mass let $E_U=\inf \sigma(H_U)$ be its ground state energy. Then
 \begin{smallenum}
 	\item \label{claim 1} $U\mapsto E_U$ is concave and continuous. 
 			It is strictly increasing on $[0,U_c]$, constant on 
 			$[U_c,\infty)$, and  
 	  \begin{align}
 		E_U <-\frac{1}{4} \quad \text{for }0\le U<U_c,\quad  E_{U_c}=-\frac{1}{4}  \, .
      \end{align}
 	  Moreover, $E_U$ is differentiable in $U<U_c$ with a positive derivative. At $U_c$ its left derivative exists and is positive.  
	\item\label{claim 2}  At critical coupling the energy $E_{U_c}=-1/4$ is 
	a simple eigenvalue at the edge of the essential spectrum. 
	The corresponding  ground state 
		$\psi_c=\psi_{U_c}\in L^2$ is unique and positive 
		up to a global phase factor.
	\item \label{claim 3} Choosing all ground states $\psi_U$ for $0\le U\le U_c$  normalized, i.e., $\|\psi_U\|_2=1$, and positive,   
	the map $[0,U_c]\ni U\mapsto \psi_U$ is continuous in $H^1$ and in  $L^q$ for all $2\le q\le \infty$.       
	Moreover, 
	there exist a constant $C<\infty$ 
		   such that for any $2\le q\le \infty$ we have 
     		\begin{align}\label{eq:basic-properties uniform Lp bound}
     			\sup_{0\le U\le U_c}\|\psi_U\|_q \le C \, .
     		\end{align}
     \item\label{claim 4} For any normalized positive ground state $\psi_U$ with 
     $\|\psi_U\|_2=1$ and any compact set $K\subset \R^6$ there exists a 
     constant $C_K>0$ such that 
    	  \begin{align}
    	  	\inf_{0\le U\le U_c} \inf_{x\in K} \psi_U(x)\ge C_K\, .
    	  \end{align}
 \end{smallenum}
\end{proposition}
\begin{proof}
  Let $H_0$ be the operator $H_U$ with $U=0$. Since 
  $H_U= H_{0}+ \frac{U}{|x_1-x_2|}$ is linear in the 
  perturbation, one sees that 
  $E_U= \inf_{\|\psi\|=1} \la\psi, H_U\psi \ra $ is an infimum of linear functions, hence concave. Since $-\infty<E_U<\infty$ for all $U\in\R$, it is continuous. 
  
  At critical coupling and above, the energy $E_{U}$ is equal to 
  the infimum of the essential spectrum, which by the HVZ theorem is 
  given by the ground state energy of hydrogen. Hence   $E_{U}= -1/4$ 
  for all $U\ge U_c$. 
  
  When $U<U_c$, perturbation theory \cite{ReeSim4} applies. First 
  order perturbation theory, the so-called Feynman--Hellmann formula, see Remark (3.5.18) in \cite{Thi02}, shows 
  \begin{align}
  	\partial_U E_U 
  		= \la \psi_U, \frac{dH_U}{dU}\psi_U \ra 
  		=  \la \psi_U, \frac{1}{|x_1-x_2|}\psi_U \ra >0 
  \end{align}
  for all $U<U_c$. Moreover, this formula also holds, as soon as a normalizable ground $\psi_U$ state 
  exists, for critical coupling $U=U_c$, when one replaces the 
  derivative $ \partial_U E_U $ with the left derivative 
  $\partial_U^-E_U= \lim_{U\nearrow U_c} \frac{E_U-E_{U_c}}{U-U_c}$, 
  see \cite{Sim77}.   This proves claim \ref{prop:basic properties}.\ref{claim 1}, once a normalizable 
  ground state exists at critical coupling. This is known from 
  \cite{HofOstHofOstSim83, FraLieSei12} but we will give an alternative proof for this, which uses the isotropic upper bound from Theorem \ref{thm:isotropic upper bound}. 
  
  \smallskip 
  The Coulomb potential $V_U$ is in the so--called Kato--class, for a definition see \cite{AizSim82, CycFroKirSim87, Simon-semigroups}, 
  for any coupling $U\in\R$. Moreover, $V_U\ge V_0$ for all $U\ge 0$, so 
  \cite[Theorem B.1.1]{Simon-semigroups} and monotonicity of the Schr\"odinger semigroup in the potential gives 
  \begin{align*}
  	\|\psi_U\|_p = \|\e^{tE_U} e^{-tH_U}\psi_U\|_p
  		\le e^{tE_U}\|e^{-tH_U}\|_{2\to p }\|\psi_U\| 
  		\le e^{tE_U}\|e^{-tH_0}\|_{2\to p }\|\psi_U\| \, . 
  \end{align*}
  So for fixed $t>0$ and with $C_p= e^{tE_U}\|e^{-tH_0}\|_{2\to p } $ 
  we have $ \|\psi_U\|_p\le C_p\|\psi_U\|$ for all $2\le p\le \infty$. 
  The Riesz--Thorin interpolation theorem also shows 
  $C_p\le C_2^{\Theta} C_\infty^{1-\Theta}$ for any $1/p = \Theta/2 + (1-\Theta)/\infty= \Theta/2$ and $0\le \Theta\le 1$. 
  Hence $C_p\le \max(C_2,C_\infty)\eqqcolon C $ and we arrive at 
  \begin{align}\label{eq:supremum-1}
  	\sup_{0\le U<U_c}\|\psi_U\|_p\le C
  \end{align}
  when $\|\psi_U\|=1$ for $0\le U <U_c$.

  Take a sequence $1<U_n<U_c$ which converges to $U_c$ and consider the corresponding 
  sequence of normalized bound states $\psi_n$ of $H_n=H_{U_n}$ with energies $E_n<-1/4$.    
  The Coulomb potential $V_U$ is relatively form bounded 
  with respect to the Laplacian $-\Delta$ in $\R^6$ with relative form bound zero, thus 
  \begin{align} \label{eq:form tiny 1}
	|\la \psi, V_U\psi\ra |\le \delta \|\nabla\psi\|^2 + b_\delta\|\psi\|^2
  \end{align}
  and, in particular,  
  \begin{align} \label{eq:form tiny 2}
	|\la \psi, \frac{1}{|x_1-x_2|}\psi\ra |\le \delta \|\nabla\psi\|^2 + b_\delta\|\psi\|^2
  \end{align}  
  for all $0<\delta<1$ and some $b_\delta>0$ uniformly in $0\le U\le U_c$ for all $\psi\in H^1(\R^6)$. 
  So for $\psi_n$ we get
  \begin{align*}
  	E_n= 
  			\la \psi_n, H_n \psi_n\ra = \la \nabla\psi_n,\nabla\psi_n\ra + \la \psi_n, V_U\psi_n \ra 
  	   \ge 
  	   		(1-\delta)\|\nabla\psi_n\|^2 -b_\delta\|\psi_n\|^2
  \end{align*}
 which implies 
 \begin{align*}
 	\|\nabla\psi_n\|^2 \le \frac{b_\delta}{1-\delta} \|\psi_n\|^2
 \end{align*}
  since the energy $E_n$ of the minimizing sequence is bounded. 
  So 
  \begin{align*}
  	\|\psi_n\|_{H^1(\R^6)}^2 
  	  = \|\psi_n\|^2 + \|\nabla\psi_n\|^2 
  	  \le \frac{b_\delta}{1-\delta} \|\psi_n\|^2 +\|\psi_n\|^2 
  	  =  \frac{b_\delta}{1-\delta}+1<\infty 
  \end{align*}
 uniformly in $n\in\N$, since $\psi_n$ is normalized.  Thus the sequence $\psi_n$ is bounded in $H^1$, hence 
 there exists a subsequence, which we also denote by $\psi_n$, 
  which converges weakly to some $\psi_c\in H^1(\R^6)$. We claim that $\psi_c$ is a normalized ground state of 
  $H_{U_c}$, i.e., $\|\psi_c\|=1$, $\psi_c\in H^1(\R^6)$, 
  and $\la \varphi, H_{U_c}\psi_c\ra = -1/4\la \varphi,\psi_c\ra$ for all $\varphi\in H^1(\R^6)$. 
 
 Since $\psi_n$ is a bounded sequence in $H^1(\R^6)$ it is  tight in momentum space,   
 \begin{align}\label{eq:tightness momentum space}
 	\lim_{L\to\infty}\sup_{n\in\N}\int_{|\eta|\ge L} |\hatt{\psi}_n|^2\d \eta =0\, .
 \end{align}
  Moreover, the upper bound from Theorem \ref{thm:isotropic upper bound}, 
  which only needs the bound \eqref{eq:supremum-1} for $\psi_n=\psi_{U_n}$ 
  with subcritical $U_n<U_c$,  easily implies also tightness of 
  the sequence $\psi_n$ in position space,
  \begin{align}\label{eq:tightness position space}
  	 	\lim_{R\to\infty}\sup_{n\in\N}\int_{|\eta|\ge L} |\psi_n|^2\d \eta =0\,  
  \end{align}  
  and standard compactness results, for example Theorem \ref{thm:tightness}, show that $\psi_n$ converges to 
  $\psi_c$ strongly in $L^2(\R^6)$, so $\|\psi_c\|=\lim_{n\to\infty}\|\psi_n\|=1$. 
  
  We will show in a moment that $\psi_n$ also converges strongly in $H^1(\R^6)$, the form domain of all operators $H_U$. Once this is the case a bound of 
  the form \eqref{eq:form tiny 1} for some $\delta>0$ implies  
  $\la \psi_c, V_{U_c}\psi_c\ra =\lim_{n\to\infty} \la \psi_n V_{U_n}\psi_n\ra$. Once we have this, we also get 
  \begin{align}
  	-\frac{1}{4}
  	  \le \la \psi_c, H_{U_c}\psi_c \ra 
  	  = \liminf_{n\to\infty}  \la \psi_n, H_{U_n}\psi_n \ra 
  	  = -\frac{1}{4} \, .
  \end{align}
  because of the strong convergence of $\psi_n$ in $H^1(\R^6)$ and 
  $\la \psi_n, H_{U_n}\psi_n \ra = E_n\to -1/4$ in the limit $n\to\infty$, 
  where we also used that  $\psi_c$ is normalized and 
  $H_{U_c}\ge -\tfrac{1}{4}$, as quadratic forms. 
  Thus $\psi_c$ is a normalized minimizer of the energy. 
  By the variational principle,  
  $\psi_c$ is a weak eigenfunction of $H_{U_c}$: $\la \varphi, H_{U_c}\psi_c\ra = -\tfrac{1}{4}\la \varphi,\psi_c\ra$ 
  for all $\varphi\in H^1(\R^6)$. The uniqueness, up to a global phase, follows from the usual arguments, 
  \cite[Theorem XII44]{ReeSim4}, see also \cite{Goe77}, since the semigroup $e^{-tH_U}$ is positivity improving, i.e. $e^{-tH_U}f>0$ if $f\ge 0$ and $f\not=0$. 
  This proves claim \ref{prop:basic properties}.\ref{claim 2}, once we established the strong convergence 
  of $\psi_n$ in $H^1(\R^6)$. 
  
  To prove the strong convergence in $H^1(\R^6)$ we note that  
  \begin{equation*}
      H_n = H_{U_c} + \frac{U_n-U_c}{|x_1-x_2|} 
      \eqqcolon H_{U_c} +W_{U_n}
  \end{equation*}
  so for all $\varphi\in H^1(\R^d)$ we have 
  \begin{equation*}
      \begin{split}
          \la \varphi, H_{U_c} (\psi_n-\psi_m)\ra 
            & = 
                \la \varphi, H_{n} \psi_n\ra 
                - \la \varphi, H_{m} \psi_m\ra \\
            & \phantom{==} 
                - \la \varphi, W_{U_n} (\psi_n-\psi_m)\ra 
                + \la \varphi, W_{U_m} (\psi_n-\psi_m)\ra \\
            & = E_n\la \varphi,\psi_n\ra 
                - E_m\la \varphi, \psi_m\ra \\
            & \phantom{==} 
                - \la \varphi, W_{U_n} (\psi_n-\psi_m)\ra 
                + \la \varphi, W_{U_m} (\psi_n-\psi_m)\ra \\
            &\le \|\varphi\|\big(
                    |E_n|\|\psi_n\| +  |E_m|\|\psi_m\|
                 \big) \\
            & \phantom{==} 
                - \la \varphi, W_{U_n} (\psi_n-\psi_m)\ra 
                + \la \varphi, W_{U_m} (\psi_n-\psi_m)\ra \, .
      \end{split}
  \end{equation*}
  Choosing $\varphi=\psi_n-\psi_m$, using also \eqref{eq:form tiny 2} 
  to bound the terms $\la \psi_n-\psi_m, W_U(\psi_n-\psi_m)\ra$, and 
  the normalization $\|\psi_n\|=1$ we get 
  \begin{equation}\label{eq:form domain convergence 1}
      \begin{split}
          \la \psi_n-\psi_m, & H_{U_c} (\psi_n-\psi_m)\ra   \\
            & \le \big(|E_n| +  |E_m|\big) \|\psi_n-\psi_m\| \\
            &\phantom{==}     
                 + (2U_c-U_n-U_m)
                  \big( 
                    \delta\|\nabla(\psi_n-\psi_m)\|^2
                    + b_\delta\|\psi_n-\psi_m\|^2   
                  \big) \, .
      \end{split}
  \end{equation} 
  On the other hand, due to \eqref{eq:form tiny 1} we also have 
  the lower bound 
  \begin{equation*}
      \begin{split}
         \la \psi_n-\psi_m, H_{U_c} (\psi_n-\psi_m)\ra 
            \ge (1-\delta) \|\nabla(\psi_n-\psi_m)\|^2
                 - b_\delta \|\psi_n-\psi_m\|^2 
      \end{split}
  \end{equation*} 
  and using this in \eqref{eq:form domain convergence 1} and rearranging 
  terms one gets
  \begin{align*}
      \big(1-\delta(1+(&2U_c-U_n-U_m)\big)  \|\nabla(\psi_n-\psi_m)\|^2 \\
      &\le 
        (|E_n| + |E_m|) \|\psi_n-\psi_m\|
         + b_\delta(1+2U_c-U_n-U_m) \|\psi_n-\psi_m\|^2
        \, .
  \end{align*}
  Since $\delta<1$  and $U_n,U_m\to U_c$, this shows 
  \begin{align}
      \limsup_{n,m\to\infty} \|\nabla(\psi_n-\psi_m)\|^2 
      \lesssim \limsup_{n,m\to\infty}\|\psi_n-\psi_m\| =0 
  \end{align}
  since $\psi_n$ converges to $\psi_c$ strongly in $L^2$. 
  Thus $\psi_n$ is also Cauchy in $H^1(\R^6)$, hence it converge strongly in $H^1(\R^6)$. 
  This finishes the proof of \ref{prop:basic properties}.\ref{claim 2}. 

\smallskip

  Having the existence of $\psi_U$ at $U=U_c$, the same argument 
  which proved \eqref{eq:supremum-1} can be used to prove 
  \eqref{eq:basic-properties uniform Lp bound}.   
  Continuity of $\psi_U$ in $U<U_c$ with respect to the $L^2$ norm 
  follows from perturbation theory. We will give an alternative proof, which also yields continuity in $H^1$ in the whole interval of parameters $U$ for which a ground state exits.   
  
  Take an arbitrary $U_0\in [0,U_c]$ and any  sequence $U_n\in [0,U_c]$  
  converging to $U_0$. Let 
  $U^1_j= U_{n_j}$ be an arbitrary subsequence. Then the above argument shows that this sequence is tight. Since $\|\psi_{U^1_j}\|=1$  there 
  exists a further subsequence $U^2_k= U^1_{j_k}$ such that 
  $\psi_{U^2_k}$ converges weakly in $L^2$ to $\psi_{U_0}$, which is the 
  unique ground state of $H_{U_0}$. By tightness, 
  $\psi_{U^2_k}$ converges strongly in $L^2$ and the above argument then shows that it also converges strongly in $H^1$.  
  
  Thus any subsequence of 
  $\psi_{U_n}$ with $U_n\to U_0\in [0,U_c]$ has a 
  further subsequence which converges to the same 
  limit $\psi_{U_0}$, i.e., $\psi_{U_n}$ converges in $H^1$ to $\psi_{U_0}$ for any sequence $U_n\to U_0$.   
  Hence $[0,U_c]\ni U\mapsto \psi_U $ is continuous in $H^1$.

  The Coulomb potential 
  $V_U$ is in the Kato--class and it is continuous in $U$ in the Kato--norm (for definitions, see \cite[Section A.2]{Simon-semigroups}).  
  Thus \cite[Theorem B.10.1]{Simon-semigroups} shows that as operators from $L^2$ to $L^q$  
  \begin{align*}
  	\|e^{-tH_U}- e^{-tH_{U_0}}\|_{2\to q}\to 0
  \end{align*}
  as $U\to U_0$ for any fixed $t>0$ and $2\le q\le \infty$. 
  Define $\wti{H}_U= H_U-E_U$. Then continuity of $E_U$ in $U$ 
  shows that also  
  $\|e^{-t\wti{H}_{U'}}- e^{-t\wti{H}_{U}}\|_{2\to q}\to 0$ as 
  $U'\to U$  for any  fixed $t>0$ and $2\le q\le \infty$. 
  Using $\wti{H}_U\psi_U=0$,  we have $\psi_U= e^{-t\wti{H}_U}\psi_U$ 
  for all $U\le U_c$. Hence  
  \begin{align*}
  	\psi_{U'}-\psi_{U}  
  		&=  e^{-t\wti{H}_{U'}}\psi_{U'} -  e^{-t\wti{H}_U}\psi_U 
  			= \Big( e^{-t\wti{H}_{U'}} -  e^{-t\wti{H}_U}\Big)\psi_{U'} 
  				+   e^{-t\wti{H}_U}\big( \psi_{U'}-\psi_U\big)
  \end{align*}
  and with the $L^2$ normalization $\|\psi_U\|=1$ and the $L^2$ 
  continuity of $\psi_U$ in $U\in [0,U_c]$ we have
  \begin{align*}
  	\|\psi_{U'}-\psi_{U}\|_q  
  		&\le  \|e^{-t\wti{H}_{U'}} -  e^{-t\wti{H}_U}\|_{2\to q}  
  				+   \|e^{-t\wti{H}_U}\|_{2\to q} \| \psi_{U'}-\psi_U\| 
  			\to 0
  \end{align*}
  as $U'\to U$ for any fixed $t>0$ and $2\le q\le \infty$. 
  This finishes the proof of claim \ref{prop:basic properties}.\ref{claim 3}. 

  \smallskip
  Let $\psi_U$ be the strictly positive ground state which, as we just proved,  
  exists for any $ U\le U_c$. Since $V_U$ is in the Kato--class, 
  any eigenfunction of $H_U$ is continuous 
  \cite[Theorem B.3.1]{Simon-semigroups}. So for any 
  compact set $K\subset\R^6$  and each $0\le U\le U_c$ 
  \begin{align}
  	c(K,U)\coloneqq \inf_{x\in K} \psi_U(x)>0 \, .
  \end{align}
  By continuity of the $L^q$ norms in $U$, there exists an open set 
  $O_U\subset\R$ such that 
  \begin{align}
  	\|\psi_{U'}-\psi_U\|_\infty<\frac{1}{2} c(K,U)
  	\quad \text{for all } U'\in [0,U_c]\cap O_U\, . 
  \end{align}
  Thus 
  \begin{align*}
  	\inf_{x\in K} \psi_{U'}(x) 
  		\ge  
  			\inf_{x\in K} \psi_U(x) - \|\psi_{U'}-\psi_U\|_\infty
  			\ge \frac{1}{2} c(K,U)
  \end{align*}
  for all $U'\in [0,U_c]\cap O_U$. Clearly, since $U\in O_U$, the open sets 
  $O_U$ cover   
  $[0,U_c]$, i.e., 
  $[0,U_c]\subset \cup_{U\in[0,U_c]}O_U$.  
  Since $[0,U_c]$ is compact, there exist finitely many $O_j= O_{U_j}$, $j=1,\ldots, N$ such that $[0,U_c]\subset \cup_{j=1}^N O_j$. 
  So with $c_j(K)= c(K,U_j)$, we get  
  \begin{align*}
  	\inf_{0\le U\le U_c}\inf_{x\in K} \psi_{U}(x) 
  		\ge  
  			\frac{1}{2} \min_{j=1,\ldots, N} c_j(K)\eqqcolon C_K >0\, ,
  \end{align*}
  which proves the last claim. 
\end{proof}

\end{appendices}

\bibliography{references}

\end{document}